\documentclass[12pt,preprint]{aastex}

\slugcomment{Submitted to ApJ}

\shorttitle{Linking Star Formation to Its Gas Reservoir:  Sampling Effects}
\shortauthors{Calzetti et al.}

\begin{document}

\title{Star Formation Laws: the Effects of Gas Cloud Sampling}

\author{D. Calzetti\altaffilmark{1}, G. Liu\altaffilmark{1,2}, J. Koda\altaffilmark{3}}

\altaffiltext{1}{Dept. of Astronomy, University of Massachusetts, Amherst, MA 01003; calzetti@astro.umass.edu}
\altaffiltext{2}{Dept. of Physics and Astronomy, The Johns Hopkins University, Baltimore, MD}
\altaffiltext{3}{Dept. of Physics and Astronomy, State University of New York at Stony Brook, New York, NY}

\begin{abstract}
Recent observational results indicate that the functional shape of the spatially--resolved star~formation--molecular~gas~density relation depends on the spatial scale considered. These results may indicate a fundamental role of sampling effects on scales that are typically only a few times larger than those of the largest molecular clouds. To investigate the impact  of this effect, we construct simple models for the distribution of molecular clouds in a typical  star-forming spiral galaxy, and, assuming a power--law relation between SFR and cloud mass, explore a range of  input parameters. We confirm that the slope and the scatter of the simulated SFR-molecular gas surface density relation depend on the  size of the sub-galactic region considered, due to stochastic sampling  of the molecular cloud mass function, and the effect is larger for steeper relations between  SFR and molecular gas. There is a general trend for all slope values to tend to $\sim$unity for  region sizes larger than 1--2~kpc, irrespective of the input SFR--cloud relation. The region size of 1--2~kpc corresponds to  the area where the cloud mass function becomes fully sampled. We quantify the effects of selection biases in data tracing the SFR, either as thresholds (i.e., clouds smaller than a  given mass value do not form stars) or backgrounds (e.g., diffuse emission unrelated to current star formation is counted towards the SFR). Apparently discordant observational results are brought into agreement via this simple model, and the comparison of our simulations with data for  a few galaxies supports a steep ($>$1) power law index between SFR and molecular gas.
\end{abstract}


\section{Introduction}

The relation that links star formation to its gas reservoir is at the foundation of the evolution of the baryonic 
component of galaxies across cosmic times. Hence, large efforts have been expended in trying to determine whether 
such a relation obeys some universal scaling that may, in turn, clue to its physical underpinning. 

Observational investigations of such relations in extragalactic environments were kick--started by a few influential works  
\citep[e.g.][]{Kennicutt1989,Kennicutt1998}, and have continued for over two decades. Major progress in the sensitivity and angular resolution of infrared  
imaging data have enabled a better handling of the dust attenuation in star formation rate indicators (SFRs), while 
increased sensitivity and mapping speed have started to yield better radio maps tracing the gas 
components of galaxies. These important advances have enabled the investigation of external galaxies on kpc or sub--kpc 
scales, and this new capability has brought new challenges. 

Evolving from the original work of \citet{Schmidt1959}, the relation between gas and star formation 
is often expressed as a power law:
\begin{equation}
\Sigma_{SFR}/ (M_{\odot}~yr^{-1}~kpc^{-2}) \propto [\Sigma_{gas}/(M_{\odot}~pc^{-2})]^{\gamma},
\end{equation}
where $\Sigma$ refers to the surface density of SFR and gas, and the value of $\gamma$ should be related to the 
main physical mechanisms that originate the relation itself \citep[e.g.,][]{Schmidt1959,Madore1977,Elmegreen1989,Kennicutt1989,Kennicutt1998,Tan2000,Krumholz2005}. 
The above equation and its variations are now customarily referred to as the Schmidt--Kennicutt Law (SK Law henceforth). 

Spatially--resolved investigations have established that the atomic (HI) gas component is usually not closely 
associated with regions of current star formation, but the denser molecular clouds, traced by CO, are  
{\citep{Wong2002, Kennicutt2007}. While there are valid considerations for including both atomic and  
molecular gas components when investigating the processes of star formation \citep{Boissier2003}, we will 
limit the analysis in this paper to the SFR--molecular~gas relation. 
Even when analyzing the direct relation between SFR and the molecular 
gas component, there are substantial differences among the results reported by several authors that warrant further analysis. 

To maintain 
clarity throughout this paper, we will term `Observed SK Law'  the power law relation:
\begin{equation}
\Sigma_{SFR}/ (M_{\odot}~yr^{-1}~kpc^{-2}) \propto [\Sigma_{H2}/(M_{\odot}~pc^{-2})]^{\gamma_{H2}}, 
\end{equation}
between the SFR and the molecular gas surface densities. 
The formal error about the best fit slope $\gamma_{H2}$ is generally a poor representation of the dispersion of the observational 
data about the best fit relation: the dispersion of the data is usually much larger than the uncertainty on $\gamma_{H2}$. It has become 
more and more common to report such dispersion as a separate measure; here we will term $\sigma_{H2}$ the r.m.s. scatter of the data 
(either observational or simulation--produced) about the best fit relation, along the $\Sigma_{SFR}$ axis in the log--log plot of equation~2. 

Azimuthally--averaged radial profiles of SFR and gas surface densities for a number of nearby galaxies have yielded values of $\gamma_{H2}$ 
in the range 0.8--2.2, depending on the galaxy, the extinction corrections applied to the SFRs, and other characteristics of the galaxies or 
the data \citep{Wong2002,Boissier2003,Heyer2004,Verley2010}, although the steeper values,  $\gamma_{H2}\gtrsim$1.2--1.4, tend to 
be more common. 

Spatially--resolved studies, where galaxy regions are divided into bins with sizes in the  range $\sim$180--2,000~pc, have yielded 
$\gamma_{H2}\sim$0.7--1.9    
\citep{Rownd1999, Kennicutt2007, Thilker2007, Bigiel2008, Blanc2009, Verley2010, Onodera2010, Momose2010,Liu2011,Rahman2011}. 
Discrepant results, however,  have been obtained even when using the same data on the same galaxy: in M51a, \citet{Kennicutt2007} 
and \citet{Bigiel2008} derive $\gamma_{H2}\sim$1.37 and 0.84, respectively. This discrepancy can be resolved when accounting for the 
different treatment of the SFR data by the two papers: while \citet{Kennicutt2007} remove the diffuse, 
low frequency emission from those data, \citet{Bigiel2008} do not. \citet{Liu2011} show that, after removal of the 
diffuse emission,  $\gamma_{H2}$ is a monotonically decreasing function 
of the spatial scale sampled in the two galaxies M51a and NGC3521. \citet{Liu2011} argue that the diffuse emission may be 
unrelated to current star formation. The exact values of $\gamma_{H2}$ 
 depend on the fitting method and the dynamical range and level of significance of the data included in the fits \citep{Blanc2009, Verley2010, Onodera2010}. The exponent of the Observed SK Law can be as large as $\sim$1.9--2.7 when measured in molecular 
clouds in the Milky Way over $\gtrsim$1~kpc scales \citep{Gutermuth2011}. This large variation may hint at a more fundamental relation of 
the denser components of the molecular gas with the SFR, which often correlate linearly \citep[$\gamma_{dense}\approx$1,][and references therein]{Gao2004,Lada2010}. The value of the power law index connecting the SFR surface density to the gas surface density may itself be 
a function of the type of gas tracer employed: high density tracers (e.g., HCN) will yield $\gamma\approx$1, while low density tracers (e.g., 
$^{12}$CO(1-0)) may be more likely to give $\gamma>$1 \citep{Narayanan2008,Juneau2009}. 

The large range displayed by observational measures of $\gamma_{H2}$ has prevented so far the pinning down of the underlying 
mechanism that drives star formation. While an exponent $\gamma_{H2}$=1.5 belies the dominance of gravitational instability as 
the main mechanism for star formation for constant galaxy gas scaleheight, other models predict lower ($\gamma_{H2}=$1) or higher ($\gamma_{H2}$=2) exponents 
\citep[e.g.,][]{Schmidt1959,Scoville1986,Wyse1989,Elmegreen1989,Elmegreen1994,Silk1997,Kennicutt1998,Tan2000,Tan2010,Wong2002,Krumholz2005}, which are still within the range of observed values. Understanding the origin of the wide 
range of $\gamma_{H2}$ values found in the literature will provide fundamental guidance to models of gas--to--star conversion \citep[e.g.][]{Elmegreen2002,MacLow2004,Krumholz2009,Monaco2011,Dobbs2011}.

Similarly important for understanding star formation, the spatial scale at which the Observed SK Law is no longer 
recovered would indicate the transition into a regime where the feedback from star formation or the differential motions between stars and 
gas become an important source of scatter. \citet{Kennicutt2007} and \citet{Thilker2007} noted that the scatter $\sigma_{H2}$ of the 
data about the best fit relation in the $\Sigma_{SFR}$--$\Sigma_{H2}$ plane decreases when the spatial scale sampled increases from 
$\sim$400--500~pc to 1--2~kpc. \citet{Onodera2010} and \citet{Momose2010} indicate break--down scales of 80~pc and 250~pc for M33 and 
NGC4303, respectively. \citet{Momose2010} recovers a SK Law on scales no smaller than $\sim$1~kpc. Finally, \citet{Liu2011} reports that 
$\sigma_{H2}$ is a monotonically decreasing function of the spatial sampling scale, between $\sim$200~pc and $\approx$1~kpc, in M51a and 
NGC3521. \citet{Feldmann2011} have investigated the impact of environmental parameters, such as the UV radiation field or the shielding 
offered by the presence of metals as a reason for a smooth trend of $\sigma_{H2}$ with spatial scale. 

The main difficulty of most spatially resolved investigations is to disentangle effects due to the underlying 
physical mechanisms from those of statistical variations due to the sampling and fitting procedures. 
Here we employ Monte Carlo simulations to explore the impact on both $\gamma_{H2}$ and $\sigma_{H2}$ of  
the sampling region's size as affected by the stochastic sampling of molecular clouds within the region, and 
other geometrical factors.  We will also investigate the impact of the fitting procedure and of the data dynamical 
range and censoring. These effects are generally driven by the small dynamical range of the molecular 
gas observations, typically spanning about 1 order of magnitude, implying that CO measurements 
reach low detection limits ($\sim$1~$\sigma$) at a level where measurements of SFR are still highly significant 
($>$3~$\sigma$). We will not be concerned with the intercept of the SK Law, which requires the knowledge of how each cloud, 
and with what efficiency,  forms stars. Our models are purposely very simple, so that the effect of each parameter and its 
variations can be quantified.

While we will attempt to account for general trends in observational data as published in the recent literature,  we will 
make only modest attempts to force an agreement between the data and our models by choosing ad--hoc parameters. We will, however,  
discuss the impact of each parameter on the observed trends and the direction of change that may produce an 
(or a better)  agreement. 

All logarithms in this paper are in base 10.

\section{Model Description}

We implement a very simple prescription for our model `galaxy', in terms of its molecular cloud content, using  
some standard results for the mass function and other properties of the clouds \citep[e.g., see summary in][]{Blitz2004}.  
We term this model our Default Model, and it will be our starting point for investigating stochastic and geometrical effects.

In this section we provide the scaling relations and assumptions that we have used in our Monte Carlo 
simulations. For the molecular clouds, the scaling relations are mainly those of our own Milky Way, which we consider 
an adequate assumption for many of the late--type, star--forming spirals published in the literature. 
We will also assume that our model galaxy is observed face--on, since many published results have 
been de--projected for inclination. 

The mass function of molecular clouds in the Milky Way and the less than a handful nearby galaxies that have been 
observed with sufficient resolution and mapping size can be described as:
\begin{equation}
dN/dM_{H2} \propto M_{H2}^{\alpha},
\end{equation}
with $\alpha\sim (-$1.5)--($-$2.9) \citep{Solomon1987,Heyer2001,Mizuno2001,Engargiola2003,Rosolowsky2005,Rosolowsky2007,Blitz2007,Wong2011,Gratier2011}. 
We choose a mean value $\alpha$=$-2$ as representative for our Default Model \citep[see][]{Gratier2011}; we will, however,  investigate later in this paper the effects 
of changing $\alpha$, since  the differences  in $\alpha$ 
from galaxy to galaxy appear to be real, and not driven by measurement uncertainties \citep{Rosolowsky2005}.

The mass of each cloud is related to its radius via the relation \citep{Solomon1987}:
\begin{equation}
M_{H2} / (3\times 10^4 M_{\odot}) = [R / (10~pc)]^{2},
\end{equation}
in the mass range 500--3$\times$10$^7$~M$_{\odot}$. Equation~4 is equivalent to a constant cloud surface density, and is 
one of Larson's Laws  \citep{Larson1981}. The clouds mean surface density of $\sim$100~M$_{\odot}$~pc$^{-2}$ implied by equation~4 is in the range observed for Milky Way clouds \citep{Heyer2009}. Higher values for the cloud mean surface density \citep[e.g.,][]{Solomon1987,Bolatto2008} 
would simply produce a 
rigid shift of the simulated data toward larger values of $\Sigma_{H2}$. We do not attempt to model the variations in cloud surface density observed by 
\citet{Heyer2009}, although these are likely to be real. The cloud mass range is chosen to encompass not only the observed 
range in our own Milky Way and other Local Group galaxies \citep[that have masses typically not larger than $\sim$3$\times$10$^6$~M$_{\odot}$, ][]{Wilson1990,Rosolowsky2005,Engargiola2003,Blitz2007}, but also to account for the potential presence of giant molecular associations with masses as large as 
10$^7$--10$^8$~M$_{\odot}$ in some star--forming galaxies 
\citep{Koda2009}. We will explore later the impact of lower high--mass cut--offs for the GMCs. The proportionality constant between the 
mass and radius of the GMC is derived from MW studies; changing it causes a change in the gas surface density of the Observed 
SK Law. For instance, decreasing the radius by 10\% at constant cloud mass causes the final gas surface densities to be on average  
20\% higher. Our Default model includes a $\sigma_{R}\sim$60\% gaussian dispersion in the mass--radius relation, consistent with 
what has been observed in the MW and neighboring galaxies \citep{Blitz2004}.  

Since we do not model the gas density profile within each cloud, we cannot follow the customary approach of relating the density of the SFR to the density of the gas \citep[e.g.,][]{Schmidt1959,Wyse1989,Kennicutt1998}. However, a power law relation between SFR and gas densities implies a power law relation between SFR and 
cloud mass. This is immediately seen for a linear correlation between $\rho_{SFR}$ and $\rho_{H2}$, i.e., $\rho_{SFR}\propto\rho_{H2}$ implies SFR$\propto$M$_{H2}$. 
For power--law correlations with slope $>$1, we seek guidance from observational data.  
For the clouds listed in \citet{Heiderman2010}, we derive $\rho_{SFR}\propto\rho_{H2}^{1.7}$ and SFR$\propto$M$_{H2}^{1.3}$ for simple 
assumptions of the clouds' geometry. Furthermore, a direct relationship between SFR and dense (n(H$_2$)$>$10$^4$~cm$^{-3}$) gas mass appears to be implied by data on 
molecular clouds within 
0.5 kpc of the Sun \citep{Lada2010}.  We thus relate each cloud to a star formation rate via the relation:
\begin{equation}
SFR/(M_{\odot}~yr^{-1})  \propto (M_{H2}/M_{\odot})^{\beta},
\end{equation}
where we will test cases with $\beta$=1.0, 1.5, and 2.0. The normalization for equation~5 is arbitrary in our model, and we use the results 
of \citet{Kennicutt2007}, \citet{Bigiel2008} and \citet{ Liu2011} to set a reasonable range for the observed SFR surface density.  In 
particular, for $\beta$=1, we assume the star formation efficiency, SFR/M$_{H2}$=5.25$\times$10$^{-10}$~yr$^{-1}$ \citep{Leroy2008,Bigiel2011}. 

The exponent $\beta$ of equation~5 is not the 
same as $\gamma_{H2}$ in the SK Law, the latter being the result of area--averages in both gas and SFR surface densities. 
In order to ensure some variation in the SFR--gas~mass relation above, we insert an artificial dispersion given by a gaussian distribution 
with $\sigma(log SFR)=$0.3~dex in equation~5, as a starting assumption. This means that any cloud with expected SFR as given by equation~5 will be  attributed a value between roughly 1/2~SFR and 2~SFR (1~$\sigma$). The actual dispersion in the SFR--cloud~mass relation 
is basically unknown at this stage, although \citet{Enoch2007} find about a factor 3 dispersion in the dense core fraction of three Milky Way molecular 
clouds, thus justifying our $\pm$2x choice. \citet{Lada2010} reports the case of two nearby molecular clouds with similar gas masses and a factor 10 
difference in SFR; this case corresponds to a 3~$\sigma$--5~$\sigma$ event for our parameter choice. We will later explore the impact of our assumption. 

The observed SFR will be assumed to be extinction--free, and we do not introduce effects of dust attenuation in our 
simulations. This is reasonable, especially as more recent studies have employed SFR indicators that combine optical/UV and infrared 
tracers, and should therefore provide a fully extinction--corrected view of star formation in galaxies \citep{Kennicutt2007,Bigiel2008,
Verley2010}.

We simulate our `sub--galactic regions' as square areas with sides in the range 200~pc--5~kpc, thus covering the majority of sizes probed in 
the literature \citep{Kennicutt2007,Bigiel2008,Blanc2009,Verley2010,Onodera2010,Momose2010}. Although we probe sizes as small as  
200~pc, we should note that the radius of the largest cloud in our model is about 300~pc (and the smallest is a little over 
1~pc in radius, equation~4). However, the largest measured CO cloud radii are smaller 
than 100~pc both in the Milky Way and M33 \citep{Wilson1990,Engargiola2003,Rosolowsky2003}.

The boundary for the smallest region size, 200~pc, in our simulations aims at avoiding 
stochastic sampling effects of the stellar Initial Mass Function (IMF). The total mean H$_2$ mass in a  square area of 200~pc side 
obtained from our simulations is around 2.5$\times$10$^6$~M$_{\odot}$, corresponding to a 
SFR$\sim$1.3$\times$10$^{-3}$~M$_{\odot}$~yr$^{-1}$ for the star formation efficiency given above. This  SFR corresponds 
to the minimum value below which the stochastic sampling of the IMF significantly affects the ionizing photon rate of a stellar 
cluster, and, therefore, any SFR indicator based on the ionizing photon rate \citep[e.g., extinction--corrected 
H$\alpha$,][]{Lee2009, Corbelli2009}. Another way to look at this limit is to recall that the above SFR produces a 
$\sim$1.3$\times$10$^4$~M$_{\odot}$ stellar 
cluster over a 10$^7$~yr timescale. This timescale is roughly the crossing time of our smallest region, for stellar velocities 
$\gtrsim$10~km~s$^{-1}$, and is equivalent to or 
smaller than a typical cloud lifetime \citep[$\sim$a few~$\times$10$^7$~yr,][]{Hartmann2001,Scoville2004,Kawamura2009}.  
For a $\approx$10$^4$~M$_{\odot}$ mass in stars, the scatter on the ionizing photon rate due to stochastic IMF sampling is 
around 20\% \citep{Villaverde2010}, for a Kroupa IMF \citep{Kroupa2001}. If the UV stellar continuum is used to trace the SFR, the limiting 
SFR below which stochastic sampling becomes an issue is about 3 times lower  than for H$\alpha$ \citep{Lee2011}. In this case, 
however,  there may be concerns about UV stellar continuum tracing star formation over timescales $\gtrsim$100~Myr, which may be longer than 
the lifetimes of molecular clouds \citep{Hartmann2001,Elmegreen2007,Kawamura2009}.

We simulate covering factors for each area  drawn randomly in the range cf$_{min}$--cf$_{max}$, in order to approximate the range of molecular 
cloud covering factors in the disks of star--forming galaxies as a function of location (arm/inter--arm regions) and 
galactocentric distance (central and outer regions). This is in agreement with the interpretation by \citet{Leroy2009} that variations in molecular 
gas surface density over large areas are an indication of varying covering factors of the area by clouds, if the cloud surface densities are roughly 
constant \citep[see, however, \citet{Heyer2009}]{Solomon1987}.  In addition to a constant distribution of covering factors between cf$_{min}$ 
and cf$_{max}$, we attempt to simulate the observed larger frequency of low covering factors \citep{Leroy2009} via the exponential distribution:
\begin{equation}
N_{reg}(cf) = 74 \times e^{(-4.3\times cf)},
\end{equation}
where $N_{reg}$ is the number of regions with covering factor cf drawn within the range cf$_{min}$--cf$_{max}$. With 
equation~6, we simulate about 50 times more regions with cf=0.1 than regions with cf=1.0. Although the exact ratio between these two 
extremes is not characterized yet (and also depends on the region's size), we adopt the uniform distribution of covering factors and the distribution 
given by equation~6 as a reasonable bracket for conditions in external galaxies, where areas of low surface density are more common than those 
of high surface density.   

Electing to fill the area, rather than a volume (i.e., by filling the thickness of the 
molecular disk with clouds), is equivalent to assuming that molecular clouds form a single layer in a face--on galaxy. This is a reasonable 
assumption in light of results for the low--inclination Large Magellanic Cloud \citep{Fukui2010} and of the fact that the mean thickness of the 
MW molecular disk is only about 120~pc \citep{Scoville1987}. 
For completeness, however, we will also explore the case in which volumes with given area and thickness of 120~pc  are populated with clouds, 
with filling factors randomly drawn within a given range ff$_{min}$--ff$_{max}$. For the volume filling, we assume that SFR tracers can 
penetrate  the entire thickness of the molecular disk. From an operational point of view, the detection of recent star formation 
either inside a molecular cloud or through a series of clouds will likely require use of a far--infrared 
SFR indicator, thus requiring that the molecular clouds are globally (as opposed to locally) optically thin to far--infrared light. This may 
occur if the optically thick core of the clouds has a small filling factor. 

The baseline parameters outlined above form our Default Model. Our Monte Carlo simulations generate a random cloud in the 
mass range specified in equation~4 and with the mass distribution of equation~3. The cloud is then assigned  a radius and a SFR, as 
specified in equations~4 and 5, which then get randomly scattered according to Gaussian distributions with $\sigma_R$ and $\sigma_{SFR}$, respectively. The area of the cloud is subtracted from a pre--specified  area to which a random covering factor cf, either with a uniform distribution or with a distribution function described by equation~6, 
in the range cf$_{min}$--cf$_{max}$ has been attributed. Clouds are generated until cf$\times$area is fully covered. The surface density of H$_2$ mass and 
SFR is finally calculated by dividing the total H$_2$ cloud mass and total SFR in each area bin by the area itself. In keeping with standard 
conventions, we express the SFR surface density $\Sigma_{SFR}$ in units of M$_{\odot}$~yr$^{-1}$~kpc$^{-2}$ and the H$_2$ mass surface density $\Sigma_{H2}$ in units of M$_{\odot}$~pc$^{-2}$ \citep{Kennicutt1989}. Additionally, scatter is added to the gas surface densities, 
to simulate measurement errors; we set the scale such that $\Sigma_{H2}\sim$6.5~M$_{\odot}$~pc$^{-2}$  is detected at the 3~$\sigma$ 
level, and realizations with lower $\Sigma_{H2}$ are not included in the fits.  The boundary at $\Sigma_{H2}\sim$6.5~M$_{\odot}$~pc$^{-2}$  is selected 
to ensure a dynamical range along the $\Sigma_{H2}$ axis comparable to that of most observational data, which is usually one order of magnitude or more. 
In the next section, we evaluate the impact of this choice against lower or higher detection limits,  and smaller or larger dynamical ranges. We 
do not add an analogous scatter component to 
$\Sigma_{SFR}$, since the SFRs in galaxies are generally measured with significantly higher confidence levels than $\Sigma_{H2}$. 

For each set of parameter 
choices, 10,000--20,000 realizations (areas) are generated, to avoid shot noise effects. The Observed SK Law, equation~2, is then derived by applying an ordinary least--square (OLS) bi--sector linear fitting \citep{Isobe1990} to the pairs of $\Sigma_{SFR}$--$\Sigma_{H2}$ realizations 
in log--log space, thus treating our simulations in a similar fashion as most actual data. We recall that this method produces the same results, 
within the uncertainties, when fitting X versus Y or Y versus X. In section~3.3, we evaluate the results of the 
OLS bi--sector linear fitting against another often--used fitting method:  the bi--linear regression fit \citep[e.g., routines 
like FITEXY in the Numerical Recipes,][]{Press2007}. The latter method, by including uncertainties in both axis, produces also similar results in the X--Y and Y--X planes.  The results from the Default Model are 
reported in the section~3.1, followed by an analysis of the changes induced on both $\gamma_{H2}$ and $\sigma_{H2}$ 
by variations in the parameters of the Default Model.  

\section{Analysis}

\subsection{The Observed SK Law for the Default Model}

The results of our Default Model simulations are shown in Figures~\ref{fig1}-- \ref{fig3}, where we have chosen cf$_{min}$=0.10 
and cf$_{max}$=1.0. Figures~\ref{fig1} and the top panels of  Figure~\ref{fig3} report the results for a uniform distribution in covering factors, while Figures~\ref{fig2} and 
the bottom panels of Figure~\ref{fig3} report the same results obtained for the covering factor distribution of equation~6. 
In Figures~\ref{fig1} and \ref{fig2}, the three panels show the simulated data and the best fit lines in the Log($\Sigma_{H2}$)--Log($\Sigma_{SFR}$) 
plot, for selected region sizes in the range 200--1,000~pc, and for the three $\beta$=1.0, 1.5, and 2.0. There is 
an obvious increase in both the slope and the dispersion perpendicular to the mean of the scaling relations, 
going from shallow ($\beta$=1) to steeper ($\beta$=2) SFR--gas~cloud~mass relations (equation~5). A more subtle 
effect is observed within each panel, with both slope and dispersion about the mean decreasing from small to large region sizes.
These trends are more evident in Figure~\ref{fig3}, where the measured values of $\gamma_{H2}$ and $\sigma_{H2}$ are shown 
as a function of region's size from 200~pc to 5~kpc. In all cases, a 
detection `limit' at $\Sigma_{H2}=$6.5~M$_{\odot}$~pc$^{-2}$  (Log($\Sigma_{H2}$)=0.8) is applied prior to fitting the data with a straight line, again to simulate standard 
measurement approaches that typically remove data below a set detection limit. With this selection, our simulated data span a dynamical 
range of $\sim$1--1.5 orders of magnitude in $\Sigma_{H2}$, similar to the ranges measured in star--forming galaxies 
\citep[e.g.][]{Kennicutt2007,Bigiel2008,Blanc2009,Onodera2010,Momose2010,Liu2011}. As expected from the choice of a common 
low--end detection limit at all region sizes, the largest dynamical range, $\sim$30X, is spanned at the smallest region size and the range decreases 
down to about one order of magnitude at the largest size (Figures~\ref{fig1} and \ref{fig2}).  Larger dynamical ranges at fixed $\Sigma_{H2}$ detection  
limit can be obtained by requiring the cloud mean surface density to be higher than our selected value of 100~M$_{\odot}$~pc$^{-2}$ \citep[see, e.g.,][]{Solomon1987,Bolatto2008}.

At constant $\beta$, we observe a variation in the {\em measured} slope $\gamma_{H2}$: it  
decreases as a function of increasing  sampling region's size (Figure~\ref{fig3}). For instance, for $\beta$=2 and a uniform distribution of 
covering factors, the measured slope goes from $\gamma_{H2}\sim$2.6 at 200~pc down to $\gamma_{H2}\sim$1.1 in a 5~kpc size region 
(top--left panel of Figure~\ref{fig3}). A similar trend is observed for a distribution of covering factors as given by equation~6 (bottom--left panel of 
Figure~\ref{fig3}). Thus, the values of $\gamma_{H2}$
do not reproduce $\beta$, for $\beta>1$. The rate of change, 
however, decreases for decreasing $\beta$, and at $\beta$=1 the measured slope changes from $\gamma_{H2}\sim$1.05--1.1 at 200~pc 
to $\gamma_{H2}\sim$1.0 at 5~kpc, hence providing a closer estimate of the actual value of $\beta$ at virtually all sizes. 

The main effect of changing the distribution of cloud covering factors from uniform to exponentially--weighted towards low values is to increase the 
density of points at $\Sigma_{H2}\lesssim$20~M$_{\odot}$~pc$^{-2}$ (Figure~\ref{fig2}). This provides a closer resemblance to observational 
relations \citep[e.g.][]{Liu2011}: once the SFR dynamical range in data and simulations is matched, the {\em mean} SFR value in the simulations 
is also closer to what measured in actual galaxies (section~4.2). 
However, the effect on the measured slope $\gamma_{H2}$ and scatter $\sigma_{H2}$ is small, although it is 
in the direction of systematically increasing both quantities (Figure~\ref{fig3}); for instance, the slope increases by 0.04--0.14, with the exact value 
depending on both $\beta$ and the region's size,  but always corresponding to less than 6\% variation.

A trend that will remain common to the Default Model and any modification that will be analyzed in the next sections is the tendency for the 
measured slope $\gamma_{H2}$ to converge to a value of $\sim$1 at large sampling sizes, independently of the value of $\beta$. This is 
due to the fact that at large region sizes the cloud function will be fully sampled even for small values of the covering factor, thus 
decreasing the contrast from region to region, and flattening the Observed SK Law  to a linear correlation. 

Similarly to $\gamma_{H2}$, the dispersion of the data about the mean trend, $\sigma_{H2}$, is a function of both $\beta$ and the sampling 
size (Figure~\ref{fig3}, right panels). It increases for increasing $\beta$ and decreases for increasing size. The increase with $\beta$ reflects 
the larger impact  that stochastic sampling has on the larger, more strongly star--forming, clouds at small regions sizes. The decrease with increasing 
sampling size mirrors the same considerations made for $\gamma_{H2}$ above: as the region increases, the cloud mass function becomes fully sampled at all covering factors, and the scatter decreases considerably.

In order to quantify our statements about the stochastic sampling of the cloud mass function, Figure~\ref{fig4} shows the mean number 
of clouds and the mean cloud mass as a function of region size, for the exponentially decreasing cloud covering factors.  A uniform 
distribution of covering factors gives similar results. 
These are independent of $\beta$, since they only concern the molecular clouds 
and the covering factor of each region. Use of mean values is appropriate, since the cloud mass and number distributions in each region are almost symmetric about a peak value, and the latter is close to the mean value (Figure~\ref{fig4}, bottom panel). The top--left panel 
show that while the mean number of clouds clearly increases for increasing region size, the trend slightly (by 50\% or less) exceeds the expectation 
of cloud numbers increasing as the square of the region size, except for the largest sizes probed, i.e., regions $\gtrsim$1--2~kpc in size. Similarly, 
the mean cloud mass increases,  also by a maximum of 50\%,  as a function of region size, and tends to level off only for regions $\gtrsim$1--2~kpc in size. This shows that the cloud 
mass function we implement is fully sampled only for regions larger than $\sim$1--2~kpc, and stochastic sampling is an important factor for smaller regions. 

In the rest of the paper, we will preeminently report results based on the Default Model with an exponentially decreasing distribution of cloud 
covering factors (equation~6), for the following two reasons: (1) all other parameters being equal, the two distributions, uniform and exponential, produce 
generally small differences in $\gamma_{2}$ and $\sigma_{H2}$; (2) equation~6 likely better approximates the conditions in external galaxies. 
In fact, for an exponential distribution of cloud covering factors, we derive a mean SFR from  
the simulations which is within 50\% of the mean value measured in the nearby galaxy M51a \citep{Kennicutt2007}, after matching the SFR dynamical range values between model and observations (e.g., Figure~\ref{fig2}). Conversely, a uniform distribution of 
cloud covering factors  yields mean SFR values that are a factor $\gtrsim$2 too high relative to what measured in data. This provides 
a justification for using equation~6 in what follows, although the uniform case will be discussed in parallel whenever relevant.

\subsection{Scatter in the SFR--Cloud~Mass Relation}

We test  the impact of our choice of $\sigma(log SFR)$=0.3 (see section~2), by running our simulations again, but with 
two more extreme values for the gaussian standard deviation: $\sigma(log SFR)$=0 (i.e., zero scatter, implying a deterministic relation between SFR and cloud mass) 
and $\sigma(log SFR)$=0.6 (i.e., a factor 4 variation in SFR at fixed cloud mass). Figure~\ref{fig5} shows the changes in $\gamma_{H2}$ 
and $\sigma_{H2}$ as a function of both $\beta$ and region size for these two choices of $\sigma(log SFR)$. The general trend for 
$\gamma_{H2}$ to decrease as a function of increasing region size, and converge towards values $\approx$1, persists even for these other 
choices of the scatter between SFR and cloud mass. Most importantly, the trend persists even when there is no scatter in the relation between
SFR and cloud mass. The measured slope also steepens for increasing $\sigma (log SFR)$, at constant $\beta$ and region size; 
 for instance, in the case of $\beta$=1 and size=300~pc, the measured slope goes from $\gamma_{H2}\sim$1 to 1.2 for $\sigma (log SFR)$ that 
increases from 0 to 0.6. Increasing the scatter in the relation between SFR and cloud mass (equation~5), thus, causes the slope to become steeper, 
and this result persists at all `detection limits' in $\Sigma_{H2}$ and even in the absence of a limit. The steepening is the result of the interplay between 
$\sigma(log SFR)$ and the uncertainties along the x--axis ($\Sigma_{H2}$). 

The measured scatter $\sigma_{H2}$ decreases for increasing region size, similarly to what is observed in Figure~\ref{fig3}, even when varying 
the value of $\sigma (log SFR)$.  Interestingly, non--negligible values of $\sigma_{H2}$ are measured also for $\sigma (log SFR)$=0. 
This scatter  purely reflects the fact that, for most of the regions we consider, the cloud mass function is stochastically sampled, rather than 
fully sampled, especially when accounting that each region is filled with clouds according to a randomly selected covering factor between 
10\% and 100\%. This effect is exacerbated  for increasing values of $\beta$, since higher weight (i.e., SFR) is given to the 
large clouds that are most subject to the effects of stochasticity. 

The above experiment highlights an important degeneracy: $\gamma_{H2}$ and $\sigma_{H2}$ can increase both by increasing 
the slope $\beta$ and by increasing the scatter in the SFR--cloud~mass relation. However, $\sigma_{H2}$ increases  fractionally more than 
$\gamma_{H2}$ when scatter increases (compare Figure~\ref{fig3}, right panels, with Figure~\ref{fig5}), 
which can provide a useful discriminant for observational data. 

\subsection{The Origin of the Scatter in the Observed SK Law}

We have seen in the previous section that the Observed SK Law preserves a measurable scatter even in the absence of intrinsic 
scatter in the relation between SFR and cloud mass (equation~5). Here we investigate the source of the scatter in our simulations. 
To fix ideas, we will concentrate on the case with $\beta$=1.5 and region size 200~pc. As before, we consider the case of an  exponential 
distribution of cloud covering factors. For $\sigma (log SFR)$=0, the Observed SK 
Law in this case has $\gamma_{H2}$=1.76 and $\sigma_{H2}$=0.49. Throughout this section, the SFR--mass relation (equation~5) is 
set to zero dispersion. 

We begin our test by removing all sources of dispersion from the simulations, including those related to each region's covering factor, 
to the mass--radius relation, and to the simulated scatter in the $\Sigma_{H2}$ measurements. We fill each region to 100\% of the area, 
i.e., cf$_{min}$=cf$_{max}$=1. We also remove any detection limit in the $\Sigma_{H2}$ generated distribution. However, we are still 
randomly sampling the clouds within each region, according to the distribution of equation~3, with our default choice of $\alpha$=$-$2. 
With these assumptions, we generate an Observed SK Law with a negligible dynamical range in $\Sigma_{H2}$, of about 0.02 dex in 
logarithm scale, centered around log($\Sigma_{H2})\sim$2. This is a direct consequence of equation~4; since the relation between 
cloud mass and radius implies a constant cloud surface density, filling the entire area with clouds (cf=1) produces bins with the same 
$\Sigma_{H2}$ values. Despite this, we already observe a dispersion of $\sigma_{H2}$=0.33 
along the $\Sigma_{SFR}$ axis, due to the combination of the non--linear relation between SF and M$_{H2}$ and the random sampling of the 
cloud mass distribution within each region. The combination of the two effects is indeed crucial for the dispersion. The same simulation, 
but with $\beta$=1 instead of 1.5, produces a distribution of points that has the same negligible dynamical range,  
and $\sigma_{H2}$=0, i.e., zero scatter along the $\Sigma_{SFR}$ axis. 

We now proceed by adding one ingredient: random covering factor between cf$_{min}$ and cf$_{max}$. We keep cf$_{max}$=1 and 
vary cf$_{min}$ between 0.05 and 0.4. In this case, the main effect is an increase in the dynamical range of the simulated data: at cf$_{min}$=0.05, log($\Sigma_{H2}$) spans the range $\sim$0.7--2, decreasing to 1--2 for cf$_{min}$=0.1, and progressively decreasing down to 
1.6--2 for cf$_{min}$=0.4. For cf$_{min}$=0.05 to 0.15 the OLS bi--sector fit gives a slope $\gamma_{H2}\sim$1.75$\sim\beta$ within the 
1~$\sigma$ measurement error. For cf$_{min}>$0.15 the fitted slope becomes steeper, owing to the decreasing dynamical range 
spanned by the data. However, the dispersion about the best fitting line remain virtually unaffected, with $\sigma_{H2}$=0.35. The direct connection  
between dynamical range in $\Sigma_{H2}$ and covering factor is again a direct consequence of equation~4 and the constant cloud surface 
density it implies. 

The dispersion remains similarly unaltered when we fix cf$_{min}$=0.10 and vary cf$_{max}$ in the range 0.60--1. Variations in the maximum 
value of the covering factor impact the high--end values of $\Sigma_{H2}$, which decrease from log($\Sigma_{H2}$)$\sim$2 for cf$_{max}$=1 
down to log($\Sigma_{H2}$)$\sim$1.8 for cf$_{max}$=0.60. Thus, the main impact of the covering factor variation is on the dynamical range 
spanned by the `observed' $\Sigma_{H2}$. 

Keeping cf$_{min}$ and cf$_{max}$ at the Default Model values of 0.1 and 1, respectively, and adding scatter in the cloud mass--radius relation 
\citep[1~$\sigma$=60\%, similar to what is observed in actual MW molecular clouds,][]{Blitz2004} further increases the dynamical range covered 
by the simulated data, from one order of magnitude to a factor $\gtrsim$60 in $\Sigma_{H2}$ (Figure~\ref{fig6}).  Both slope and dispersion remain basically unchanged, at $\gamma_{H2}\sim$1.73 and  $\sigma_{H2}\sim$0.35. 

Re--adding into the simulations the measurement uncertainties for $\Sigma_{H2}$, so that values at  $\Sigma_{H2}\sim$6.5~M$_{\odot}$~pc$^{-2}$ are detected at the 3~$\sigma$ level, changes the $\Sigma_{SFR}$--$\Sigma_{H2}$ distribution in more than one way. The 
dynamical range increases further relative to the previous cases, but almost exclusively in the direction of small $\Sigma_{H2}$ values 
(Figure~\ref{fig6}), down to $\Sigma_{H2}$=0.4~M$_{\odot}$~pc$^{-2}$. This reflects the fact that lower $\Sigma_{H2}$ values are detected 
at increasingly lower confidence. Furthermore, the distribution becomes markedly non--symmetric relative to the best fit line, reflecting 
the non--symmetric nature of scatter in a logarithmic plot. Finally, $\gamma_{H2}$ decreases significantly, down to $\gamma_{H2}$=1.46 and 
$\sigma_{H2}$ increases to its final value of 0.49. When we add a detection limit to $\Sigma_{H2}$, $\gamma_{H2}$ increases back to 
the value 1.76  (Figure~\ref{fig5}, top--left), while $\sigma_{H2}$ remains unchanged (Figure~\ref{fig5}, top--right).

In summary, in the absence of scatter between SFR and cloud~mass, there is still considerable scatter in the data about the best fitting line 
in the $\Sigma_{SFR}$--$\Sigma_{H2}$ plane, with $\sigma_{H2}\sim$0.5 at 200~pc for $\beta$=1.5. About  60\% of this scatter is 
due to the combination of the non--linear relation between SFR and M$_{H2}$ and stochastic sampling of the cloud mass function. This effect produces a little over  80\% of the scatter $\sigma_{H2}$ for the $\beta$=2.0 case (Figure~\ref{fig5}, top--right). 
The remaining $\sim$40\% of the scatter for the $\beta$=1.5 case is due to our modeled measurement 
uncertainties; the behavior of this second portion of the scatter reflects the non--symmetric nature of errors in logarithmic scale. For $\beta$=1 
only  the latter portion of the scatter is present, implying that the values $\sigma_{H2}\lesssim$0.2 for $\beta$=1 in Figure~\ref{fig5} (top--right) are 
due mainly to the effects of measurement uncertainties along the $\Sigma_{H2}$ axis. 

\subsubsection{Volume Filling versus Area Covering}

The relative thinness of molecular disks implies that usually no more than 1--2 clouds overlap along a given line of sight for a face--on disk 
\citep{Scoville1987,Kawamura2009}. Furthermore, CO observations can often separate multiple clouds along a line of sight via 
the different velocity imprints of separate clouds. Although this helps justify the use of a bi--dimensional model for our simulated `galaxy', it 
is worth investigating the general trend obtained in the case of a three-dimensional model where the dimension perpendicular to the area 
covered with clouds is 120~pc thick. This is on account that even if clouds overlapping along a line of sight can often be separated, their 
associated SFR usually cannot, as SFR tracers do not carry the same spectral information as CO data. 

For this model, we choose a volume filling factor with ff$_{min}$=0.1 and ff$_{max}$=1; we limit our simulations to the simplest case 
in which no scatter is added to any relation or data,  although we still retain an exponential distribution of cloud covering factors. 
 For $\beta$=1.5 and 200~pc region size, we obtain $\gamma_{H2}=$1.13 with 
$\sigma_{H2}=$0.22. This should be compared 
with the analogous two--dimensional case above which gives $\gamma_{H2}=$1.75 with $\sigma_{H2}=$0.35. Filling a volume produces 
a significantly flatter slope and smaller dispersion about the mean trend in the $\Sigma_{SFR}$--$\Sigma_{H2}$ distribution than covering an area. The flattening of $\gamma_{H2}$ cannot be simply attributed to variations in the dynamical range covered by the 
simulated data in the two cases, since they have similar ranges (1.1~dex versus 1~dex). However, our simulations do not include 
 control on the shape of the volume to be filled, which affects the final result by allowing multiple clouds to be located along the line of 
 sight. Indeed, a direct comparison of the volume--filling simulation with the area--covering simulation shows that the former fills each 
 volume with about 6 times more clouds, that are about 20\% less massive, on average than what the latter does. 
 The combination of these effects produces a 
 more effective averaging of the cloud mass distribution in the volume--filling case than in the area--covering case, thus yielding an overall 
 flatter $\gamma_{H2}$ slope. 
 
 Actual observations are likely to be in--between our two cases of areal--cover and volume--filling. In the next sections we limit our analysis 
 to the two--dimensional case, although the result from this section should be kept in mind for any general conclusion. 

\subsection{Fitting Method, Range, and Data Censoring}

In the previous sections, we have presented the baseline results for our Default Model, using the OLS bi--sector linear fitting. This is the same 
fitting approach adopted by a number of observational analyses of the SK Law \citep[e.g.][]{Blanc2009,Verley2010,Liu2011,Rahman2011}. The advantage of 
the OLS bi--sector fitting is the ability to divide the locus of the data into two roughly equal--number areas; however, its disadvantage is 
that error bars on the data are not included in the fitting, and statistically insignificant data points get the same weight as significant ones 
\citep[see discussion in][]{Verley2010}. 
Another routine often used to determine the slope and intercept of the Observed SK Law is the bi--linear regression fit \citep{Kennicutt2007, Verley2010}. 

We compare the best fit slopes obtained with both the OSL bi-sector fit and the bi--linear regression fit in Figure~\ref{fig7}, for both cases 
of the Default Model (left panel) and the model in which the relation between SFR and cloud mass has zero scatter in equation~5 
($\sigma(log SFR)$=0, right panel). 
For both fitting methods, the best fit slope is a decreasing function of increasing region size, and in neither 
case the actual value of $\beta$ is systematically recovered (except for $\beta$=1.0 using the OLS fitting method). However, the bi--linear 
fit yields larger numerical values for $\gamma_{H2}$ than the OLS fit,  and the magnitude of the discrepancy is an increasing function of 
increasing $\beta$, decreasing region size, and increasing intrinsic scatter between the SFR and the cloud mass. For $\beta$=1, 1.5, and 2, 
$\delta \gamma_{H2}/\gamma_{H2}(OLS)\sim$20\%--2\%,  50\%--20\%, and 80\%--30\%, respectively, for regions in the range 200--2,000~pc.  
The bi--linear fit method yields larger 
uncertainties, by a factor  3--15, in the final uncertainty of $\gamma_{H2}$ than the OLS fit, which is a better reflection of the scatter 
in our simulated data points. 

Next we test the impact of measurement uncertainties. For most studies, the dominant source of uncertainty is the gas (CO or other molecular 
gas tracers) measurement, as tracers of SFR are usually detected at high significance even when 1~$\sigma$ detection limits are reached 
in CO. We can approach the problem of measurement uncertainties in two different ways: (1) by  changing the significance level of our detection `limit' in 
$\Sigma_{H2}$; and (2) by varying the value of the detection limit. The former tests changes in $\gamma_{H2}$ induced by the varying significance of the 
data, but preserving the full dynamical range of the simulated data points. The latter mimics variations in the 
dynamical range of observational data due to varying detection limits, and should test the impact  of common data censoring procedures.  
In order to highlight trends, we concentrate our analysis on three representative region sizes: 200~pc, 500~pc, and 1~kpc. We use, from now 
on, the OLS bi--sector fitting as our default. 

When varying the significance of the detection limit   without changing the numerical value of the limit (set to 
$\Sigma_{H2}\sim$6.5~M$_{\odot}$~pc$^{-2}$ in the Default 
Model), we measure an increase in $\gamma_{H2}$ for increasing 
detection threshold between 1~$\sigma$ and 5~$\sigma$, with most of the change concentrated between 
1~$\sigma$ and 3~$\sigma$ (Figure~\ref{fig8}). The change is generally modest,  
$\delta \gamma_{H2}/\gamma_{H2} \lesssim$15\%  for $\beta$=2 and $\delta \gamma_{H2} /\gamma_{H2}\lesssim$5\% for $\beta$=1, 
at all sizes. The values of 
$\gamma_{H2}$ become roughly constant for thresholds $\ge$3~$\sigma$, in our model, although they level off at different values 
for different region sizes, and recover $\beta$ systematically only for the largest size (1~kpc). 

Variations of the detection limit that vary the numerical value of the  lowest valid data  
change the dynamical range of the data. Higher limits than our selected 3~$\sigma$ at $\Sigma_{H2}\sim$6.5~M$_{\odot}$~pc$^{-2}$ will decrease the 
data range; vice versa for lower limits. For instance, going from a  3~$\sigma$ threshold to a 5~$\sigma$ one reduces the dynamical range by 0.45~dex  
in log($\Sigma_{H2}$) 
in our simulations. This attempts to mimic observational situations in which different detection thresholds may be chosen for the same data, and also addresses the issue of data censoring, especially for $\Sigma_{H2}$, which tends to have lower data significance than $\Sigma_{SFR}$ for most observational cases. 
Figure~\ref{fig9} shows that varying the dynamical range of the data, 
together with their significance  exacerbates variations in the observed slope, with larger changes in $\gamma_{H2}$ between 1~$\sigma$ and 5~$\sigma$ than in the 
previous case (cf. Figure~\ref{fig8} with Figure~\ref{fig9}).  As before, the largest variations are observed for the largest values of $\beta$, 
with  $\delta \gamma_{H2}/\gamma_{H2} \lesssim$50\% for $\beta$=2 and $\delta \gamma_{H2}/\gamma_{H2} \lesssim$35\% for $\beta$=1. 

\subsection{Variations in the H$_2$ Clouds Mass Function}

The maximum mass of a molecular cloud in a galaxy is not necessarily a constant, although variations from galaxy to galaxy do not appear to be large. 
In the MW, M31, and LMC  the maximum measured H$_2$ masses are about 1--3$\times$10$^6$~M$_{\odot}$ 
\citep{Wilson1990,Rosolowsky2005,Engargiola2003,Blitz2007}, 
but the largest clouds could be smaller in the LMC, depending on  identification techniques \citep{Wong2011}. Giant Molecular Associations, as 
massive as 10$^7$--10$^8$~M$_{\odot}$, could be present in some galaxies \citep{Koda2009}. Conversely, the 
slope $\alpha$ of the cloud mass function 
(equation~3) varies considerably from galaxy to galaxy, with values that range from $\sim-$1.5 in the inner parts of the MW to $\sim-$2.6/$-$2.9 in M33 and the LMC \citep{Engargiola2003,Rosolowsky2005,Wong2011}. 
The actual range of $\alpha$ might be smaller than the one just quoted, though; shallow values of $\alpha$ may be the result of cloud blending, while 
steep values may be due to fitting the mass function beyond the clouds high--end  cut--off \citep{Rosolowsky2005}. Nevertheless, the existence of a range of 
values for $\alpha$ is considered both physical and significant. Our Default Model uses $\alpha=-$2.0 and a maximum H$_2$ cloud mass of 
3$\times$10$^7$~M$_{\odot}$, the latter to accommodate the Giant Molecular Associations identified by \citet{Koda2009} in the M51a 
galaxy. The normalization is set to allow one cloud to be formed with the maximum cloud mass M$_{cloud}$(max). 

The impact of variations of M$_{cloud}$(max) and $\alpha$ is summarized in Figures~\ref{fig10} 
and \ref{fig11}. M$_{cloud}$(max)  is changed from 3$\times$10$^5$~M$_{\odot}$ to  3$\times$10$^7$~M$_{\odot}$ (our Default Model 
value) in factor 10 increments of mass. Values of $\alpha$ considered are $-1.5$, $-2.0$ (Default Model value), and $-2.3$, which spans 
a range close to the observed one. Figure~\ref{fig10} shows the measured slope $\gamma_{H2}$ and the dispersion $\sigma_{H2}$ about 
the best fit line of the Observed SK Law in both cases. 
Variations in M$_{cloud}$(max) have a larger impact, in terms of fractional changes in $\gamma_{H2}$, than variations in $\alpha$  
when $\beta>1$, at least in the range explored here. As the maximum 
cloud mass decreases, $\gamma_{H2}$ tends to values of unity  and $\sigma_{H2}$ decreases monotonically; for any value of the 
SFR--cloud~mass power law index, $\gamma_{H2}\le$1.2 
for M$_{cloud}$(max)=3$\times$10$^5$~M$_{\odot}$ and region sizes $\gtrsim$500~pc. This behavior is in agreement with the expectation that 
for decreasing values of M$_{cloud}$(max), the cloud mass function is fully sampled for smaller region sizes and, for $\beta>$1, less efficient star--forming clouds are included in the SFR accounting. Figure~\ref{fig11}  supports this conclusion:  
for decreasing M$_{cloud}$(max), the mean number of clouds in a region increases,  albeit modestly, while the mean cloud mass decreases. 

An intuitive result can be obtained by setting all clouds to have the same mass, such that at least one cloud is sampled by the smallest 
region simulated, i.e., M$_{cloud}<$3$\times$10$^5$~M$_{\odot}$. In this case, $\gamma_{H2}$=1$\pm$0.1, irrespective of region size, cloud size, 
or $\beta$. Also $\sigma_{H2}$ has values that are typically a factor 3 or more smaller than the equivalent values obtained with a distribution of 
cloud sizes. If we set the clouds to have identical mass  M$_{cloud}>$3$\times$10$^5$~M$_{\odot}$, implying that the clouds are larger 
than the smallest simulated area, we obtain $\gamma_{H2}>$1, but  
$\gamma_{H2}<\beta$ (for $\beta>1$) with little variation as a function of region size. This is because, for identical clouds, all regions sample the 
same SFR surface density values, which then simply 
scales as the H$_2$ surface density. This result may help explain the trend of $\gamma_{H2}$ in the bottom panels of Figure~\ref{fig10}, which 
show a generally decreasing function for 
increasing $\alpha$. A flatter cloud mass distribution ($\alpha$=$-$1.5) than that of our Default Model gives more 
uniform weight to all masses, which then  flattens the trend in $\gamma_{H2}$. 

Changes in the normalization of the cloud mass function have a small impact of the values of $\gamma_{H2}$. If we allow 
the normalization of the cloud mass function to be such that 100 clouds with value M$_{cloud}$(max)  are produced, rather than 1 cloud, 
all values of $\gamma_{H2}$ steepen by 0.03--0.04, for M$_{cloud}$(max)=3$\times$10$^7$~M$_{\odot}$ and $\alpha$=2.0. The steepening 
is still within 1--3~$\sigma$ of the measured uncertainty in the slope, thus a small variation. 

In summary, variations in the cloud mass function have a large impact on the resulting parameters of the Observed SK Law,  for both changes 
in  M$_{cloud}$(max) and $\alpha$.  Changes  in the normalization of the cloud mass distribution 
produce small--to--negligible effects at all region sizes and $\beta$ values explored in this paper.

\section{Effects of Selection Biases in Star Formation}

Our Default Model attributes star formation to each cloud that is sampled into a region according to equation~5, independently of the 
characteristics of the cloud. This is a simplistic approach that we will try to refine in this section.

As clouds become smaller and less massive their ability to condense into high density cores may decrease, and so 
will be their capability to form stars. This can be translated into a minimum cloud mass threshold for star formation in our
simulations. Additionally, low--mass clouds will be less likely to form massive stars than larger mass clouds, thus will not contribute to SFR measurements that 
rely on massive star tracers (UV, H$\alpha$, ...). We can estimate the fraction of SFR lost to this effect, by assuming that clouds less 
massive than $\sim$3,000~M$_{\odot}$ are unlikely to form UV--bright or ionizing stars. This implies that 16\% of the SFR is not 
detected for $\beta$=1.0, but the fraction decreases to less than 1\% for $\beta=$1.5 or higher. This reasoning, however, only applies if there is 
a direct correlation between cloud mass and the most massive star formed \citep[e.g.][]{Weidner2006}. If star formation is a purely stochastic 
process \citep[e.g.][]{Calzetti2010}, the stellar IMF will be fully sampled for large numbers of clouds, even if these are low--mass ones. 

There are other reasons for implementing thresholds for star formation to cloud masses, and quantifying their effects on the Observed SK Law. 
One of the procedures  employed in some recent papers, including \citet{Liu2011}, prescribes the removal of an 
extended/diffuse component from the data that are used to measure the SFRs. This extended component can 
represent up to 50\% of the total stellar emission (either as direct  or as dust--reprocessed stellar light). Much debate is currently 
underway about whether the extended component is effectively unrelated to current star formation (i.e., the emission is tracing 
light from intermediate/old stellar populations), or whether it represents some 
low--level, unresolved star formation. In the former case, its removal prior to SFR measurements is justified, but it would not in the 
latter case. 

If the diffuse component is unrelated to current star formation, studies that do not remove that component 
\citep[e.g.][]{Bigiel2008} add an artificial contribution to the SFR measurements. The addition of this background will affect 
measurements of both $\gamma_{H2}$ and $\sigma_{H2}$, and we will quantify its impact in section~4.2.

\subsection{Cloud Mass Threshold for Star Formation}

Using the Default Model as a starting point, we model the SFR thresholds by removing a fraction of the SFR from the simulations. We 
specifically assign SFR=0 to H$_2$ clouds that are below a given mass threshold,  M$_{cloud}$(thr). Clouds below this threshold are still 
accounted for in the final census for the gas surface density, but they do not contribute to the SFR surface density. 

As we can intuitively infer that selective removal of star formation from the small clouds will steepen the Observed SK Law, we quantify this effect 
for the flattest SFR--M$_{H2}$ relation we analyze, i.e., $\beta$=1.0. We consider the case in which 1/3, 1/2, and 60\% of the total 
SFR is removed, on average, from each region, which corresponds to threshold cloud masses M$_{cloud}$(thr)=10$^{4.3}$~M$_{\odot}$, 
10$^{5.1}$~M$_{\odot}$, and 10$^{5.6}$~M$_{\odot}$, respectively (Figure~\ref{fig12}). 

As expected, a steepening of the Observed SK Law is 
observed,  which is already significant when 33\%  of the  
SFR is removed,  especially for region sizes smaller than $\sim$1~kpc.  Similar trends  
are observed for the scatter of the data about the mean trend, $\sigma_{H2}$. 

The derivative of $\gamma_{H2}$ becomes increasingly more negative as the SFR removal increases up to $\sim$50\% (Figure~\ref{fig12}, left), 
and then it flattens again for small region sizes. 
The derivative changes sign below 400~pc for the case with 60\% of the star formation removed. For this case, a large 
fraction of bins do not contain any detectable star formation at small region sizes; this fraction is about 50\% at 300~pc and 
70\% at 200~pc. This is due to the combined effects of region size and distribution of cloud covering factors, 
that tend to give preference to small (non star forming) clouds. Regions with small covering factors {\em and} a non--zero SFR, thus, crowd in  
the upper envelope of the $\Sigma_{SFR}$--$\Sigma_{H2}$ distribution, which results in smaller $\gamma_{H2}$ values (flatter trends in the 
$\Sigma_{SFR}$--$\Sigma_{H2}$ plane). Larger size regions suffer progressively less from this selection problem, as do lower thresholds.  
The trend for the derivative of $\sigma_{H2}$ is similar to that of $\gamma_{H2}$, except that a flattening of the trend with region size occurs 
when $\ge$60\% of the SFR is removed. 

For a uniform distribution of cloud covering factors, the overall behavior is similar, but the effect is less pronounced: the derivative of $\gamma_{H2}$ starts 
flattening again below 400~pc at 60\% star formation removal (50\% of the 200~pc bins contain star formation in this case). 
In general, a uniform distribution of covering factors gives 
values of $\gamma_{H2}$ on average 0.05--0.1 smaller than reported in Figure~\ref{fig12}, left panel, and values of $\sigma_{H2}$ on average 0.02--0.05 smaller. 

The changes in the derivative of the $\gamma_{H2}$ and the $\sigma_{H2}$ trends with region size for increasing values of the threshold are present for $\beta>$1 as well. 
It should be noted that a given cloud mass threshold corresponds to a smaller fraction of the SFR removed in the case of $\beta>$1; for instance,
M$_{cloud}$(thr)=10$^{5.1}$~M$_{\odot}$ removes only about 6\% of the total SFR for $\beta$=1.5, as opposed to 50\% for $\beta$=1. 

\subsection{Background Addition to Star Formation}

We simulate the presence of a diffuse, uniform background in SFR data by adding a constant value to the SFR in each region of the Default 
Model, corresponding to a set fraction of the mean SFR value from all regions. Since the addition of a constant value will result in a 
general flattening of the Observed SK Law, we quantify this effect for the $\beta$=1.5 case. 

The results are summarized in Figure~\ref{fig13}.  A constant background addition larger than 10\%  produces  values of 
$\gamma_{H2}<\beta$, for any region size. For backgrounds equal or 
larger than 50\%, the recovered $\gamma_{H2}\le$1.05 for all region sizes, and the slopes are flatter even than those for the case of $\beta$=1.0 with 
no background added  (Figure~\ref{fig3}, bottom panels). 
Analogous considerations apply to $\sigma_{H2}$. In the case of a uniform distribution of cloud 
covering factors, the qualitative trends are similar to those in Figure~\ref{fig13}, with quantitative offsets 
in both $\gamma_{H2}$ and $\sigma_{H2}$ similar to those discussed in section~4.1.

The background levels used in Figure~\ref{fig13} are calculated as fractions of the mean SFR measured in the 10,000 realizations 
of each region size. As the mean SFR values reflect the distribution of covering factors, the background fractions correspond 
to higher absolute SFR values for the uniform distribution of cloud covering factors than for the exponentially decreasing one.  
One general consequence is that, at constant background fractions, the values of $\gamma_{H2}$ generated with the former 
distribution are generally a little smaller, by $\Delta \gamma_{H2}\le$0.05, than those generated with the latter one.

\section{The Azimuthally--Averaged SK Law}

Many analyses derive the SK Law using azimuthally--averaged radial profiles of galaxies \citep[e.g.,][]{Wong2002,Boissier2003,Heyer2004,Verley2010}, rather than regions within galaxies. The advantage of radial profiles is the higher signal--to--noise ratio than can be obtained on individual 
measures. The main disadvantage is that the cloud mass function is averaged in regions that increase progressively in area with 
radius.

We modify the Default Model to simulate an azimuthally--averaged SK Law. We require that regions be annuli that grow in radius between 
R$_{min}$ and R$_{max}$, in steps of R$_{step}$. For each annulus, we generate 200 independent realizations, to ensure sufficient 
statistical averaging. We also require that the lowest gas surface density value generated by our model is 
detected at the 5~$\sigma$ level, to mimic the higher signal--to--noise ratio typical of such analyses. This translates into requiring that  
$\Sigma_{H2}$=3~M$_{\odot}$~pc$^{-2}$ is detected at the 5~$\sigma$ level in our simulations.  No other parameter of the Default Model 
is changed, in order to facilitate comparison with the individual region simulations. 

The presence, within a single simulation, of regions of different total area helps increasing the dynamical range spanned by $\Sigma_{H2}$: 
it covers from $\Sigma_{H2}$=3~M$_{\odot}$~pc$^{-2}$ at the low--end to $\Sigma_{H2}$=100~M$_{\odot}$~pc$^{-2}$ at the 
high end, almost irrespectively of $\beta$ and R$_{max}$. The dynamical range of $\Sigma_{H2}$ for the azimuthally--averaged simulation is thus 2--3  
times larger than in the case of individual regions,  especially when compared with large region sizes. For simplicity, our simulations assume R$_{min}$=R$_{step}$=200~pc \citep[e.g.,][]{Heyer2004,Verley2010} and a variable R$_{max}$ between 4~kpc and 14~kpc. 

The best fit slopes $\gamma_{H2}$ for the Observed SK Law and the parameter choices above are shown in Figure~\ref{fig14} for the three 
cases of $\beta$=1.0, 1.5, and 2.0, and a range of R$_{max}$ values. Results for both the uniform and exponential distributions 
of cloud covering factors are shown. As may be expected from the results obtained in the previous sections, 
the azimuthally--averaged $\gamma_{H2}$ is always lower than $\beta$ for $\beta>1.0$, and decreases for increasing R$_{max}$. This 
result may be better understood with a specific example: for R$_{min}$=200~pc and R$_{max}$=4~kpc, the smallest area simulated is 
0.13~kpc$^2$ and the largest is about 5~kpc$^2$ (corresponding to the largest annulus). The convolution of the contribution from all the 
annuli in a given simulation has the general effect of flattening the $\gamma_{H2}$ slope relative to $\beta$. 
The case $\beta$=1.0 is an exception, with $\gamma_{H2}\approx$1 at all R$_{max}$ (Figure~\ref{fig14}). 

Our simulations do not include a dependence of the cloud covering factor on galactocentric distance, which has been suggested by 
\citet{Leroy2009}. However, the comparison, at fixed $\beta$, between the slopes $\gamma_{H2}$ of  the uniform and exponential distributions of cloud 
covering factors in Figure~\ref{fig14} suggests that changes in the overall distribution of covering factors has a small impact on the measured slope:   
$\delta \gamma_{H2}/\gamma_{H2}\lesssim$8\% at $\beta=$1.5 for the two distributions we consider.

\section{Discussion}

\subsection{Summary of Simulation Results}

Despite the inevitable simplifications our models contain, a number of general trends can be garnered from the above analysis. There is 
a strong dependence of the slope $\gamma_{H2}$ of the Observed SK Law on the physical size of the region sampled; specifically, at fixed exponent $\beta$ 
of the intrinsic relation between SFR and cloud mass (SFR$\propto$M$_{H2}^{\beta}$, equation~5), $\gamma_{H2}$
decreases for increasing region's size. The derivative of the trend is larger, in absolute value, for higher $\beta$ (e.g., Figures~\ref{fig3} and \ref{fig5}). 
In fact, for $\beta$=1, $\gamma_{H2}$ rarely deviates from a value of $\approx$1, except when the scatter between 
SFR and M$_{H2}$ is a factor $\sim$4 or larger (Figure~\ref{fig5}, bottom panels). For $\beta>$1, $\gamma_{H2}$
usually transitions from $\gamma_{H2}>\beta$ at small scales ($\lesssim$500~pc) to $\gamma_{H2}<\beta$ at large scales ($\gtrsim$1,000~pc), 
again with the exception of scenarios in which the scatter between SFR and M$_{H2}$ is large. 
This transition scale is strongly dependent on the maximum cloud mass, M$_{cloud}$(max), that can form in the galaxy, and less strongly dependent 
on the power law exponent $\alpha$ of the cloud mass function  (Figure~\ref{fig10}). For  M$_{cloud}$(max)$\lesssim$10$^{6.5}$~M$_{\odot}$, 
i.e., a factor 10 or more lower than our default assumption, $\gamma_{H2}<\beta$ at $\sim$500~pc and larger sizes. 

The values of $\gamma_{H2}$ are influenced by the fitting procedure and by the  dynamical range of the data, as already remarked by other authors 
on the basis of observational results \citep[e.g.][]{Blanc2009,Verley2010,Liu2011}. If instead of using the OLS bi--sector fitting, we employ the FITEXY 
routine, i.e., a $\chi^2$--minimization procedure that weights the result by the uncertainty along both the X and Y axis, the resulting values of 
$\gamma_{H2}$ are systematically higher than those produced by the OLS bi--sector method \citep[Figure~\ref{fig7}; see, also][]{Verley2010}, and the 
discrepancy increases for decreasing region sizes. For $\beta$=1, 1.5, and 2, 
$\delta \gamma_{H2}/\gamma_{H2}(OLS)\sim$20\%--2\%,  50\%--20\%, and 80\%--30\%, respectively, for regions in the range 200--2,000~pc. The 
general trend is for the OLS and FITEXY slopes to converge for both increasing region sizes and for decreasing values of $\beta$, since in 
both cases $\gamma_{H2}$ tends to values of unity. 
The formal uncertainty on $\gamma_{H2}$(FITEXY) is, however, much larger than that produced by the OLS bi--sector 
fitting routine, by factors 3--15 for the three $\beta$ values considered. 

The dynamical range along the molecular gas surface density axis has a major influence on the value of $\gamma_{H2}$, in the sense that 
smaller dynamical ranges (e.g., stronger censoring or higher thresholds for the significance of the observational data) produce steeper values of 
$\gamma_{H2}$ (Figure~\ref{fig9}). The relation between the values of $\gamma_{H2}$ and $\beta$ is a complex function of both the dynamical 
range and the region size considered. However, a common result is for uncensored data to yield $\gamma_{H2}<\beta$ on of both the dynamical 
range and the region size considered. However, a common result is for uncensored data to yield $\gamma_{H2}<\beta$ at virtually any region 
size and any $\beta$. This is partly due to the non--symmetric nature of data scatter when projected along logarithmic axes. 

Another notable influence to the values of $\gamma_{H2}$ is given by the scatter in the SFR--cloud~mass relation (Figure~\ref{fig5}). The larger the scatter, 
the larger the value of $\gamma_{H2}$: for instance, for $\beta$=1, $\gamma_{H2}$=1 for zero scatter between SFR and cloud mass and $\gamma_{H2}\sim$1.2 for a factor 4 scatter (1~$\sigma$), at region sizes$\lesssim$700~pc,  and an exponential distribution of cloud covering factors. The fractional difference is higher for $\beta$=1 ($\sim$25\%) than for higher values of $\beta$ ($\lesssim$10\%--15\%). 
The same result holds for $\sigma_{H2}$, which increases both for increasing $\beta$ and increasing scatter in the SFR--cloud~mass 
relation. This produces a potential degeneracy between high--$\beta$ and high--scatter, although $\sigma_{H2}$ increases faster for 
large scatter than $\gamma_{H2}$. 

The slope $\gamma_{H2}$ converges towards values of 1 at large region sizes, for all values of $\beta$ analyzed in this paper 
($\beta$=1,1.5, 2).   This is a reflection of the fact that, as the region's size increases, the cloud mass function becomes better sampled. 
Thus, any difference between regions in the same realization (`galaxy') will mostly be due 
to differences in the covering factor, and the `data' will start lining up along a line with slope=1 in the Log($\Sigma_{SFR}$)--Log($\Sigma_{H2}$) plane. 
The convergence occurs at region sizes$\gtrsim$2~kpc  if M$_{cloud}$(max)=10$^{7.5}$~M$_{\odot}$, 
but quickly decreases down to 1~kpc  for M$_{cloud}$(max)$<$10$^{6.5}$~M$_{\odot}$ (Figure~\ref{fig10}). 
Our simulation results agree well with the observational results of \citet{Schruba2010}, where they find that regions centered on CO peaks produce similar 
gas depletion timescales as regions centered on H$\alpha$ peaks in M33 once such regions have diameters of about or larger than 1~kpc. The \citet{Schruba2010} 
result implies that 1~kpc is about the scale where both the HII region luminosity function and the cloud mass function are fully sampled; 
the latter is reproduced by our simulations once we take into account that M$_{cloud}$(max)$\sim$10$^{6}$~M$_{\odot}$ in M33 \citep{Engargiola2003}. 

The scatter of the `data' about the best fitting line, $\sigma_{H2}$, is always significant even when there is no scatter in the SFR--cloud~mass relation 
(Figure~\ref{fig5}, top--right), and is due to a combination of effects: (1) measurement uncertainties along the $\Sigma_{H2}$ axis, which 
affects `data' for any value of 
$\beta$; (2) the combination of a non--linear SFR--M$_{H2}$ relation with random sampling of the cloud mass function, which affects only cases with 
$\beta>$1. The first contribution produces a minimum plateau of $\sigma_{H2}\sim$0.1--0.2 at $\beta$=1, while the second 
contribution increases $\sigma_{H2}$ with increasing $\beta$. Expectedly, all values of $\sigma_{H2}$ increase at fixed region size and $\beta$ if the 
scatter in the SFR--M$_{H2}$ relation increases. Like $\gamma_{H2}$, $\sigma_{H2}$ decreases for increasing region's size, implying that, as the cloud 
mass function becomes better sampled at larger sizes, so does the SFR. 

Additions of SFR thresholds, as a consequence of either physical mechanisms (e.g., the lowest mass clouds cannot form massive enough stars 
to ionize the gas or emit in the UV) or selection biases, has the general effect of steepening $\gamma_{H2}$ and increasing $\sigma_{H2}$ (Figure~\ref{fig12}). 
Conversely, the addition of a uniform contribution to the SFR axis (e.g., a background emission which, although unrelated to the current star formation, is included in the census of SFR) has the general effect of flattening  $\gamma_{H2}$ and reducing $\sigma_{H2}$. 

Finally, analyses of azimuthally--averaged data should generally recover $\gamma_{H2}<\beta$ for $\beta>1$ and $\gamma_{H2}\sim\beta$ for $\beta$=1 
(Figure~\ref{fig14}). The `flattening' of $\gamma_{H2}$ is an effect of averaging regions with a range of sizes when annuli of increasing radius are considered in such analyses.

\subsection{Comparison with Observations}

Given all the factors that can influence the observed slope of the SK Law and the scatter about the best fit line, it is perhaps not surprising that 
analyses in the literature have yielded many different, and sometimes contrasting, results. In this section, we compare our simulations against 
a subset of observational results, explicitly those results for which we are familiar with or can reconstruct the assumptions and choices built into the analysis (e.g., 
detection limits for the data, treatment of the background in the SFR data, fitting routines, etc.). A caveat is that our pixel--based simulations can only be compared with observations treated in a similar fashion, and are not immediately adaptable to measurements 
that use specific selection criteria for the galactic regions (e.g., regions selected to be peaks of SFR).

\subsubsection{Individual Galaxies: M51a and NGC3521}

We begin by comparing our simulations to the observational results obtained by our group. \citet{Liu2011} derived the Observed SK Law in 
the nearby galaxies M51a and NGC3521, for a variety of region sizes and 
detection limits. The authors derive the SFR maps for those two galaxies from a combination of both H$\alpha$ and 24~$\mu$m emission 
and far--UV (FUV) and 24~$\mu$m emission, following the calibrations of \citet{Calzetti2007} and \citet{Bigiel2008}, which account for both the 
obscured and unobscured portions of the SFR. \citet{Liu2011} also remove a low--frequency emission from the H$\alpha$, FUV, and 24~$\mu$m images 
of both galaxies, at the level of about 50\% of the total emission in each band. Their stated reason for this removal is that the diffuse, low--frequency 
emission is likely not directly associated with the SFR traced by the molecular gas in a given region. We report the data from \citet{Liu2011} in Figure~\ref{fig15}, 
with the region sizes given by the square root of the de--projected area of each galaxy bin. For each of the two galaxies, we 
adopt two extreme values of the inclination (20$^{\circ}$--42$^{\circ}$ for M51a and 65$^{\circ}$--73$^{\circ}$ for NGC3521), to account for uncertainties in this parameter as well. 
For each value of $\beta$, we show in Figure~\ref{fig15}  a modification of the Default Model that most closely accounts for the observed trend in both 
$\gamma_{H2}$ and $\sigma_{H2}$, using the exponentially decreasing cloud covering factor distribution (equation~6). The models shown 
{\em are not fits to the data}, but simply a close approximation of the observations. In all cases, 
the models have been generated with  a similar dynamical range in Log($\Sigma_{H2}$) at the  3~$\sigma$ detection limit, as 
used by \citet{Liu2011} for their analysis. 

For M51a, the observed trends of both $\gamma_{H2}$ and $\sigma_{H2}$ are more closely, and simultaneously, reproduced by $\beta$=1.5 with 
about 1\% of the SFR removed (meaning that the lowest mass clouds,  M$_{cloud}\le$10$^{3.8}$~M$_{\odot}$, that produce the bottom 1\% 
 of the star formation are 
assigned SFR=0, top panels of Figure~\ref{fig15}). A small excess in the predicted $\gamma_{H2}$ is, however, found for regions $>$700~pc. 
 Best fitting models for $\gamma_{H2}$ with $\beta$=1.0 and 2.0 tend to produce a flat trend for region sizes $<$400~pc; $\beta$=2.0 models also 
predict steeper slopes than observed at sizes$>$700~pc. However, the main discrepancy between $\beta$=1.0 and 2.0 models and observations 
is in the $\sigma_{H2}$ trend (Figure~\ref{fig15}, top--right panel): both models  predict larger--than--observed scatter about the best fit lines for regions 
$<$700~pc. Changing parameters in the Default Model is unlikely to help improve the agreement between models and observations: parameter 
choices that would decrease the values of $\sigma_{H2}$ (e.g., decrease in the scatter of the SFR--cloud~mass relation or decrease in the SFR threshold) 
would also decrease the values of $\gamma_{H2}$ at fixed region size, thus introducing a discrepancy between models and observations for this 
quantity. 
Using a homogeneous distribution of cloud covering factors provides similar results that only minimally differ in the quantification of the background or 
thresholds to implement (typically by  5\%). 

There is a large degree of degeneracy among the models that can reproduce the data for $\gamma_{H2}$ in NGC3521 
(Figure~\ref{fig15}, bottom--left panel). Within the 1--1.5~$\sigma$ uncertainty of the data, the observed trend and amplitude of $\gamma_{H2}$ 
are reproduced by: a $\beta$=2.0 model with a 25\% added background to the SFR; a  $\beta$=1.5 
with a 10\% background;  and a $\beta$=1.0 with a cloud mass threshold that removes 40\% of the SFR. For $\sigma_{H2}$, 
the $\beta$=1.5 reproduce the observed trend more accurately than the $\beta$=1.0 
and $\beta$=2.0 models, although none of the models matches the data for sizes$<$400~pc. The high inclination of NGC3521 is likely to 
complicate any comparison with models.

If M51a and NGC3521 share a common underlying relation between SFR and cloud mass, we conclude that the Observed SK Law slope and data scatter can more 
closely be  reproduced by an exponent $\beta$=1.5. In the case of M51a, \citet{Liu2011} appear to have  slightly 
(by about  1\%) over--subtracted the low--frequency background in the SFR maps, and to have slightly under--subtracted it (by about 10\%) in the case of NGC3521. To test this picture, we compare the trend in $\gamma_{H2}$ of both galaxies as a function of the dynamical range 
(expressed as a detection threshold, 1~$\sigma$ to 5~$\sigma$, for $\Sigma_{H2}$), as derived by \citet{Liu2011}. We choose two representative 
projected region sizes, 300~pc and 700~pc, from Tables~3 and 4 of those authors, which correspond to de--projected region sizes of 330~pc and 770~pc 
in M51a and 500~pc and 1200~pc in NGC3521. Figure~\ref{fig16} shows that the models reasonably, albeit not perfectly,  reproduce the observational trends; 
the largest deviations ($>$2~$\sigma$ along the vertical axis) are observed in M51a  for the highest detection limits, 
where the slopes are systematically over--predicted 
by  $\delta \gamma_{H2}/\gamma_{H2}\lesssim$10\%. This disagreement is  likely driven by the decreasing dynamical range of the data 
for increasing detection limit. However,  the comparison in 
Figure~\ref{fig16} stresses the similarity in the trend as a function of detection limit for model and data, despite the numerical discrepancy.   

Within the picture above, we can attempt to interpret results on the same two galaxies obtained by other authors. Among the pixel--based analyses, \citet{Bigiel2008} obtain 
$\gamma_{H2}$=0.84 and 0.95 for M51a and NGC3521, respectively, with $\sigma_{H2}\sim$0.2 over region sizes =750~pc \citep[could be 1~kpc for 
NGC3521 if the galaxy is closer than what those authors assumed, see][]{Liu2011}. \citet{Bigiel2008} use a combination of FUV and 24~$\mu$m 
data to derive spatially--resolved SFRs, and do not remove any low--frequency background from their SFR maps, on account that both FUV and 
24~$\mu$m emission should be mostly tracing current SFR. The low frequency background removed by \citet{Liu2011} 
corresponds, as already remarked, to about 50\% of the total luminosity tracing the SFR; in our models, re-adding this background to the simulations corresponds to 
adding a contribution equivalent to $\sim$100\% of the mean SFR. \citet{Bigiel2008} also analyze their data above a 3~$\sigma$ significance level for 
$\Sigma_{H2}$. Our Default Model, using $\beta$=1.5 {\em and} a uniform background contribution in the range 60\%-80\%, yields $\gamma_{H2}$ 
in the range  0.90--0.81 at 750~pc  and  0.85--0.76 at 1~kpc, and a scatter about the best fit lines $\sigma_{H2}\sim$0.25, 
for $\Sigma_{H2}\ge$3~$\sigma$. The values of the model slopes approach the results obtained by \citet{Bigiel2008}, similar to the conclusion reached by \citet{Liu2011} using observational data. Even when adopting a more modest background level of 20\% of the total, as recently suggested by \citet{Leroy2012}, which corresponds to 30\% in our convention, the models with $\beta$=1.5 yield overall low values of $\gamma_{H2}$, in the range 1.10--1.03 
for region sizes in the range 750--1,000~pc.

Still for M51a, using region sizes $\sim$170~pc, \citet{Blanc2009} derive a slope $\gamma_{H2}$=0.82$\pm$0.05 and a scatter about the 
mean trend $\sigma_{H2}$=0.43$\pm$0.02. \citet{Blanc2009} use the extinction--corrected H$\alpha$ emission to trace the SFR, and remove a 
uniform background at the level of 11\% of the total H$\alpha$, on the basis that the diffuse H$\alpha$ component may be due to contributions from adjacent regions, and, therefore, not measure the local SFR. These authors fit their data including upper limits to both the molecular gas and SFR surface density components. We attempt to reproduce this scenario with our Default Model, by adopting  a 50\%--60\% contribution from a uniform 
background \citep[10\%--20\% less than in the case of][to simulate the background removal of \citet{Blanc2009}]{Bigiel2008}, $\beta$=1.5, same region sizes as \citet{Blanc2009}, and no detection limit 
to $\Sigma_{H2}$ (see Figure~\ref{fig9}). Our simulations yield  $\gamma_{H2}$=0.85--0.80 for the two background choices, and 
$\sigma_{H2}\sim$0.32. Varying the fraction of background added to the SFR can lead to better or worse agreement with the observational data. Finally, the results by \citet{Kennicutt2007}, where a slope of 1.37 is obtained for the molecular SK Law, are difficult to interpret within 
the framework of our simulations, as those authors center their 500~pc apertures on peaks of star formation.

Independently of the exact details of each measure, the wealth of independent measures for M51a offers the opportunity to investigate some 
 trends that should be applicable to galaxies in general. The steep slopes obtained by \citet{Liu2011} can be reconciled with the much shallower 
 slopes obtained by \citet{Bigiel2008} 
and \citet{Blanc2009} if: the relation between SFR and cloud mass is super--linear, specifically $\beta\sim$1.5; a uniform background at the level of $\approx$60\%--80\% of the SFR (i.e., 30\%--40\% of the galaxy luminosity)  is present in the  
SFR maps of \citet{Bigiel2008} and \citet{Blanc2009}; a slight over--removal (by about  1\% of the SFR) of this 
background is present in the SFR data of \citet{Liu2011}. This scenario can also account for the trend in $\sigma_{H2}$ at fixed region size; at 750~pc,  it decreases from a value 
$\sim$0.45  \citep{Liu2011} to $\sim$0.2  \citep{Bigiel2008} (and from $\sim$0.44 to 0.25 in our models), if a uniform 
contribution to the SFR is present  in 
the measurements from the latter authors. Finally, while the slope measured by \citet{Blanc2009} is similar to that of \citet{Bigiel2008}, their scatter is about a factor of 2 
larger; the shallow slope in \citet{Blanc2009} is likely a combination of residual background in the SFR maps, together with 
the use of low significance data in their fits; the large scatter $\sigma_{H2}$ in \citet{Blanc2009}  reflects the impact of stochastic sampling of the 
molecular cloud mass function in their small regions \citep[170~pc in size versus 750-1,000~pc of][]{Bigiel2008}. 

\subsubsection{Other Galaxies}

We can attempt to interpret other results within this scenario. One of the best studied galaxies in this regard is our neighbor M33 \citep{Wong2002, Heyer2004, Verley2010, Onodera2010}, which, at only 840~kpc distance and with modest inclination (54$^{\circ}$), offers one of the most unimpeded 
views of star formation in an external spiral galaxy. We concentrate on the results of \citet{Verley2010}, who perform both azimuthally averaged and spatially resolved analyses, in addition to discriminate between OLS and least square fitting techniques. \citet{Verley2010} attempt to 
account for the contamination of the SFR maps from an unrelated diffuse emission in the infrared (at the level of 30\%), and use data above 2~$\sigma$ significance. We compare the results from \citet{Verley2010} to our Default Model with log[M$_{cloud}$(max)]=6.5, to approach the actual 
situation in M33, where molecular clouds have maximum masses around 10$^6$~M$_{\odot}$ \citep{Engargiola2003}. Our $\beta=$1.5 simulation of azimuthally--averaged bins, with maximum radius of 6~kpc and radial step of 240~pc, as adopted by \citet{Verley2010} for their data, yields  $\gamma_{H2}$=1.30$\pm$0.10, when using the bi--linear regression fit routine FITEXY, to be compared with the observed value $\gamma_{H2}$=1.1$\pm$0.1. \citet{Heyer2004} obtain a larger value for $\gamma_{H2}$=1.36$\pm$0.08,  more similar to our model results. 
 These authors do not remove a low frequency background and should, 
in principle, obtain a shallower slope than that of \citet{Verley2010}. Their steeper slope could be  partly due to an insufficient accounting of the unobscured SFR, since the authors trace the SFR with infrared emission only. The direct correlation between intensity of the star formation and dust attenuation \citep[e.g.,][]{Calzetti2007} gives larger weight to the brighter star--forming regions when infrared SFR tracers are used, and yields a steeper SK Law, while use of UV and/or optical (i.e. subject to dust attenuation) SFR tracers will 
yield shallower--than--expected SK Laws. Within this framework, we can easily interpret \citet{Wong2002}'s results, which use H$\alpha$ as a SFR tracer and obtain steeper slopes when galactocentric--dependent dust attenuation corrections are applied to the SFR data.  

For the spatially resolved analysis, \citet{Verley2010} adopt a scale of 180~pc, and use both FITEXY and the OLS bi--sector fitting, obtaining $\gamma_{H2}$=2.22$\pm$0.07 and 1.46$\pm$0.34, respectively; these are  to be compared with our simulation's $\gamma_{H2}$=2.47$\pm$0.18 and 1.61$\pm$0.02, respectively, and $\sigma_{H2}$=0.43. \citet{Onodera2010} derive a slope $\gamma_{H2}$=1.18$\pm$0.11, on scales of 1~kpc, using a 2~$\sigma$ detection limit for the data and a least--square fit routine. They 
do not remove a low--frequency background from their SFR data. Assuming a background level similar to that determined by \citet{Verley2010} (corresponding to 60\% 
background level in our convention),  we determine $\gamma_{H2}$=0.99$\pm$0.08 for M33 at 1~kpc scale using our bi--linear regression fit routine, in  
good agreement 
with the value from \citet{Onodera2010}. In general, our Default Model, adapted to the smaller clouds in M33, reproduces the observed trends in the SK Law slope fairly well, and the actual $\gamma_{H2}$  typically within 1--2~$\sigma$ uncertainties, for $\beta$=1.5. 

\citet{Thilker2007} perform a pixel--based analysis of NGC7331, a highly inclined galaxy (77$^{\circ}$) at a distance of 14.7~Mpc. The authors' 6$^{\prime\prime}$ bins correspond, once projection effects are removed, to areas with $\sim$900~pc size. 
The authors also separate between regions dominated by H$_2$ and regions dominated by HI; the two regimes are distinguished by the H$_2$--dominated 
regime having Log($\Sigma_{H2}$)$\gtrsim$0.8 and the HI--dominated regime 
characterized by  Log($\Sigma_{H2}$)$\lesssim$1.1. For the first regime, a plot of $\Sigma_{SFR}$ versus $\Sigma_{H2}$ gives a slope of 1.64, 
while in the second case the authors recover a shallower slope, $\gamma_{H2}$=1.20, thus yielding a difference in slope of $\Delta\gamma_{H2}\sim$ 0.45. 
Using $\beta$=1.5 and dividing the simulations into two regimes at Log($\Sigma_{H2}$)=0.9, we get $\gamma_{H2}$=1.58 for the high $\Sigma_{H2}$ range and $\gamma_{H2}$=1.16 for the low $\Sigma_{H2}$ range, with  $\Delta\gamma_{H2}\sim$0.4, similar to that derived by \citet{Thilker2007}. 

\citet{Rahman2011} present a pixel--based analysis for a sample of 14 nearby star--forming late--type spirals, using interferometry CARMA data for the molecular 
gas, with 1~$\sigma$ detection limit around 3~M$_{\odot}$~pc$^{-2}$, and the 24~$\mu$m emission to trace the SFR. To avoid potential contamination 
from low--surface--brightness quiescent regions, they limit their analysis to pixels with gas surface density $\Sigma_{H2}\ge$20~M$_{\odot}$~pc$^{-2}$. Analysis 
using both pixels at native resolution (6$^{\prime\prime}$) and at common 1~kpc resolution yield an overall slope $\gamma_{H2}\sim$1.1$\pm$0.1 
and a galaxy--average slope $\gamma_{H2}=$0.96$\pm$0.16 (OLS bisector fit). We use the best fit model for M51a, which includes a 80\% background contribution, 
simulated at 1~kpc region size to compare with the results of \citet{Rahman2011}; for log($\Sigma_{H2}$)$\ge$1.3, we find $\gamma_{H2}$=1.10, 
with $\sigma_{H2}$=0.22. When combining simulations in a range of region sizes between 200~pc and 1,000~pc, to approximate the authors' analysis 
at fixed angular resolution of 6$^{\prime\prime}$, we obtain $\gamma_{H2}$=1.21. 
Thus, despite the unavoidable variations from galaxy to galaxy, the model 
used to account for the observational results of M51a can be effectively applied to other late--type spirals. Our result also indicates that restricting analyses to 
molecular gas surface brightnesses $\Sigma_{H2}\ge$20~M$_{\odot}$~pc$^{-2}$ is not completely effective at avoiding the influence of a diffuse background contribution to the SFR, because of the potentially large background levels. Higher thresholds for  $\Sigma_{H2}$ may, conversely, excessively restrict the dynamical range in gas surface density, thus decreasing the fidelity of the fitted relations.

In conclusion, when we have sufficient information on the observational conditions to implement those conditions in our simulations, 
we tend to reproduce both the power law exponent and scatter about the mean of the Observed SK Law,  using as a starting point a 
common relation between SFR and cloud mass. 

\subsection{Limitations}

Although our simple model can account for some of the trends observed in recent data analyses of the SK Law, it contains a number of 
simplifications that will limit its applicability. For instance, in the previous section, we remarked that SK Law analyses 
involving specific selection criteria for the sub--galactic regions, used in lieu of blind pixellation, cannot be reproduced by our simulations.

The most notable limitation, however, is the {\em constant scaling} adopted for all simulations sharing the same relation between 
SFR and M$_{cloud}$. Thus, we cannot account for situations where the scaling of gas--to--stars conversion changes. 
This includes the `global' SK Law, i.e., the relation between the SFR and gas surface densities for whole galaxies, where the 
observed trend, either as a progressive \citep{Kennicutt1998} or sudden \citep{Daddi2010, Genzel2010} change in the relation between 
the two quantities, has been interpreted in terms of a change in the star formation efficiency (SFE) as the total SFR surface density increases 
from normal disks to starburst galaxies.

The constant scaling can also affect the comparison between simulations and observations for azimuthally averaged studies, if the SFE changes 
among substructures within  galaxies \citep{Momose2010}, and/or is higher in the center than in the outer regions (e.g., the galaxy hosts a central starburst). It should be remarked, however, that there is a degeneracy between changes in SFE and a non--linear relation between SFR and M$_{cloud}$. For instance, a low SFE in interarm regions may be simply the result of these regions hosting 
small clouds, unlike their spiral arm counterparts \citep[e.g.][]{Koda2009}; this situation is simulated in our models through variations in the covering 
factor of regions. The resolution of the degeneracy between the two potential interpretations (increase in SFE versus variations in the clouds' maximum mass  in different environments) will require resolving structures in galaxies at the molecular cloud scale (e.g., with ALMA). 

Our models do not include a direct connection between SFR and dense gas \citep{Heiderman2010,Lada2010}. We use a 
very simplified and ad--hoc model for  the relation between SFR and relatively low density gas (n$\sim$10$^2$~cm$^{-3}$, as traced by low--J $^{12}$CO). 
Within this scenario, a  linear relation between SFR and high density gas could imply a non--linear correlation 
between low and high density gas. Recent investigations on a few molecular clouds in proximity of the Sun give contrasting results \citep{Gutermuth2011,Lada2011}, highlighting the need for more extensive analyses. 

Finally, we do not model in detail any physical displacement between recent star formation and peaks of cold gas emission, which is, however, 
observed in many galaxies \citep[e.g.,][]{Kennicutt2007,Momose2010}. This displacement will affect results at the smallest region sizes, and 
may be a partial cause for the large scatter observed in the data of \citet{Blanc2009}. However, we mitigate the absence 
of a displacement component in the models by assigning a scatter to the SFR--cloud~mass relation, which allows for the existence of clouds 
with much lower and much higher SFR than the mean. 

\section{Summary and Conclusions}

In an attempt to understand the large variety of results on the Observed (molecular) SK Law reported in the literature, we have simulated 
spatially--resolved galaxies using simple recipes for the relation between SFR and cloud mass, the cloud mass and radius, and 
the cloud mass function. We have added reasonable sources of scatter, such as a random covering factor for each region (uniformly or exponentially 
distributed), gaussian scatter 
between the cloud mass and radius and between the SFR and cloud mass, and detection noise in $\Sigma_{H2}$. We have then proceeded to investigate the slope $\gamma_{H2}$ and scatter about the best fit line, $\sigma_{H2}$ of the Observed SK Law. 

Our main result is that the Observed molecular SK Law, which relates the surface densities of cold gas and SFR, is a complex convolution of the intrinsic relation between SFR and gas clouds with stochastic sampling of the cloud mass function, and strongly depends on the region 
size considered. Its dynamical range, slope, and scatter 
about the mean trend are affected by the scatter between cloud mass and radius, between SFR and cloud mass, and by the dynamical range and sensitivity of the cold gas data. The slope $\gamma_{H2}$ is also a function of the fitting method adopted (Figure~\ref{fig7}), as 
already noted by \citet{Blanc2009} and \citet{Verley2010}. 

The scatter about the mean trend, $\sigma_{H2}$, is a sensitive function of the measurement uncertainties in $\Sigma_{H2}$, and of the slope 
and scatter in the SFR--M$_{H2}$ relation. For a slope $\beta$=1.0 between SFR and M$_{H2}$, and in the absence of scatter between 
these two quantities, the scatter $\sigma_{H2}$ about the mean trend of the SK Law is dominated by the sensitivity of the cold gas map coupled 
with the non--symmetric nature of error bars in logarithmic scale. As $\beta$ increases, also $\sigma_{H2}$ increases: at constant region size 
$<$1~kpc, $\sigma_{H2}$ roughly triples and quadruples for $\beta$=1.5 and 2.0, respectively, relative to the $\beta$=1.0 value. This increase of $\sigma_{H2}$ for increasing $\beta$ reflects the combination of the non--linear relation between SFR and M$_{H2}$ and stochastic sampling of 
the cloud mass function at small region sizes. Introducing a scatter between SFR and M$_{H2}$ has the general (and expected) effect of 
increasing $\sigma_{H2}$ at all $\beta$ values. 

The presence of a large contribution to $\sigma_{H2}$ from the typically shallow 
 sensitivity limit of the cold gas maps argues in favor of pursuing deep CO imaging of nearby galaxies to improve their sensitivity and increase the dynamical  range probed. Significant improvements have been brought by CO mapping with the Nobeyama 45--m millimeter telescope 
 \citep[e.g.,][]{Momose2010,Liu2011}, and more will be ushered by the Large Millimeter Telescope and other large single--dish facilities. 
 More sensitive maps than currently available will break the degeneracy among the various contributors to the scatter $\sigma_{H2}$. 

For SFR$\propto$M$_{H2}^{\beta}$ and $\beta>1.0$, the slope of the Observed SK Law $\gamma_{H2}\ne\beta$, and the SK Law slope is 
in general a decreasing function of the region size. At small region sizes, typically $<$400--500~pc, $\gamma_{H2}>\beta$, and at large region 
sizes, $>$700~pc--1~kpc, $\gamma_{H2}\lesssim\beta$. As the region size increases, $\gamma_{H2}$ tends to unity, a reflection of the increasing ability of the larger regions to sample the cloud mass function reliably.  These general trends with region size persist for a large range of 
parameter variations, including cloud mass function slope and maximum cloud mass, cloud covering factor distribution, and variation in the scatter between SFR and M$_{H2}$. 
They also persist when changing sensitivity limits and dynamical range for the cold gas maps (Figures~\ref{fig8} and \ref{fig9}), and the fitting method. While we model our `galaxies' as single cloud layers (areal covering with thickness=1~cloud), moving to multiple clouds along the 
line of sight (volume filling) has the general effect of further flattening $\gamma_{H2}$ towards values of unity, irrespective of $\beta$ and region 
size. 

The general absence of a 1--to--1 relation between $\gamma_{H2}$ and $\beta$ poses a challenge for studies of the SK Law on sub--galactic 
scales. In this respect, it becomes important that spatially--resolved studies, but which cannot resolve individual clouds, investigate 
trends as a function of region size. Furthermore, the fact that $\gamma_{H2}$ converges to unity for large region sizes, virtually irrespective 
of any condition, strongly argues against the approach of using kpc--size or larger regions to derive the SK Law, unless these  
sizes are part of a thorough investigation that includes a range of sub--kpc region sizes. 

The case of  $\beta$=1.0 provides the most stable condition for $\gamma_{H2}$ which remains close to $\beta$, within $\pm$0.1/0.2, for 
a large range of parameters. The most notable exceptions  are the cases where: (i) a large scatter is present  between SFR and 
M$_{H2}$ (Figure~\ref{fig5}, bottom--left); and (ii) a large threshold has been applied to the SFR tracers (Figure~\ref{fig12}, left). In both cases, 
large ($>$1) values of $\gamma_{H2}$ are accompanied by large values of the scatter $\sigma_{H2}$ about the mean trend. The case of a SFR
threshold is easier to control, as it is produced at the level of data handling; an artificial threshold, for instance, implies that a fraction of 
regions do not contain detectable SFR. As an example, if 60\% of the SFR at the low end is removed, about 70\% of the 200~pc regions and about 50\% of 
the 300~pc regions do not contain detectable star formation. In this respect, if SFR is linked to the cloud mass via a linear relation, measurements of the Observed SK Law 
will yield in general values close to unity. Deviations from this value should indicate a power--law index $\beta>$1.

We have compared our simulations against the Observed SK Law for three nearby galaxies, M51a, M33, and NGC3521, for which we 
have been able to reconstruct the assumptions that have entered into the analysis of the data by the original authors. In all cases, we have 
been able to reproduce, within 1--2~$\sigma$, the measured slopes and the observed scatter about the mean trend using a common relation between 
SFR and M$_{H2}$ with $\beta$=1.5. The large range of values, especially 
for $\gamma_{H2}$, obtained by different authors are a manifestation of different regions sizes, detection thresholds, fitting methods, and 
treatments of the background contamination that affects 
the emission used to trace SFRs.  Current data, especially for the two low--inclination galaxies M51a and M33, do not support $\beta$=1.0 
as the exponent linking SFR and cloud mass. However, a more extensive set of observations and analyses will be needed to place this conclusion on a firmer footing.

We have also established that including data on the cold gas below a 2--3~$\sigma$ threshold 
causes a flattening of $\gamma_{H2}$  that can result in $\gamma_{H2}<\beta$ (Figure~\ref{fig9}), 
implying that care should be taken when including low significance data in analyses of the SK Law, as already remarked by \citet{Verley2010}.  

Presence of a background unrelated to current star formation in the maps used to trace SFR will generally cause a flattening of $\gamma_{H2}$ 
relative to $\beta$. Presence of a threshold in SFR, for which the least massive clouds do not form stars, will have as general effect a steepening of $\gamma_{H2}$. Hence, a detailed understanding of the contributors to a specific band used to trace SFR will be crucial for pinning down the 
relation between SFR and cold gas. While there is likely general agreement that backgrounds exist, there is still controversy on 
what fraction of the total emission in a galaxy this background represents, and how it is distributed. Clearly, this represents a fundamental and 
necessary step to perform in order to unravel the functional form of the scaling law of star formation, and the combination of trends in $\gamma_{H2}$ and $\sigma_{H2}$ can provide valuable discriminators.

In summary, disentangling the physical relation linking star formation to gas clouds from observational and analysis imprints will require that 
future studies derive the SK Law, its slope and scatter about the mean trend, using a range of physical sizes within each galaxy, and that the 
sensitivity limit and dynamical range of the cold gas data and other conditions in the analysis are carefully taken into account as a source 
of bias for both $\gamma_{H2}$ and $\sigma_{H2}$.

\acknowledgments

This work has been partially supported by the NASA ADP grant  NNX10AD08G. 


\clearpage

\begin{figure}
\figurenum{1}
\plottwo{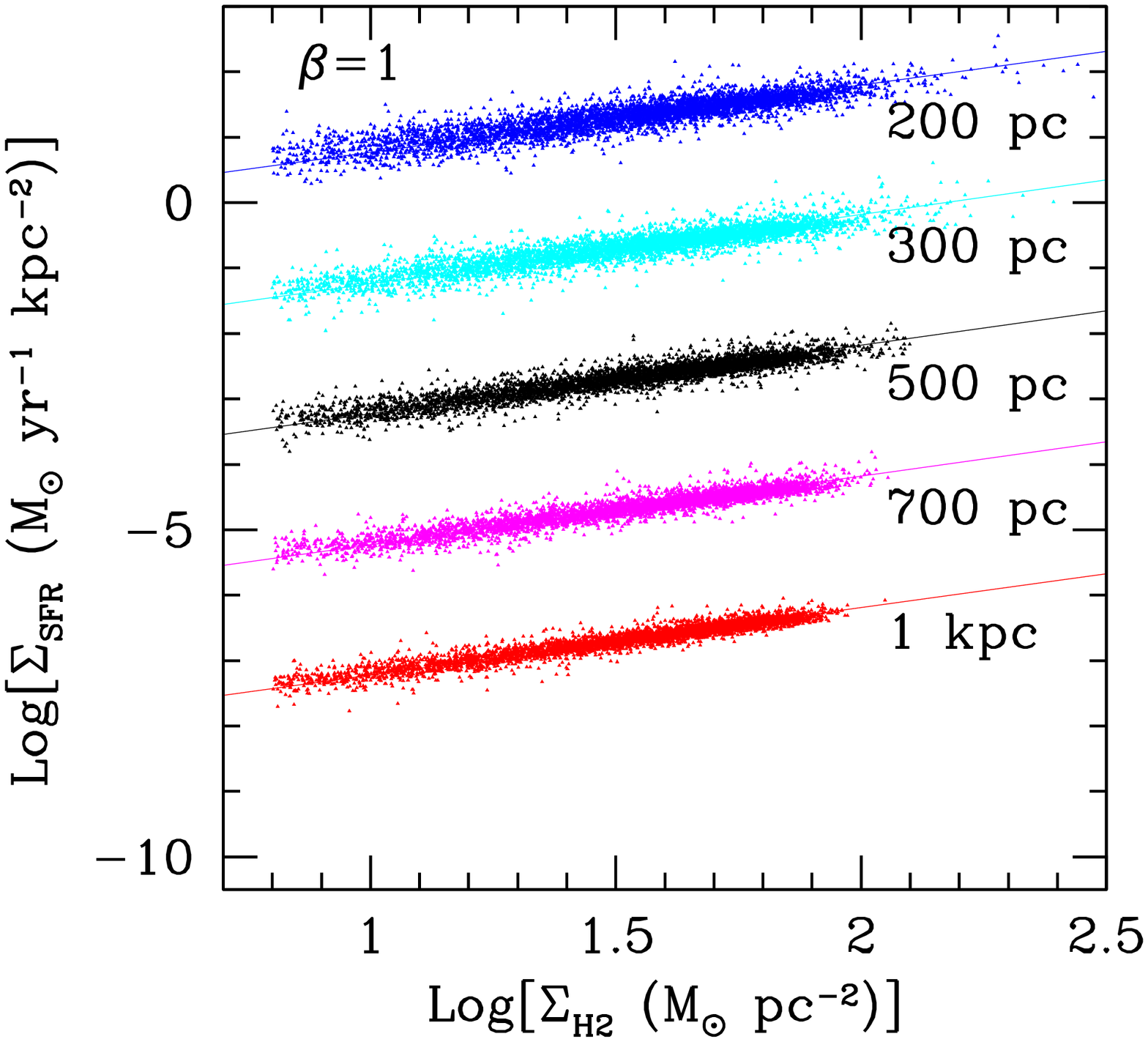}{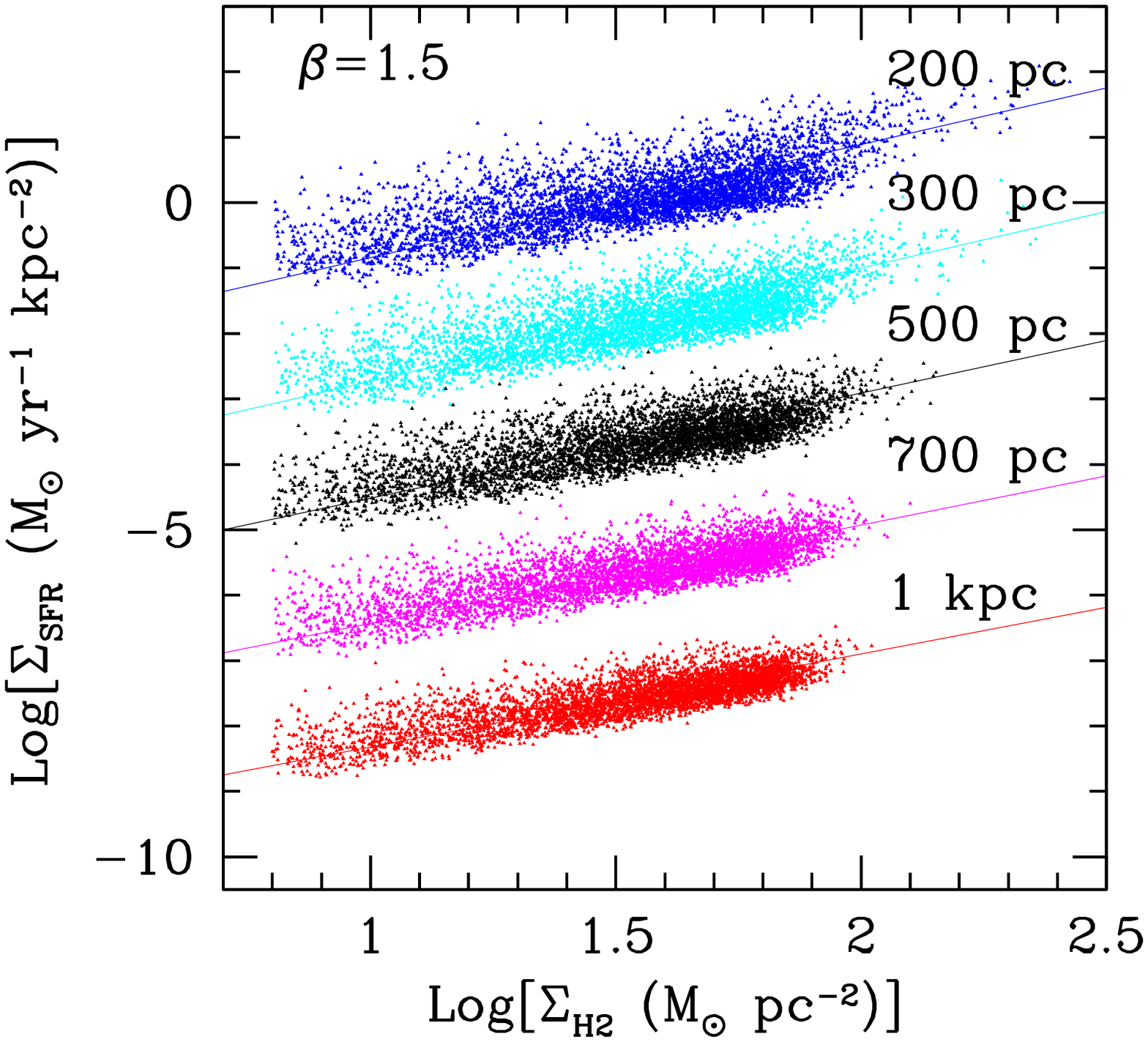}
\plotfiddle{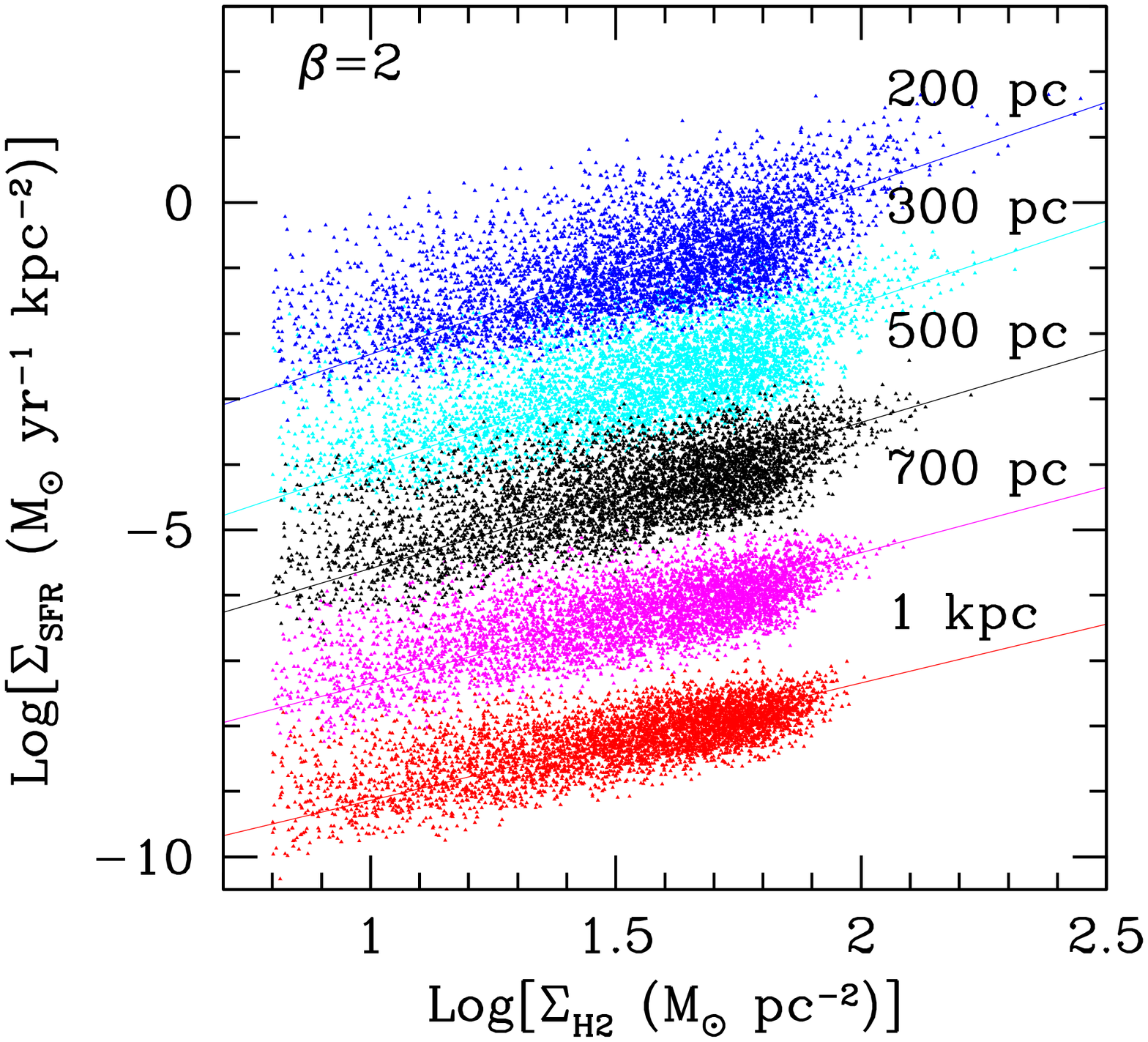}{0.01in}{0.}{215}{215}{0}{0}
\caption{The Observed SK Law, i.e., the scaling relation between the SFR surface density and the molecular gas surface 
density, for the Default Model (section~2) with a uniform distribution of cloud covering factors, and three
choices for the parameter $\beta$ (top left of each panel) that relates SFR and cloud mass: SFR$\propto$M$_{H2}^{\beta}$ (equation~5). 
`Data' from the simulations (color points) and OLS bi-sector linear best fits through the simulation 
results (color lines) are reported for a range of linear sizes of the regions used to calculate surface densities (indicated at the right--hand--side of each 
simulated dataset).  The 300~pc simulations (cyan) are shown at the original scale; the other datasets are shown shifted by $+$2 (200~pc, blue), $-$2 (500~pc, black), 
$-$4 (700~pc, magenta), and $-$6 (1~kpc, red) along the vertical direction.  
\label{fig1}}
\end{figure}

\clearpage 
\begin{figure}
\figurenum{2}
\plottwo{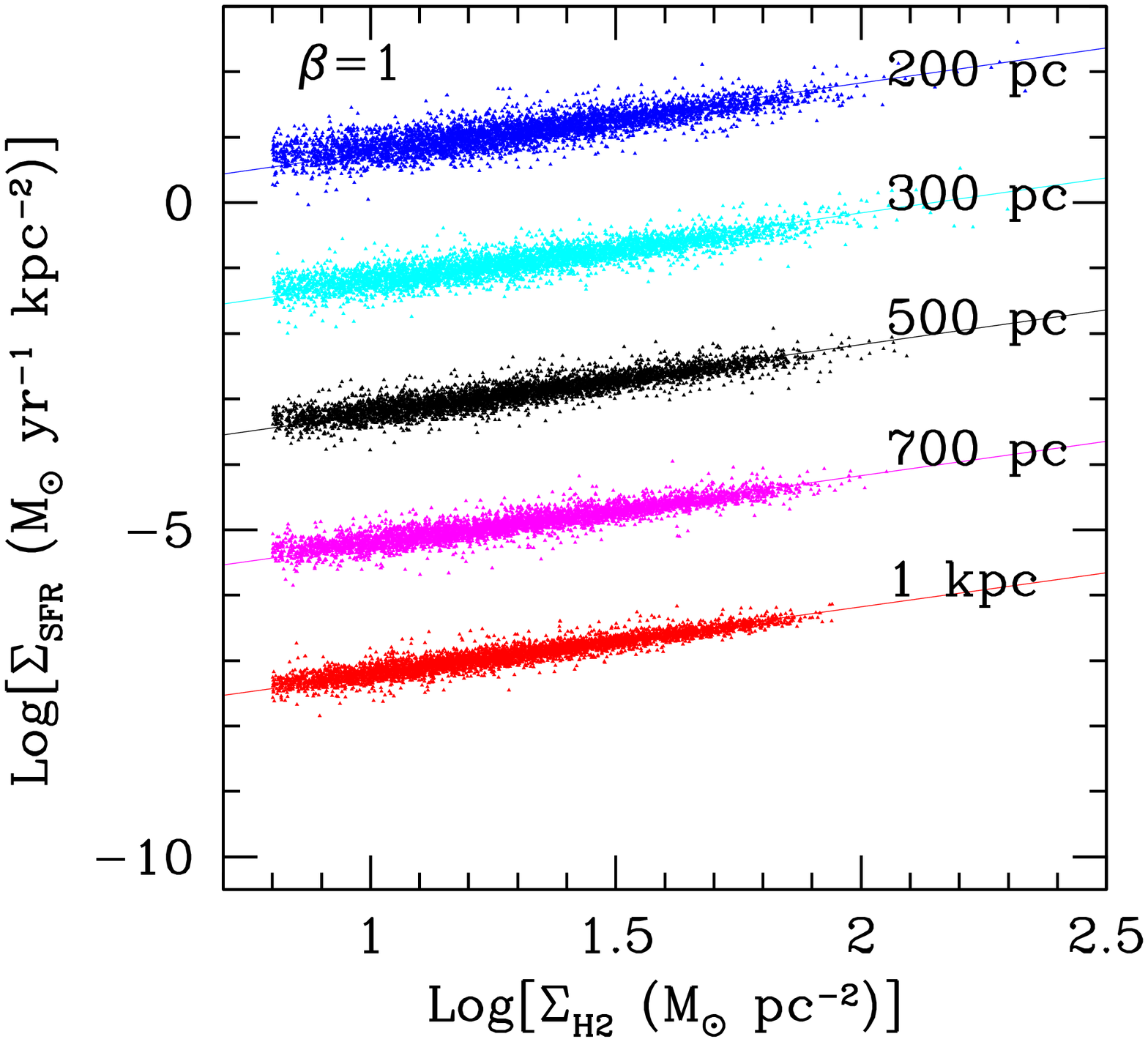}{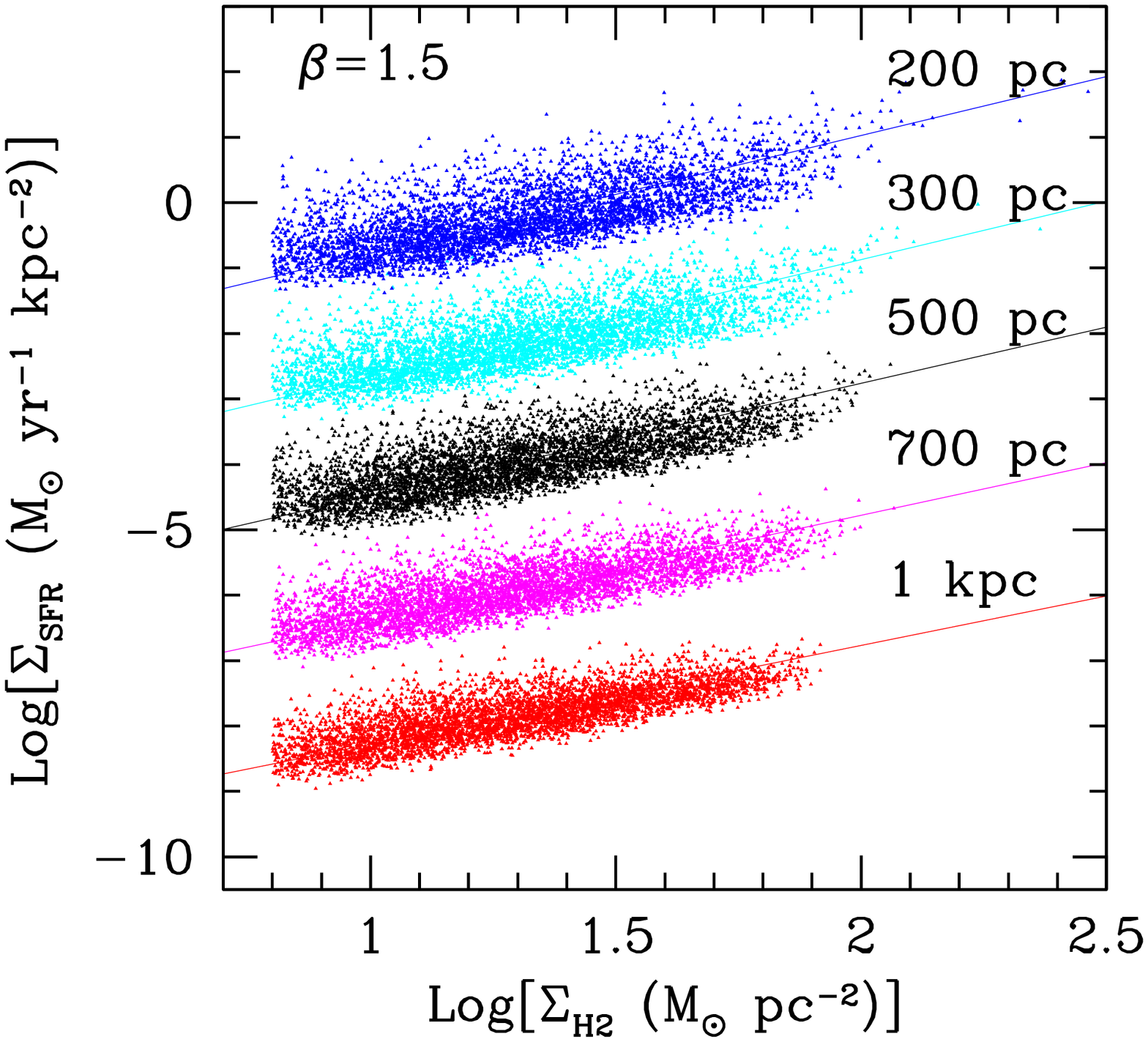}
\plotfiddle{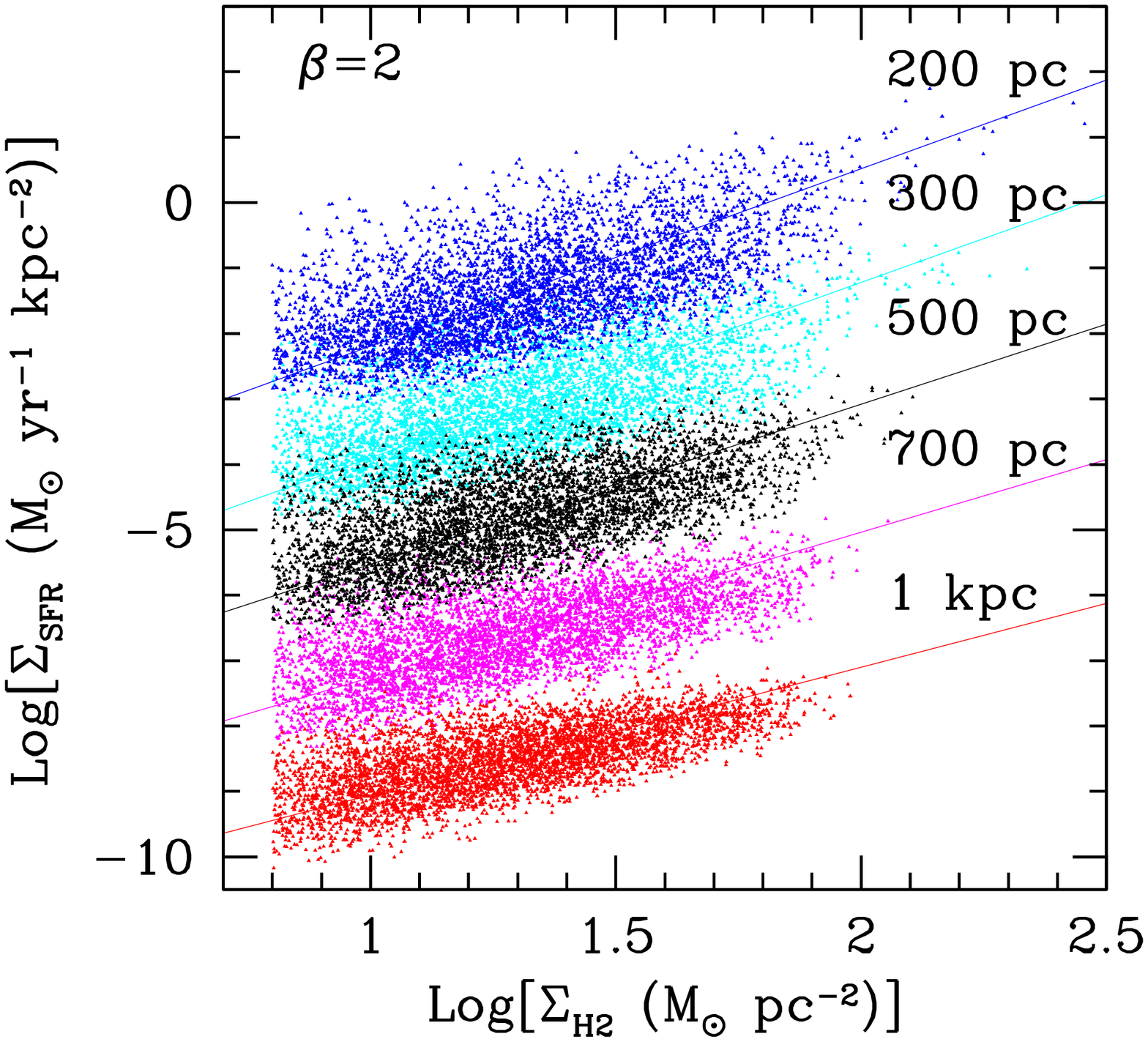}{0.01in}{0.}{215}{215}{0}{0}
\caption{The  same as Figure~\ref{fig1}, but for an exponentially decreasing distribution of cloud covering factors (equation~6). The 
region at low gas surface densities is better populated by the simulated points for this covering factor distribution than in the case 
of a uniform distribution.
\label{fig2}}
\end{figure}

\clearpage 
\begin{figure}
\figurenum{3}
\plottwo{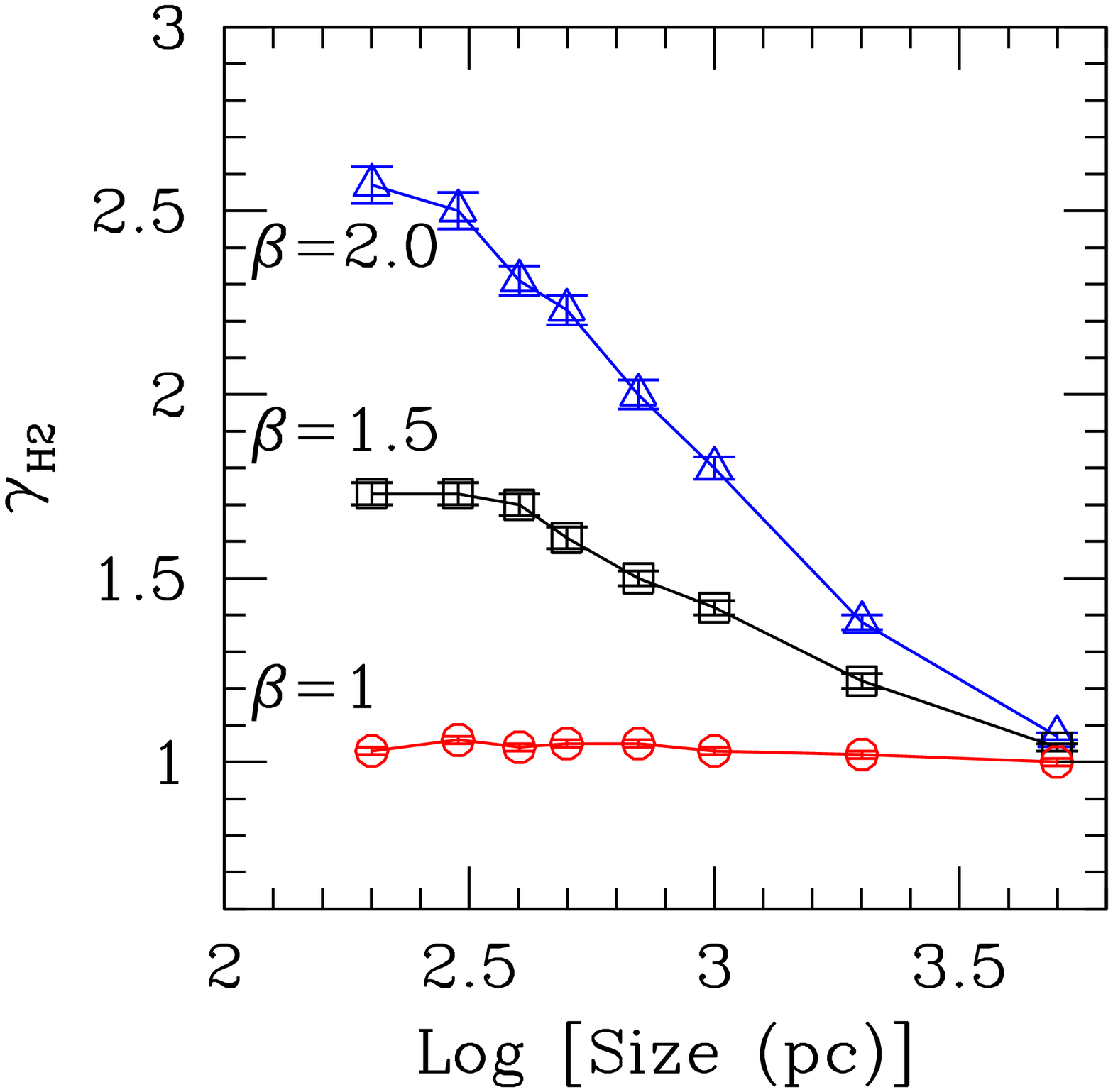}{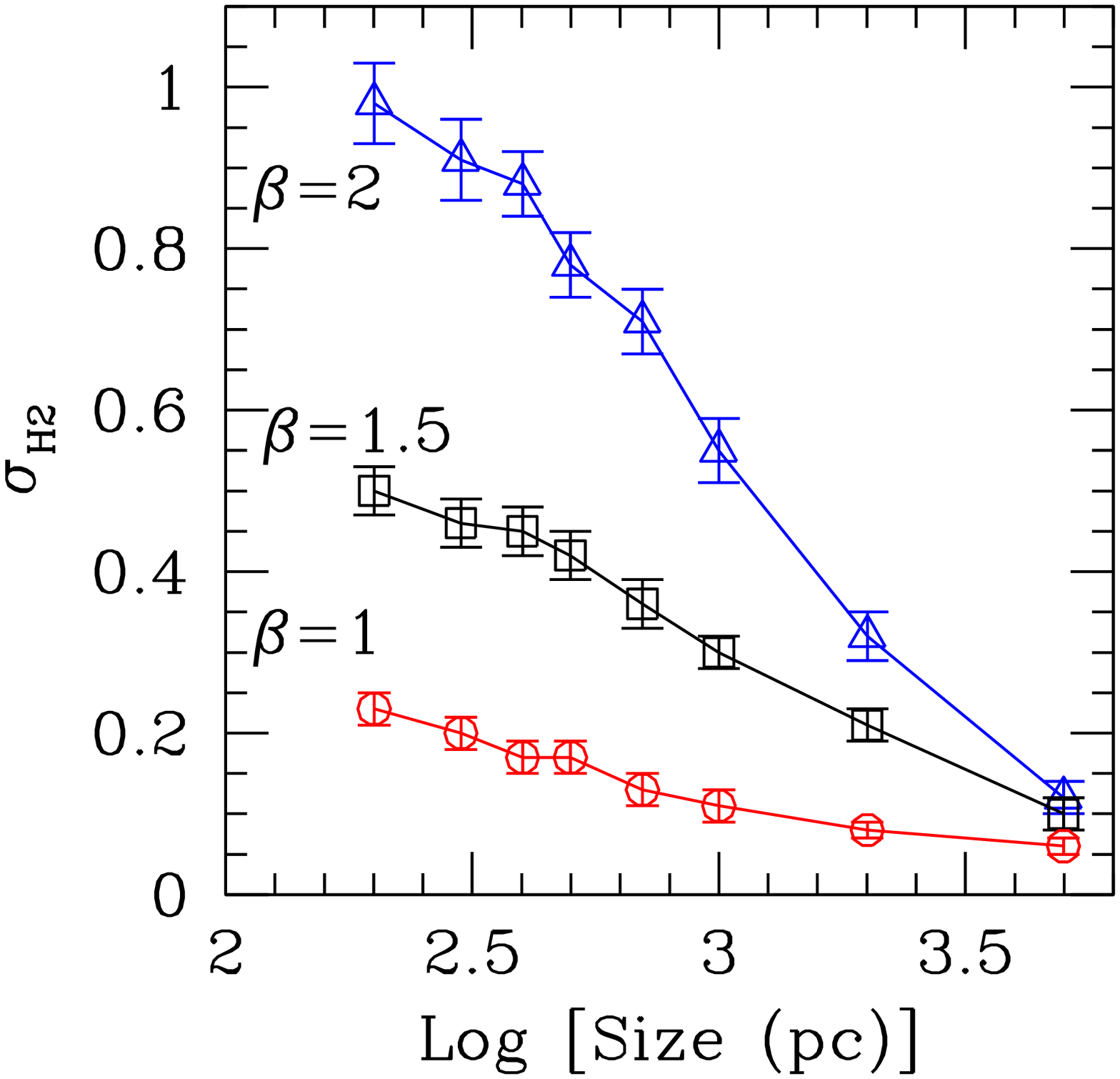}
\plottwo{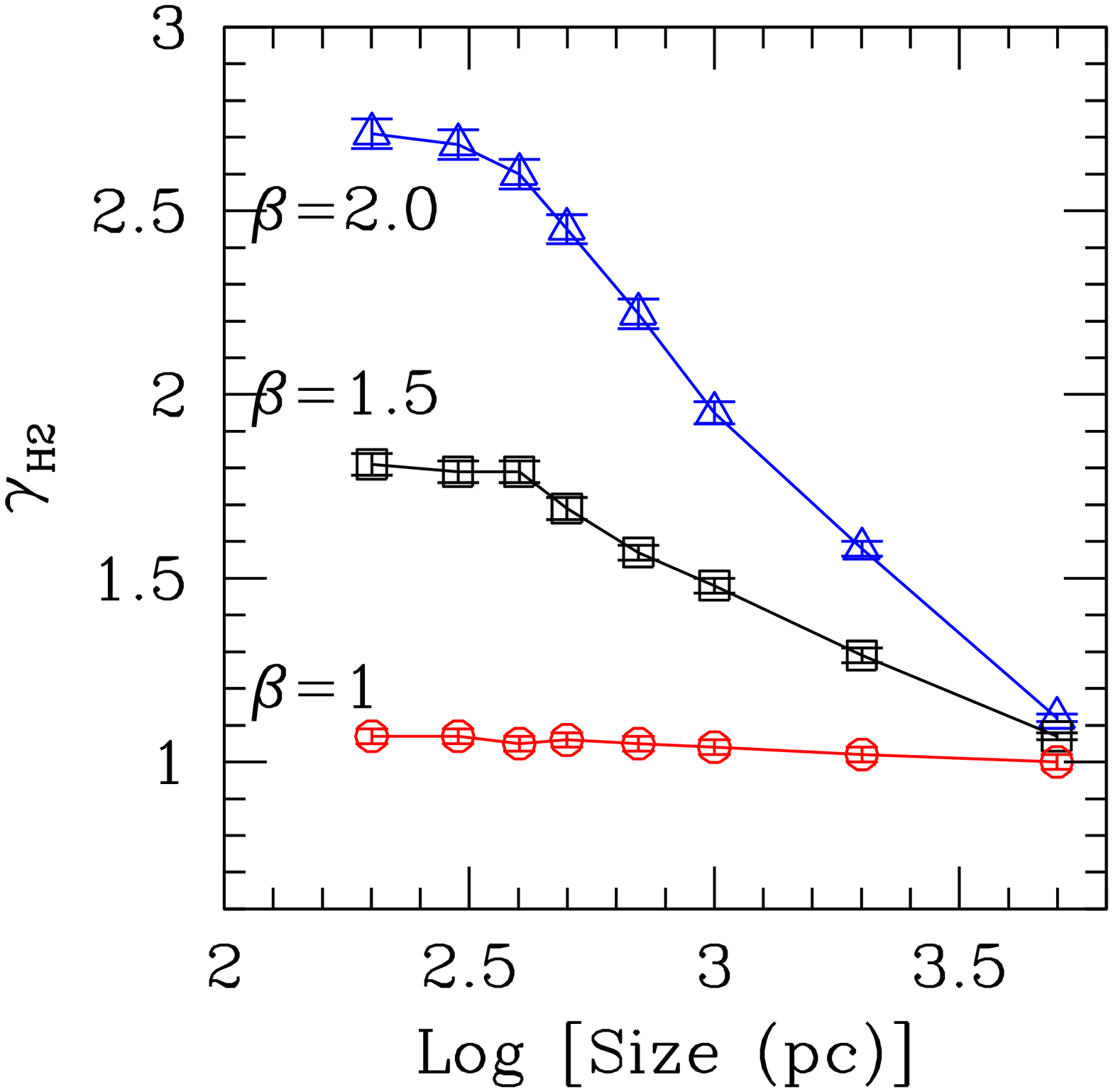}{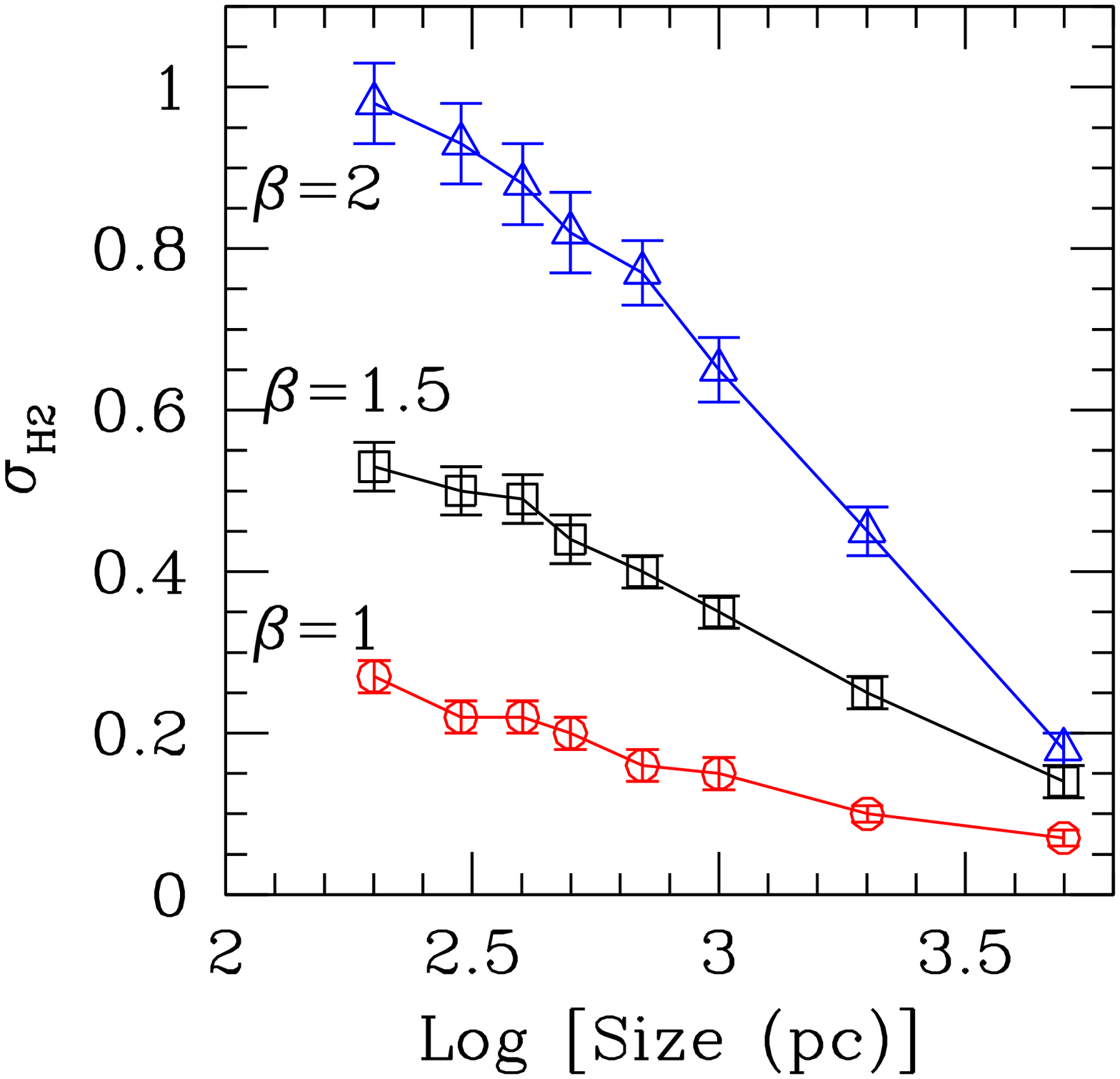}
\caption{The best fit slope $\gamma_{H2}$ (left panels) and dispersion $\sigma_{H2}$ about the best fitting line (right panels) of the 
Observed SK Law (equation~2) for our simulations with Default Model parameters and either a  uniform distribution of cloud covering factors 
(top panels) or an exponentially decreasing distribution of cloud covering factors (equation~6, bottom panels). All quantities are shown 
as a function of the sampling region's size, in the range 200-5000~pc. The symbols connected by lines refer to: 
$\beta$=2 (blue triangles), $\beta$=1.5 (black squares), and $\beta$=1 (red circles), from equation~5.  The measured exponents 
$\gamma_{H2}$ and dispersions $\sigma_{H2}$ are slightly larger in value, by $\sim$0.04--0.14 ($\le$6\%), for the case of exponentially 
decreasing cloud covering factors than for a uniform distribution. 
\label{fig3}}
\end{figure}

\clearpage 
\begin{figure}
\figurenum{4}
\plottwo{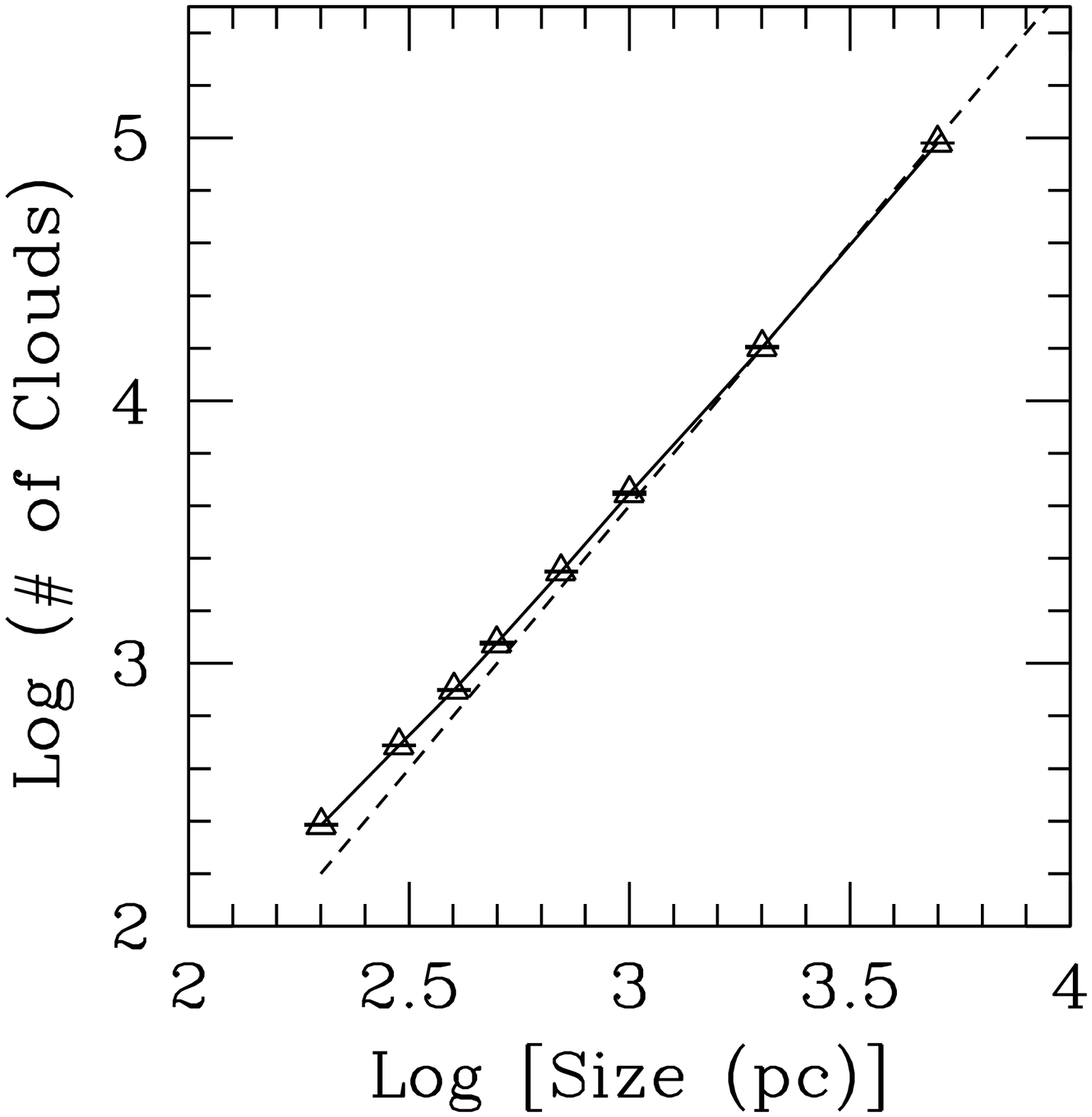}{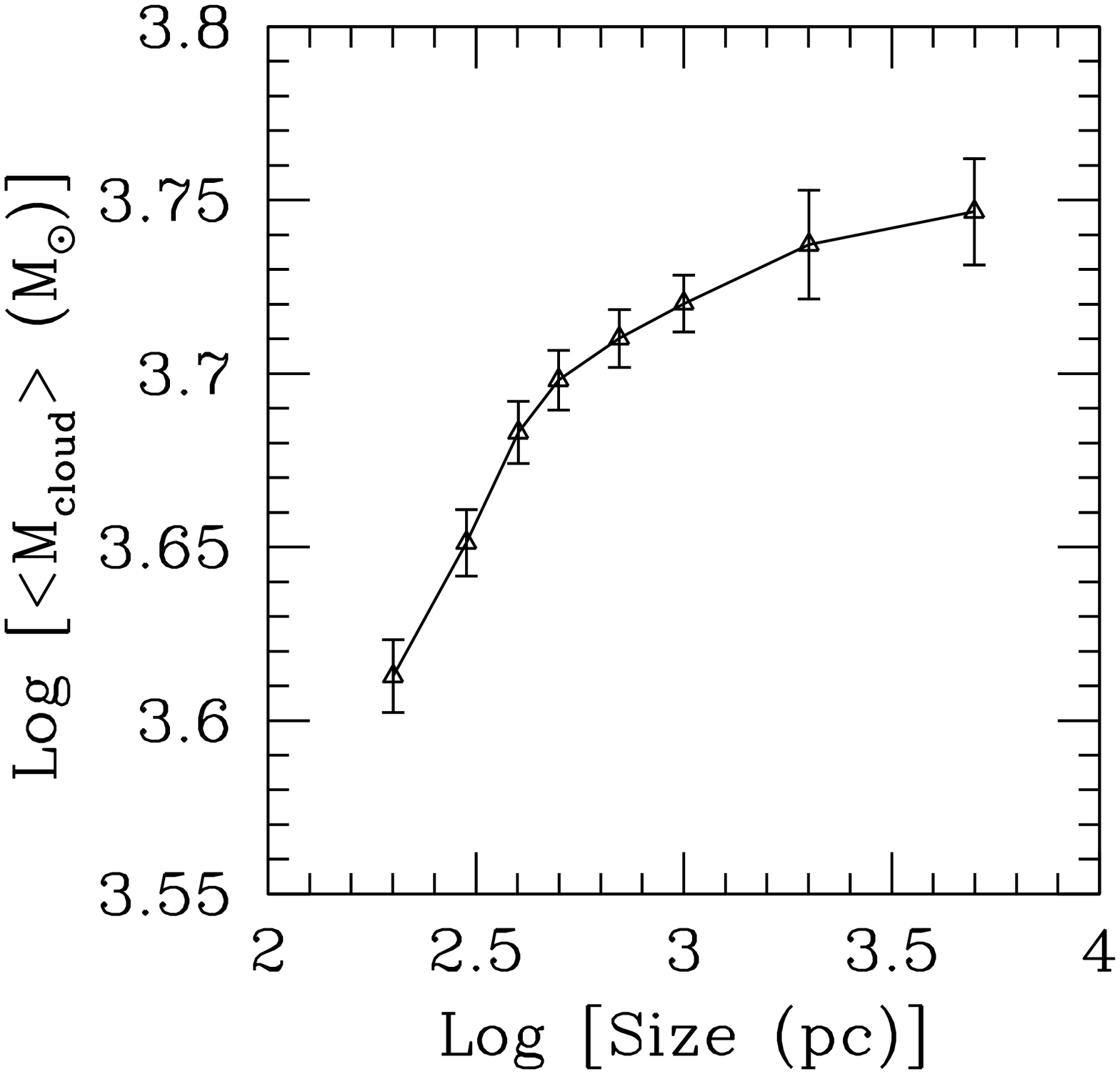}
\plotfiddle{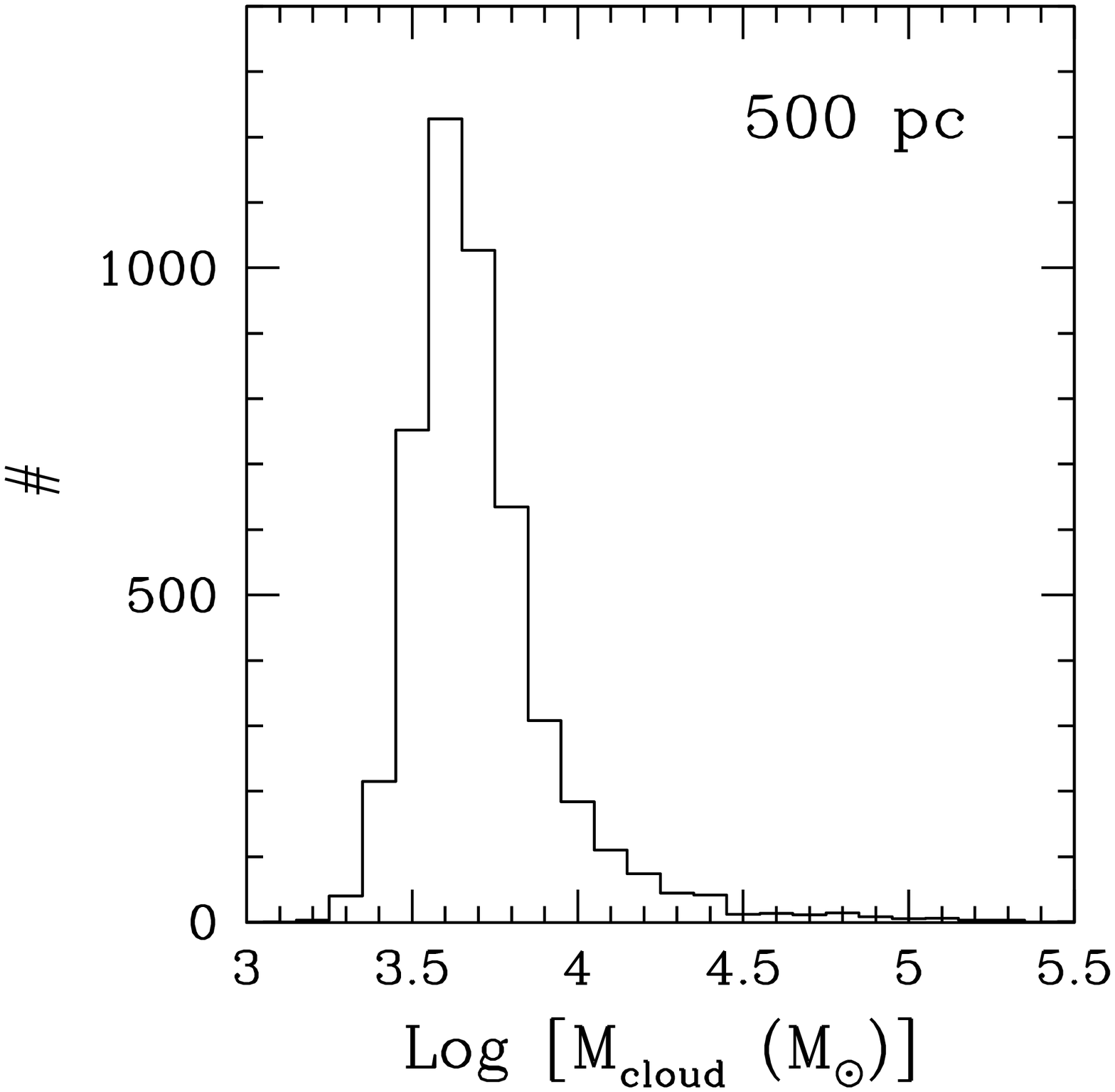}{0.01in}{0.}{215}{215}{0}{0}
\caption{The mean number of clouds in each regions (top--left panel) and the mean 
cloud mass (top--right panel)  as function of region size, for the Default Model with an exponential distribution of cloud covering factors. 
The mean number of clouds  slightly  (by 50\% or less) exceeds 
the locus (dash line) expected if the mean number of clouds grows proportionally to the region's area, up to sizes of $\sim$1--2~kpc. 
 The mean cloud mass flattens only beyond this region size. Both plots  suggest 
that the cloud mass function is not fully 
sampled until about 1--2~kpc. The peak of the distribution of cloud masses within the 500~pc region (bottom--left panel) is  
 very close in value to the mean cloud mass plotted in the top--right panel. Similar plots are obtained for a uniform distribution of cloud covering factors.
\label{fig4}}
\end{figure}

\clearpage 
\begin{figure}
\figurenum{5}
\plottwo{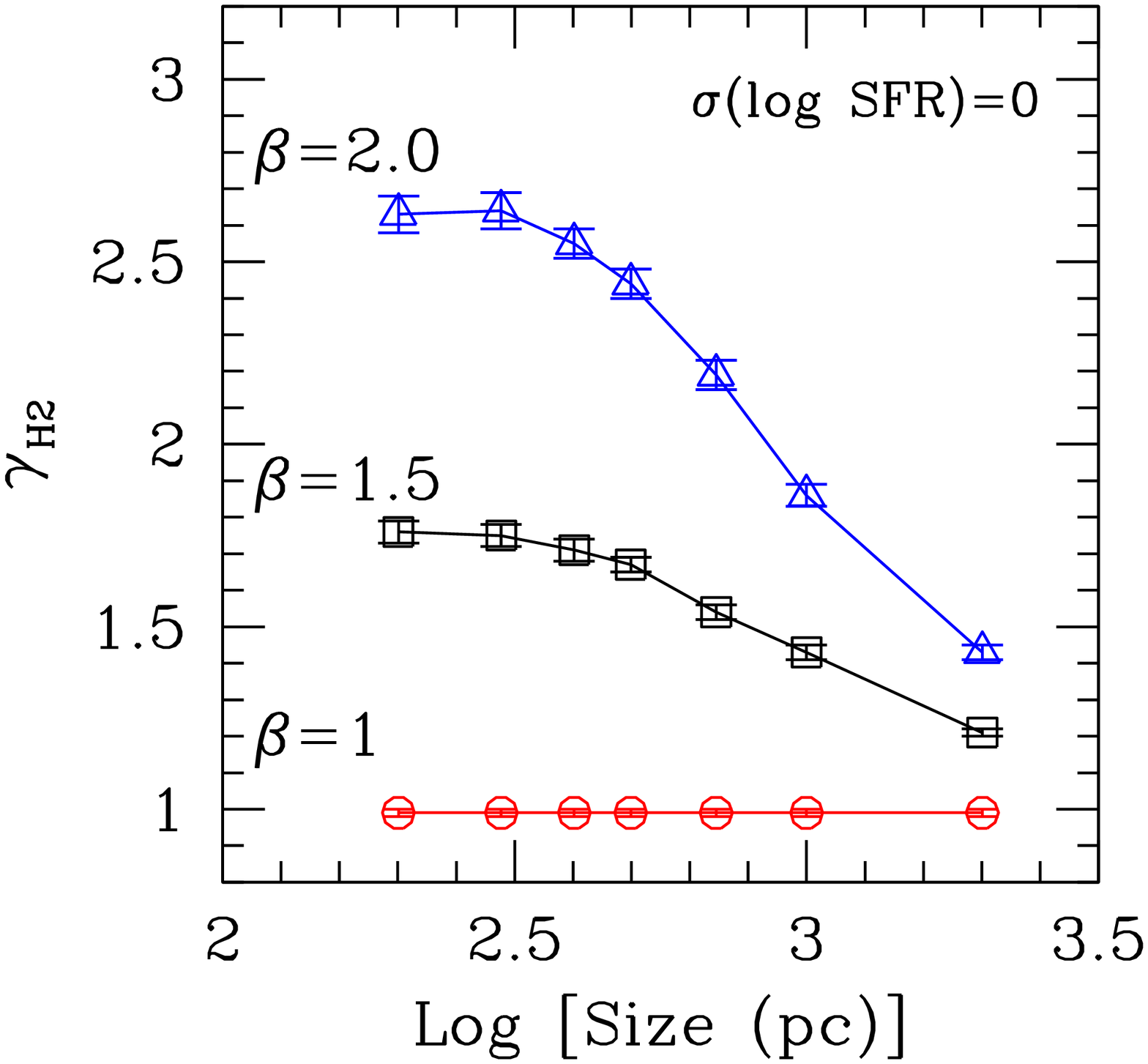}{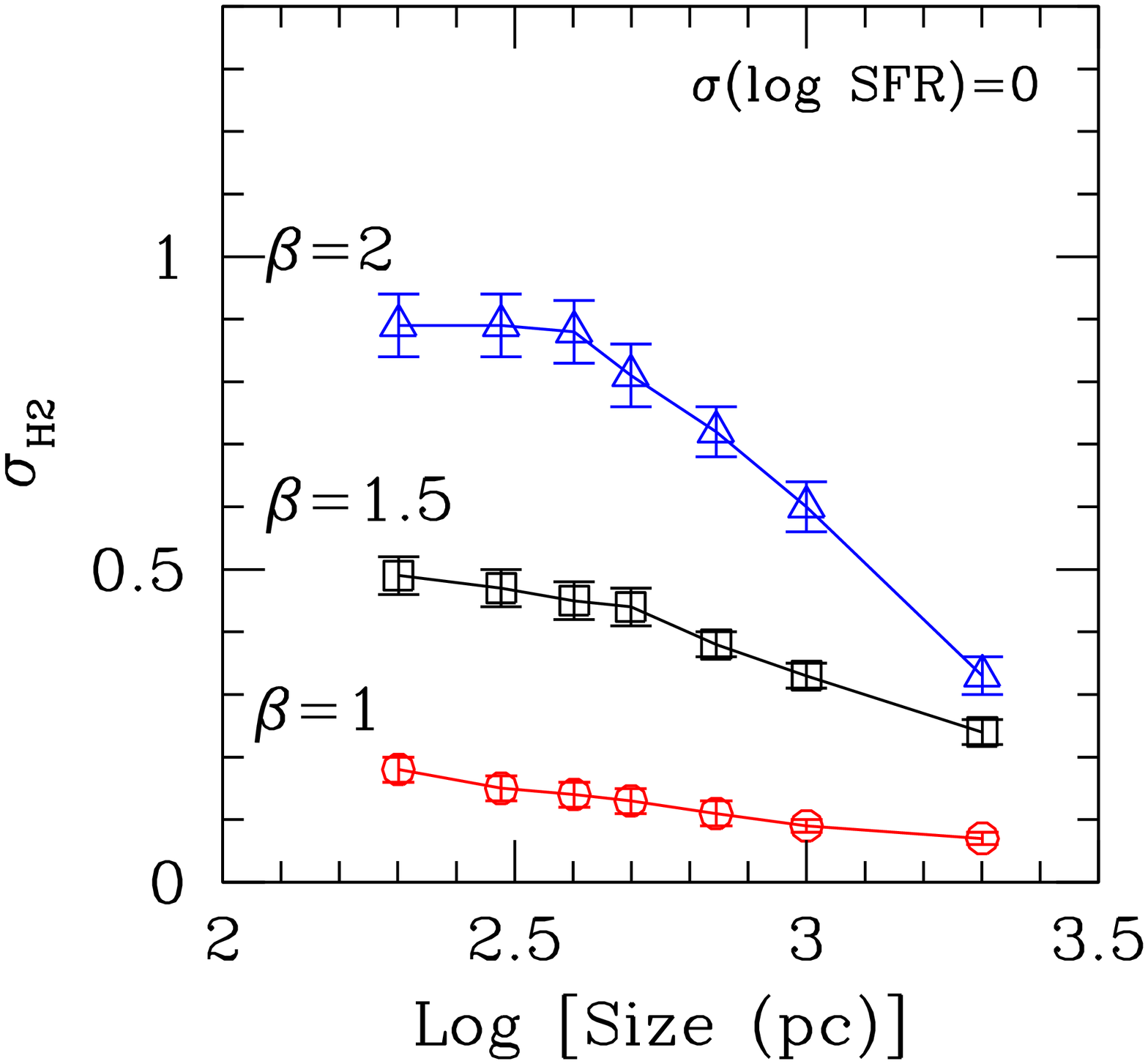}
\plottwo{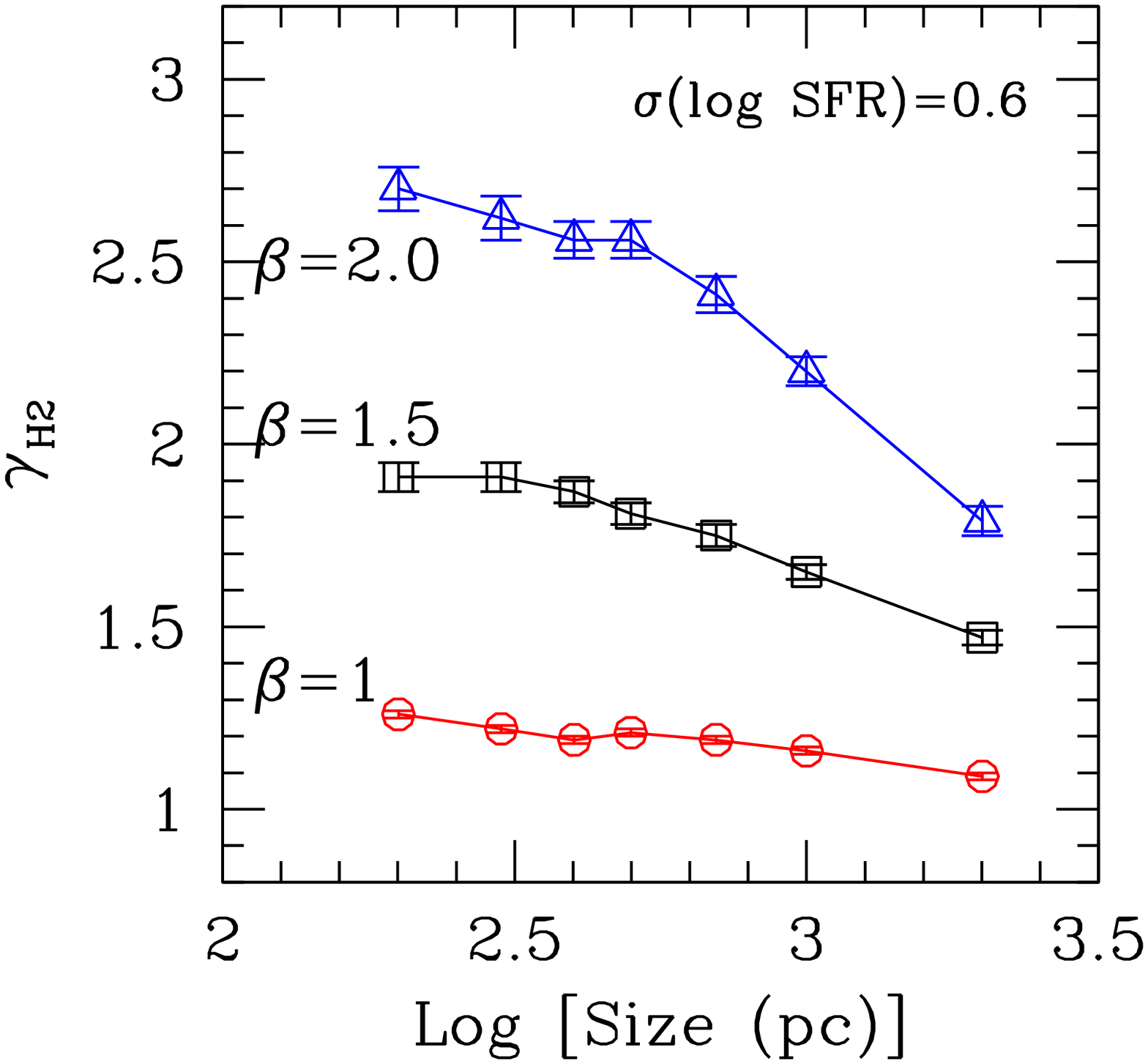}{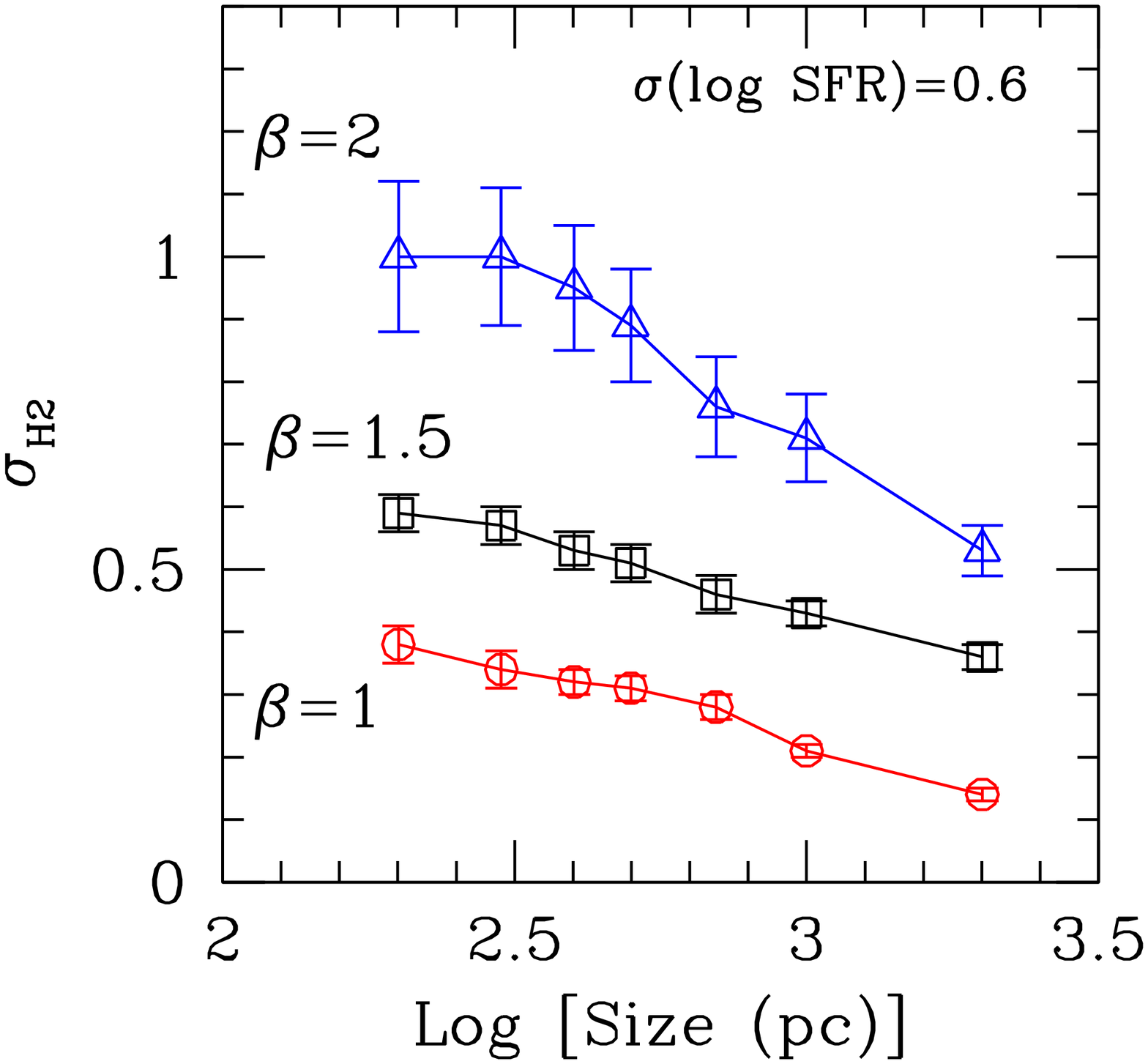}
\caption{The same as Figure~\ref{fig3},  bottom panels (exponential distribution of cloud covering factors), 
for the two cases in which the SFR is deterministically related to the cloud mass ($\sigma(log SFR)$=0, 
upper panels) and the SFR varies by a factor up to 4 (1~$\sigma$, lower panels) at fixed cloud mass. 
The latter case is a factor of 2 larger 
scatter in SFR than our Default Model. Both the measured slope $\gamma_{H2}$ (left panels) and the scatter $\sigma_{H2}$ about the mean fitting 
trend (right panels) are shown for the size range 200--2000~pc.  A non--negligible scatter in the simulated data is present even when there is no scatter between SFR and gas 
cloud mass.
\label{fig5}}
\end{figure}

\clearpage 
\begin{figure}
\figurenum{6}
\plotfiddle{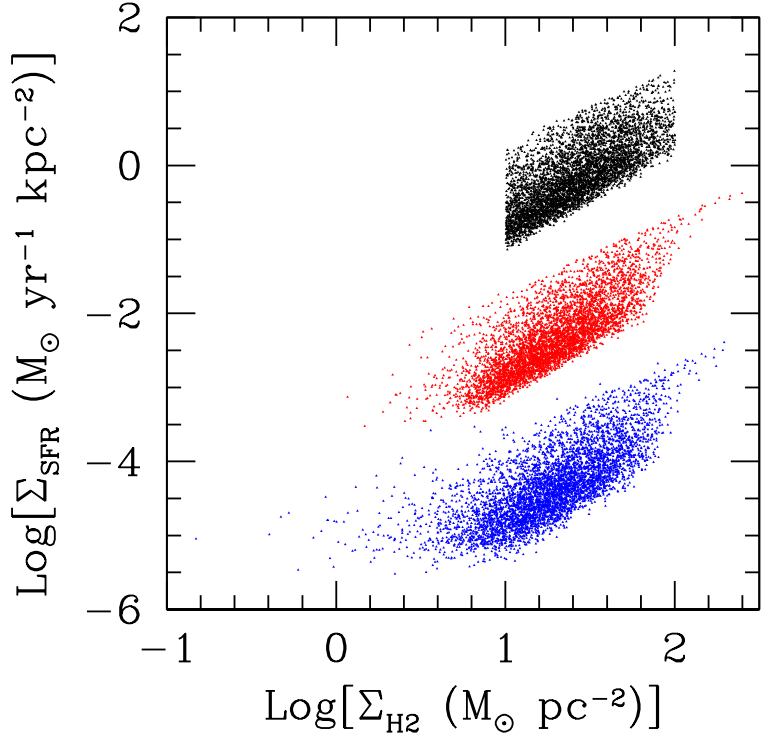}{0.0in}{0.0}{300}{300}{70}{0}
\caption{The same as Figure~\ref{fig2}, for $\beta$=1.5 and 200~pc region size, adding in sources of scatter one at the time. Black points show the distribution of simulated data  for our default range of covering factors,  exponentially and randomly distributed between 10\% and 100\%, but without any scatter for the cloud  mass-radius correlation and without noise in the measurements of $\Sigma_{H2}$. Red points show the distribution of 
the simulated data with the same assumptions as the black points, plus a 60\% scatter in the cloud mass--radius correlation. Blue points show 
the distribution of the simulated data with the same assumptions as the red points, with in addition a scatter in $\Sigma_{H2}$ aimed at simulating measurement errors; our Default Model assumes that values of $\Sigma_{H2}\sim$6.5~M$_{\odot}$~pc$^{-2}$ are detected at the 
3~$\sigma$ level. The addition of both scatters in the cloud mass--radius relation and the $\Sigma_{H2}$ measurements, especially the 
latter one, highlights the non--symmetric nature the distributions acquire in logarithmic plots. Once an additional scatter in the SFR--cloud~mass relation 
is added, the distributions of Figure~\ref{fig2} are recovered.
\label{fig6}}
\end{figure}

\clearpage 
\begin{figure}
\figurenum{7}
\plottwo{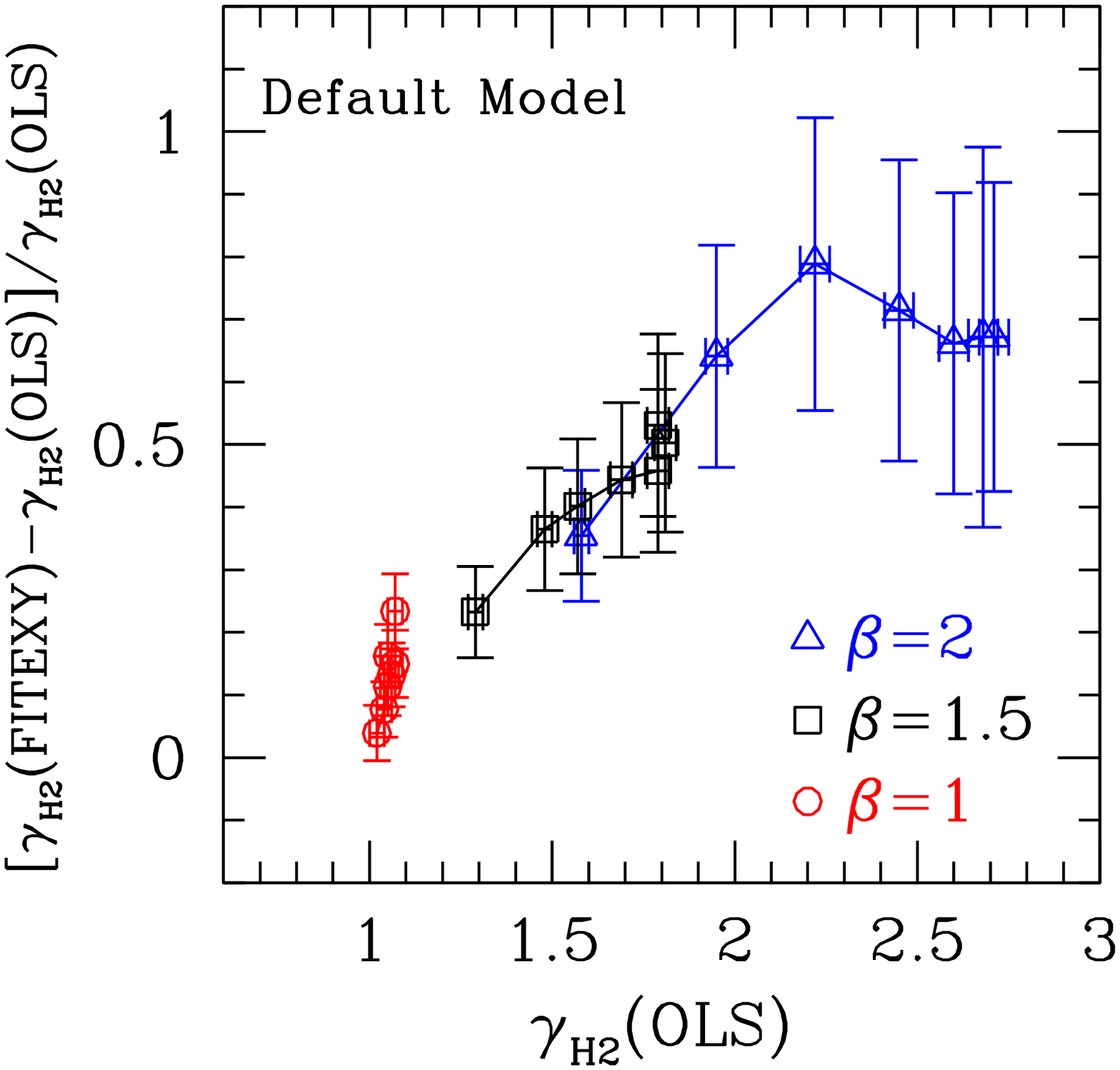}{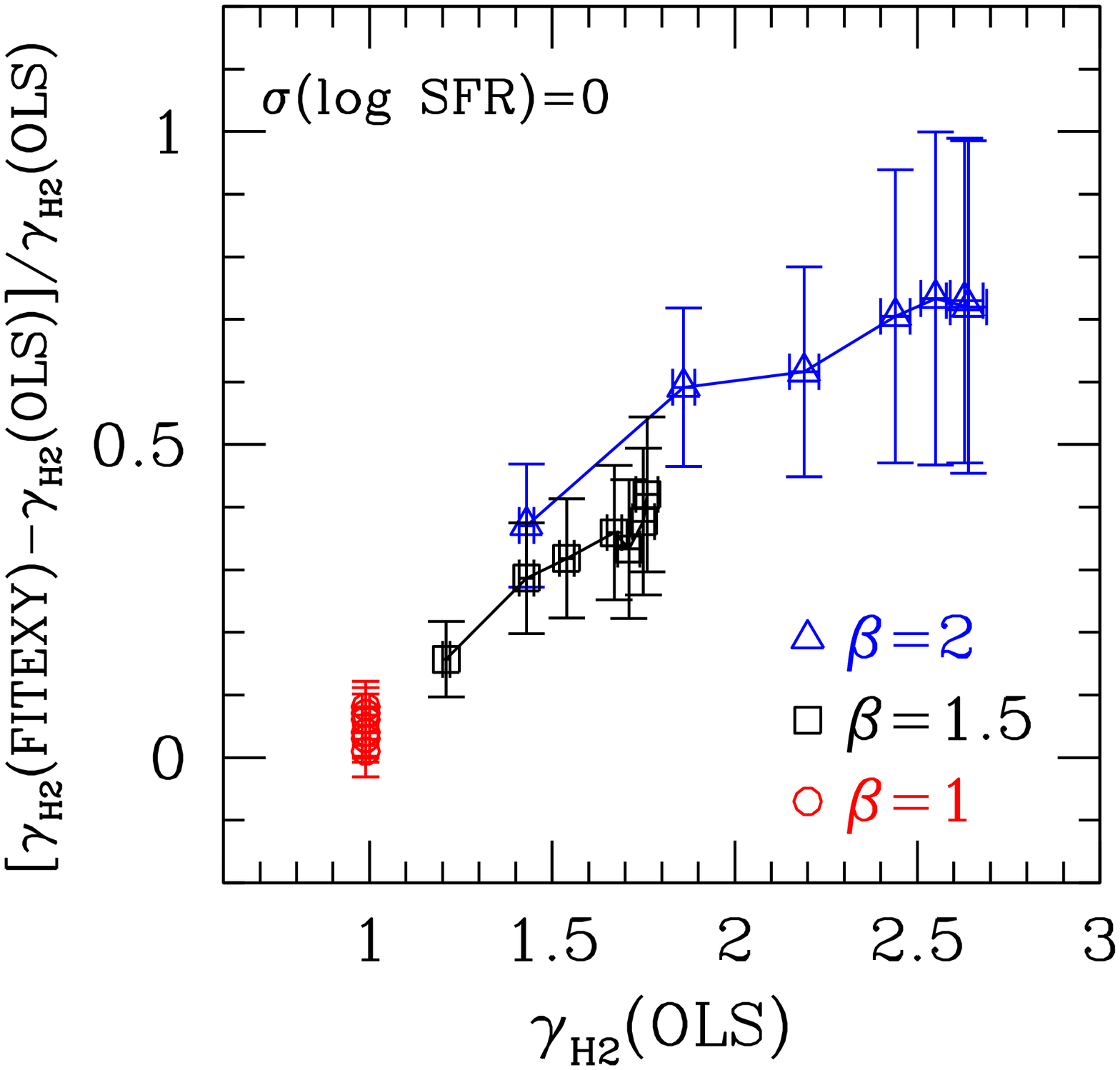}
\caption{A comparison of the best fit slopes $\gamma_{H2}$ measured by the OLS bi--sector linear fitting method and 
by the bi--linear regression fitting method (FITEXY). The fractional difference [$\gamma_{H2}$(FITEXY)$-\gamma_{H2}$(OLS)]/$\gamma_{H2}$(OLS) 
is plotted as a function of the best fit slope from the OLS fitting method. The Default Model ($\sigma(log SFR)$=0.3, left panel) and the case 
of zero scatter between SFR and cloud mass ($\sigma(log SFR)$=0, right panel; see equation~5) are shown, both for an  exponential 
distribution of cloud covering factors. For each value of $\beta$, the 
difference between the slopes determined with the two methods are shown for our range of region sizes; the sizes move from right to left  in the range 200-2,000~pc, in the sense that smaller regions produce steeper slopes with both methods. The bi--linear regression fit 
always measures values that are larger than those obtained from the OLS bi--sector fit, although in neither case the actual value of $\beta$ is 
generally recovered. The slopes obtained via  the FITEXY routine show larger error bars than those from the OLS fit, reflecting the use of 
the data uncertainties in the former fitting algorithm; the FITEXY best fit uncertainties dominate the vertical error bars in the two plots.
\label{fig7}}
\end{figure}

\clearpage 
\begin{figure}
\figurenum{8}
\plottwo{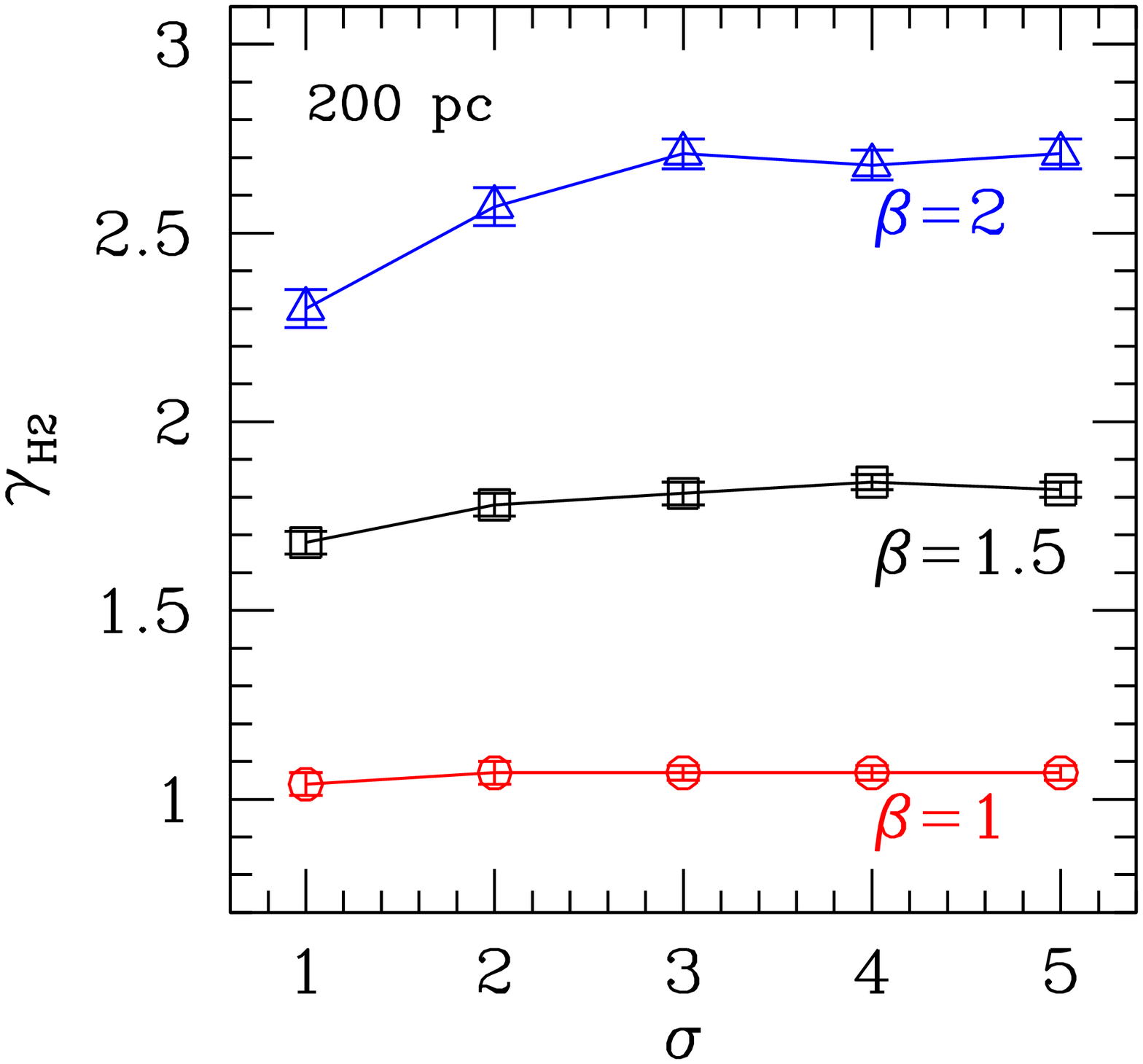}{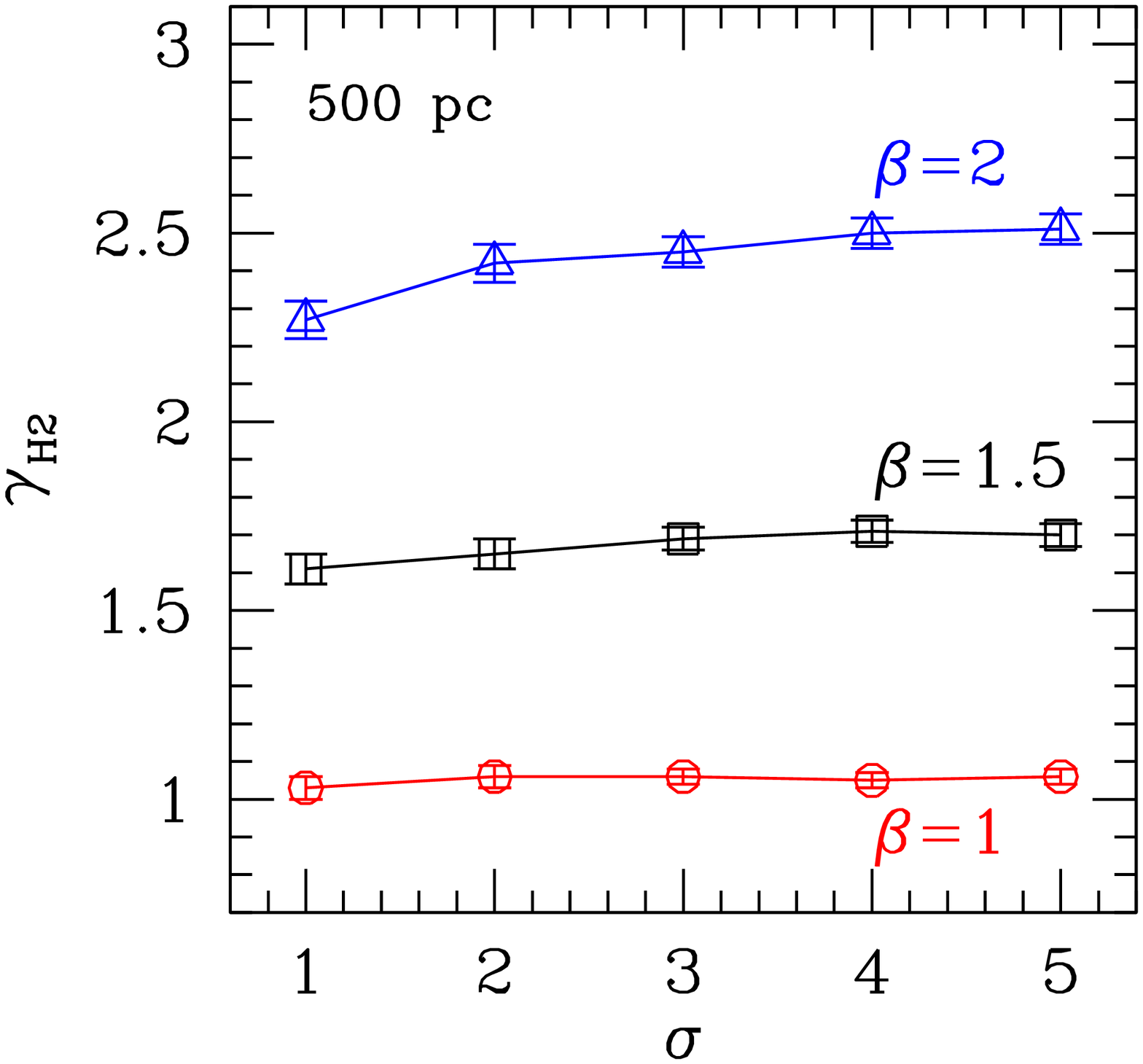}
\plotfiddle{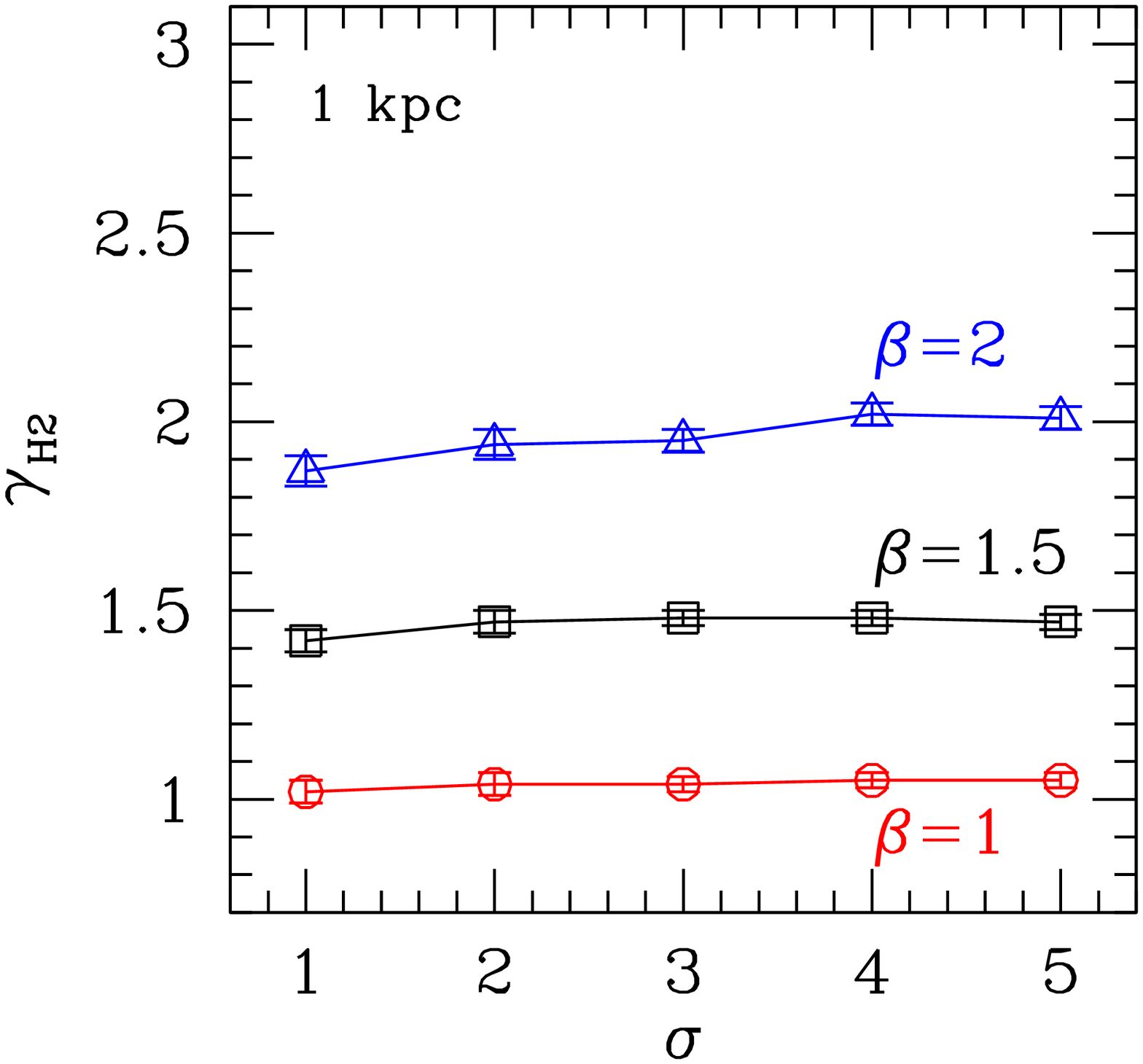}{0.01in}{0.}{215}{215}{0}{0}
\caption{The slope $\gamma_{H2}$ of the Observed SK Law for three representative region sizes (200, 500, 1000~pc), for 
data thresholds that vary between 1~$\sigma$ and 5~$\sigma$. The thresholds are varied while keeping the dynamical range of the 
simulated data unchanged. A modest increase in $\gamma_{H2}$ for increasing threshold is observed at all region sizes, with the 
largest variations present for  $\beta$=2 ($\delta \gamma_{H2}/\gamma_{H2}\lesssim$15\%) and the smallest for $\beta$=1 
($\delta \gamma_{H2}/\gamma_{H2}\lesssim$5\%). 
\label{fig8}}
\end{figure}

\clearpage 
\begin{figure}
\figurenum{9}
\plottwo{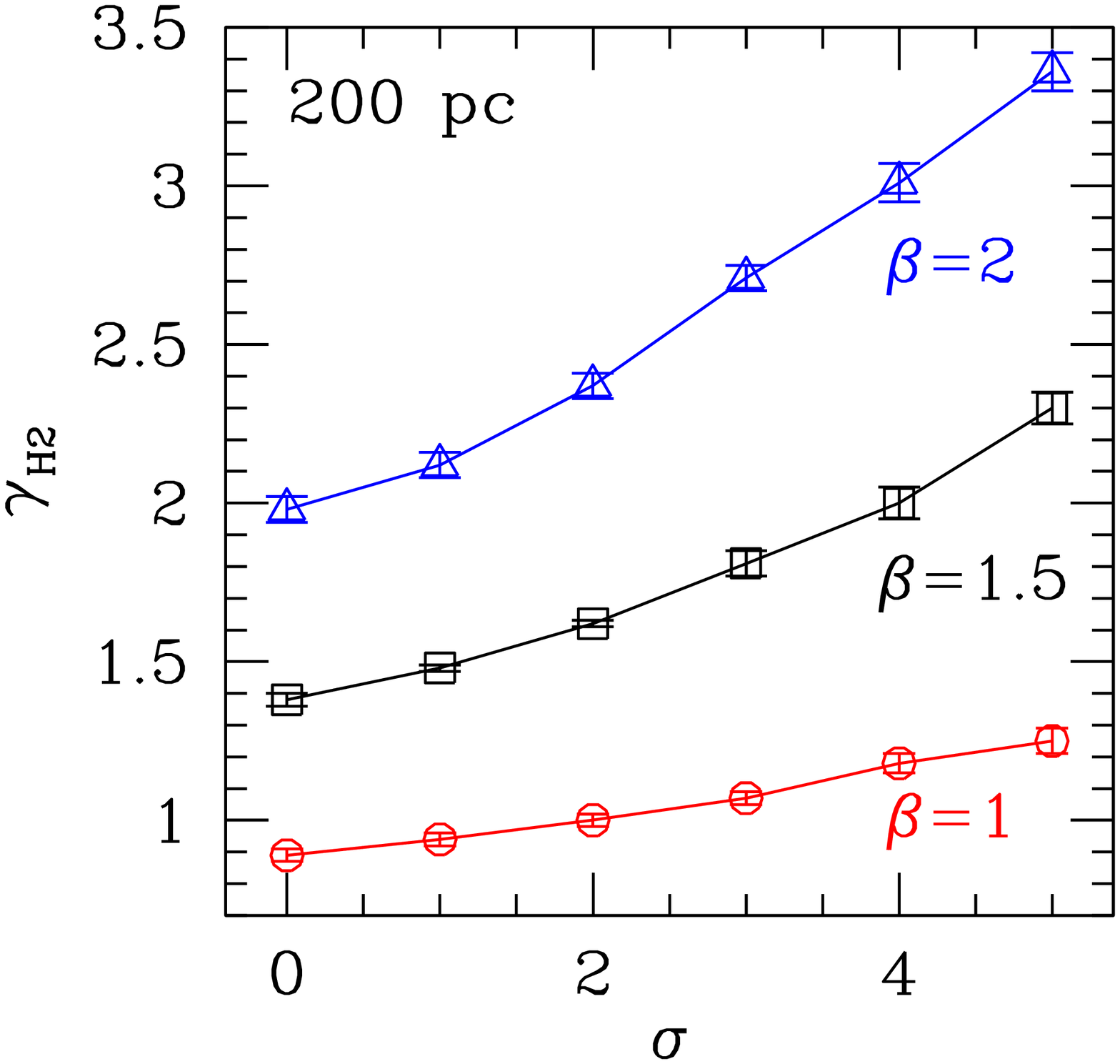}{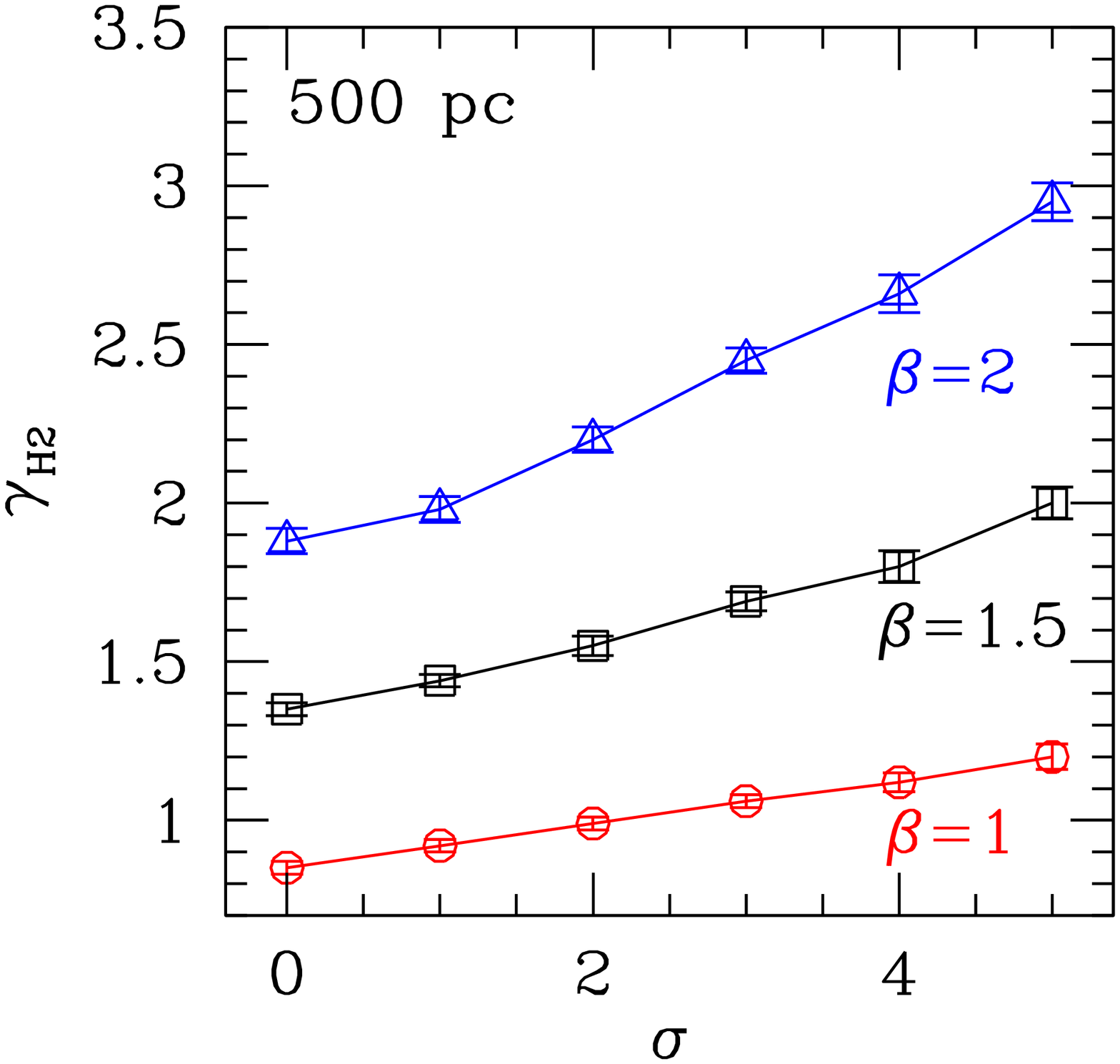}
\plotfiddle{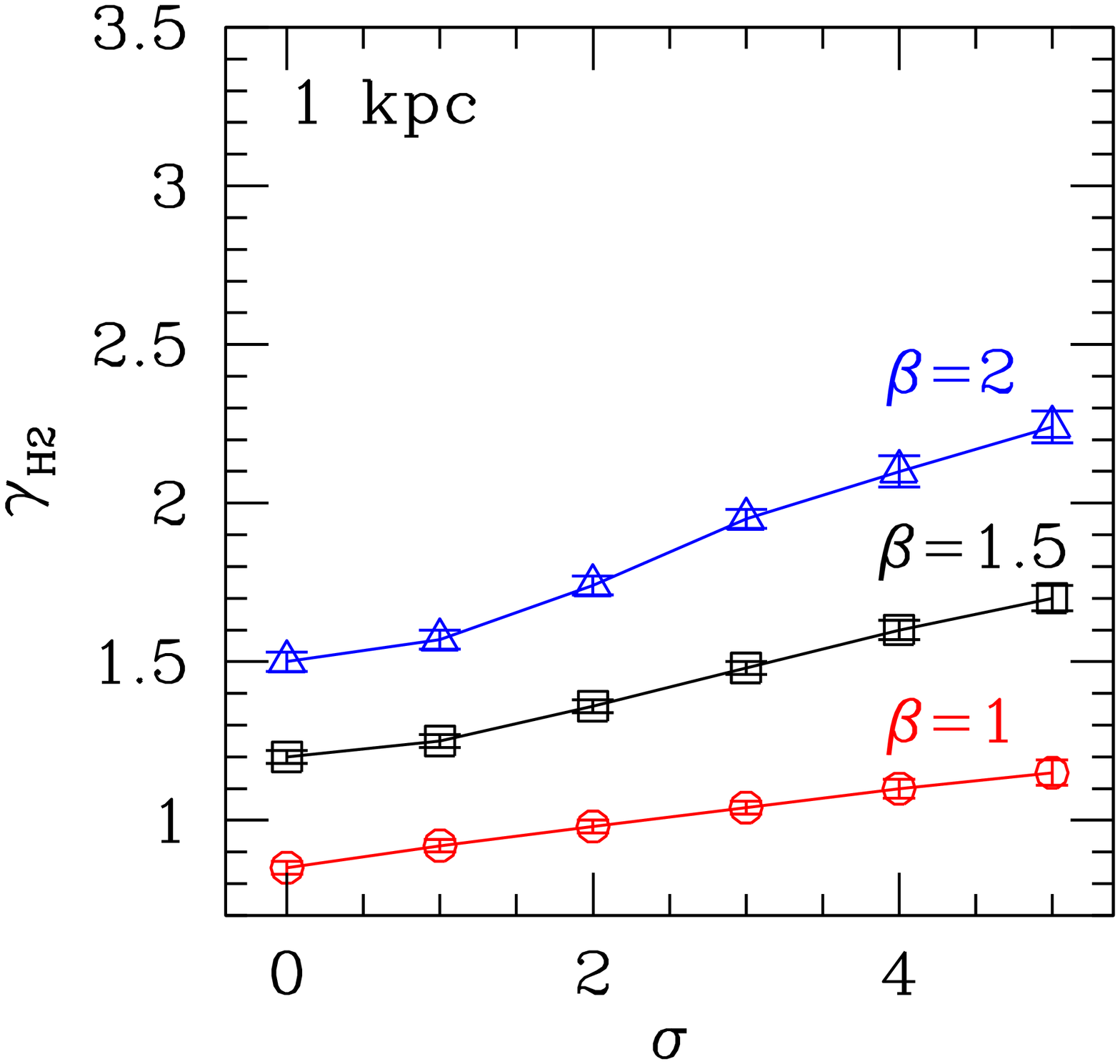}{0.01in}{0.}{215}{215}{0}{0}
\caption{The slope $\gamma_{H2}$ of the Observed SK Law for three representative region sizes (200, 500, 1000~pc), for 
data thresholds that vary between 0~$\sigma$ (no threshold; all data included) and 5~$\sigma$. The dynamical range of the data 
varies so that the highest threshold corresponds to the smallest dynamical range for the simulated data.  A pronounced increase 
 in $\gamma_{H2}$ for increasing threshold is observed at all region sizes, unlike Figure~\ref{fig8}. As in the previous figure, the 
largest variations are present for  $\beta$=2 ($\delta \gamma_{H2}/\gamma_{H2}\lesssim$45\%) and the smallest for $\beta$=1
 ($\delta \gamma_{H2}\lesssim$20\%). Note that the vertical scales of Figures~\ref{fig8} and \ref{fig9} are different. 
\label{fig9}}
\end{figure}

\clearpage 
\begin{figure}
\figurenum{10}
\plottwo{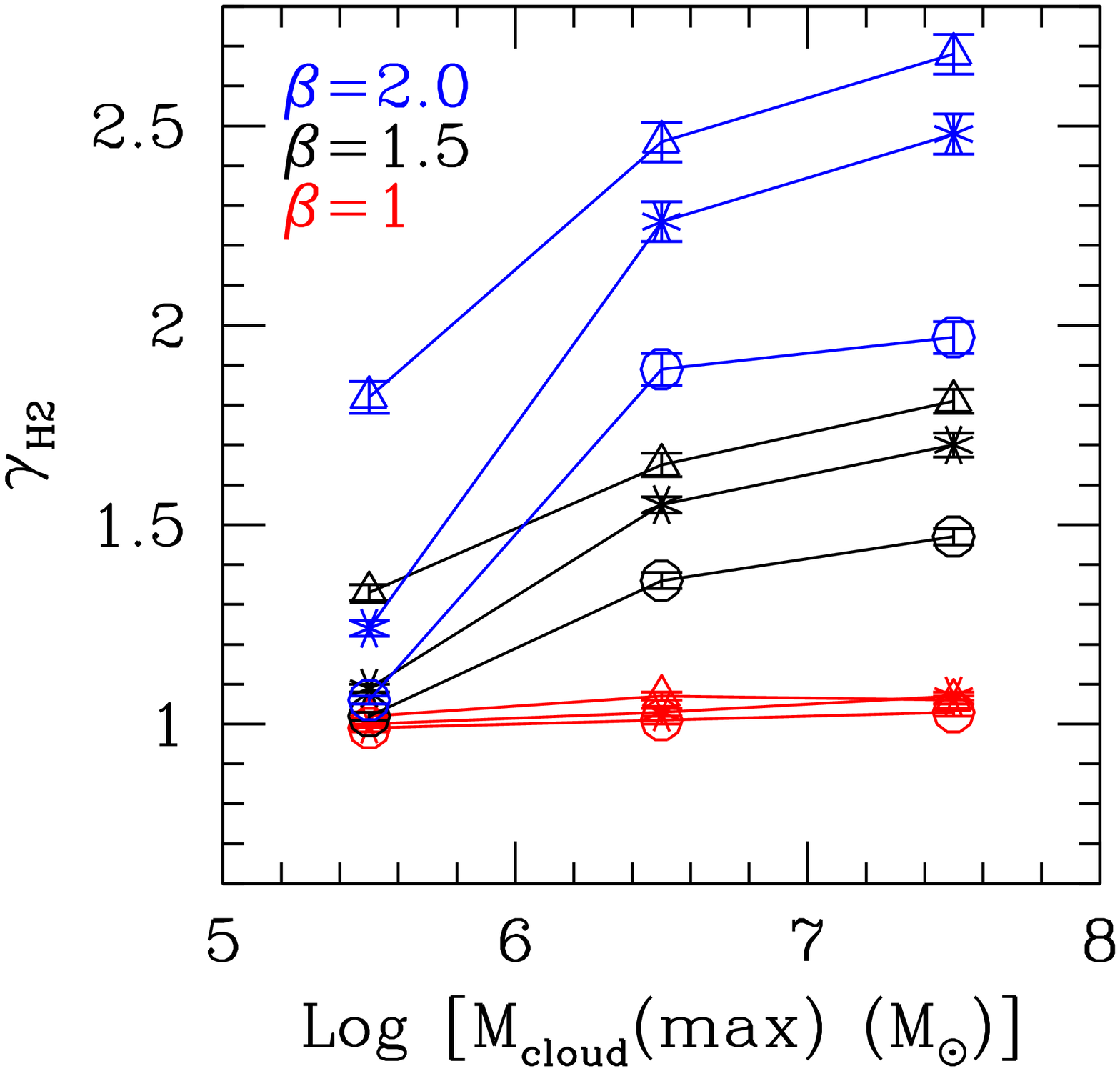}{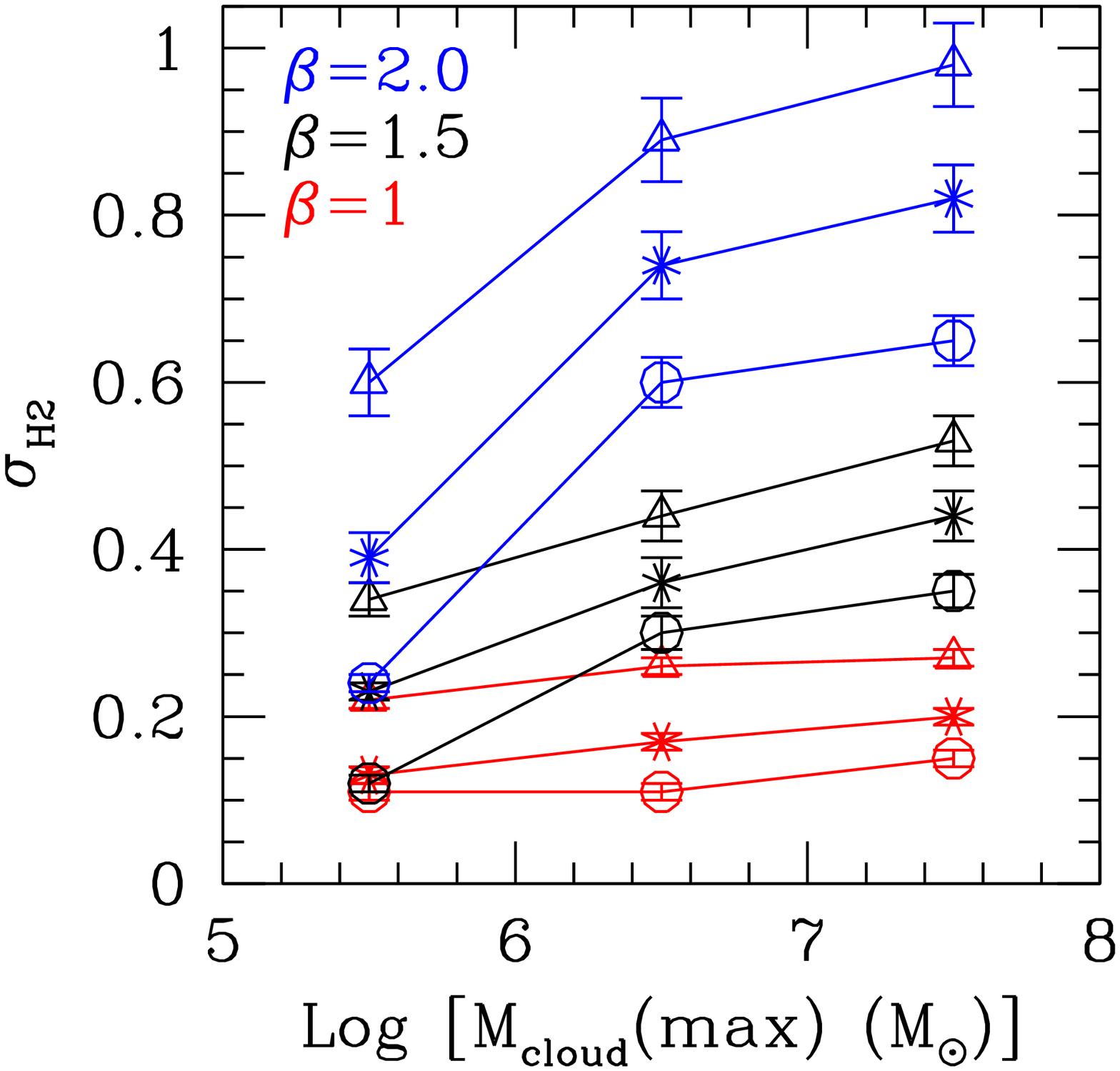}
\plottwo{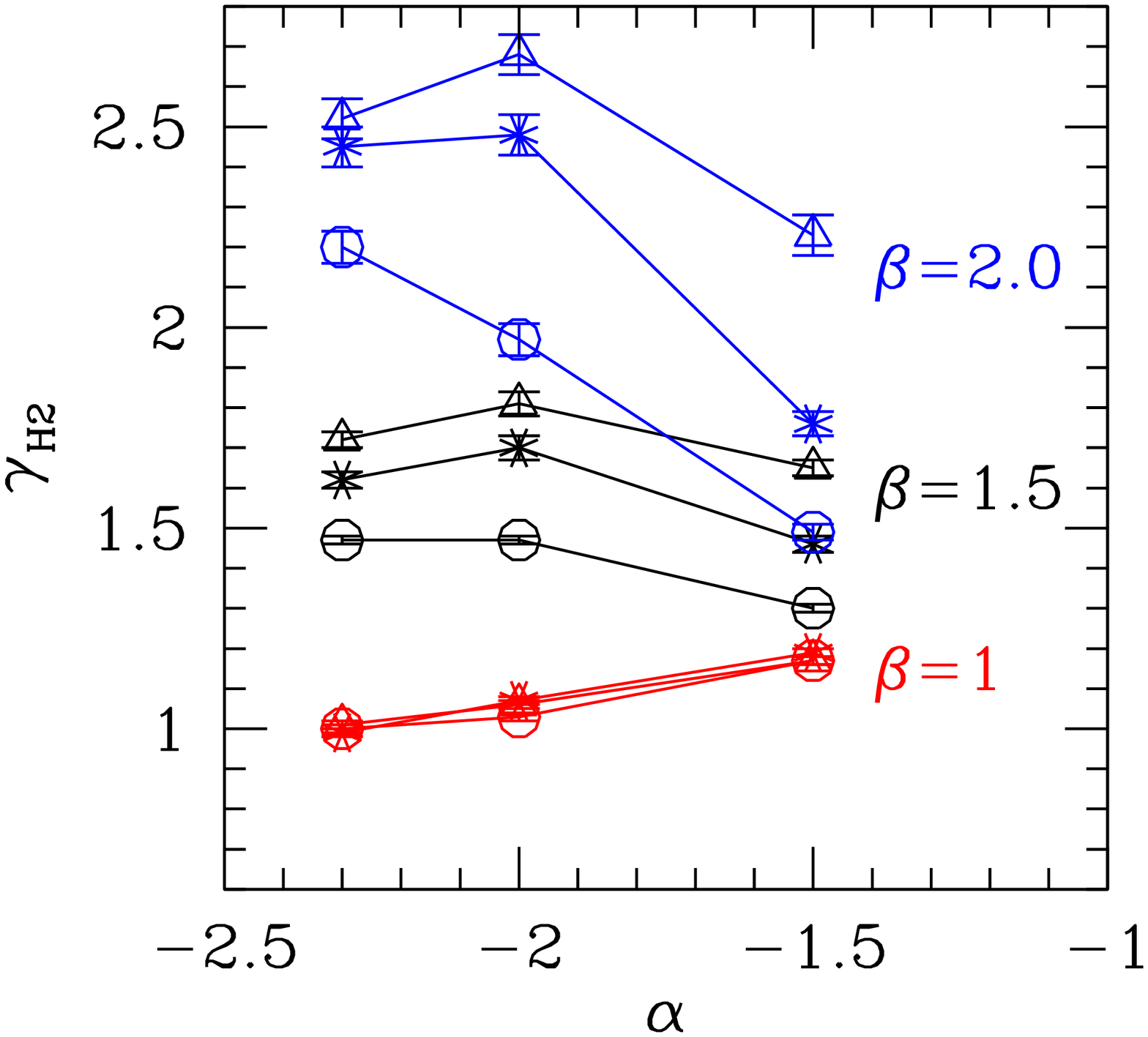}{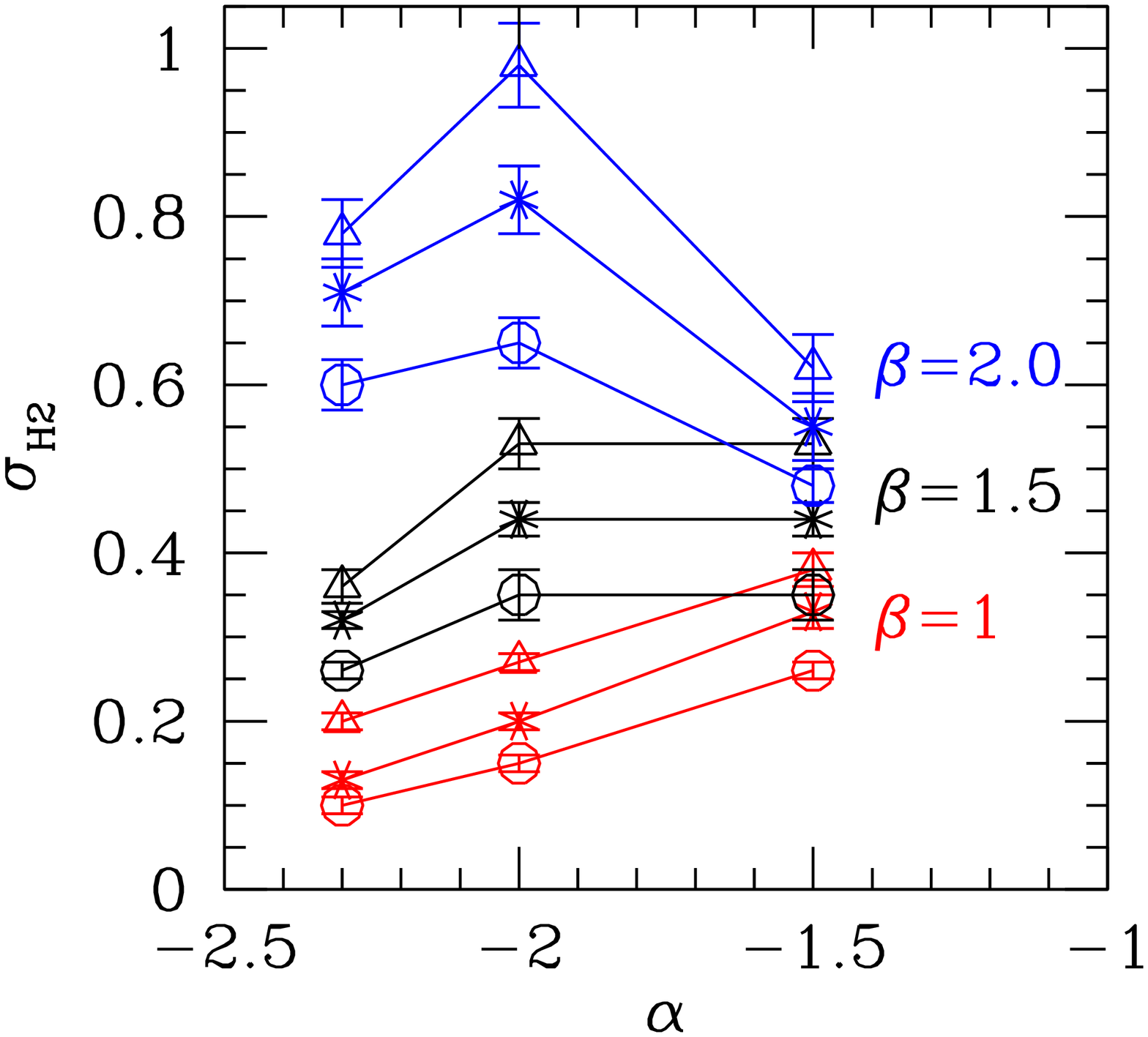}
\caption{The best fit slope $\gamma_{H2}$ (left panels)  and dispersion $\sigma_{H2}$ about the best fitting line (right panels)  of the 
Observed SK Law as a function of the maximum cloud mass M$_{cloud}$(max) (top panels) and of the 
cloud mass function power law index $\alpha$ (bottom panels; equation~3). The three values for the SFR--cloud~mass relation power index 
(equation~5) are shown: $\beta$=1.0 (red), $\beta$=1.5 (black), and $\beta$=2.0 (blue), for three representative region sizes,  
200~pc (triangles), 500~pc (asterisks), and 1~kpc (circles). 
The decreasing M$_{cloud}$(max) pushes $\gamma_{H2}$ closer to values of unity, which is what is expected when the cloud mass function is fully sampled 
in each region. A similar overall effect is observed for a flattening cloud mass function (increasing $\alpha$), when $\beta>$1. 
\label{fig10}}
\end{figure}

\clearpage 
\begin{figure}
\figurenum{11}
\plottwo{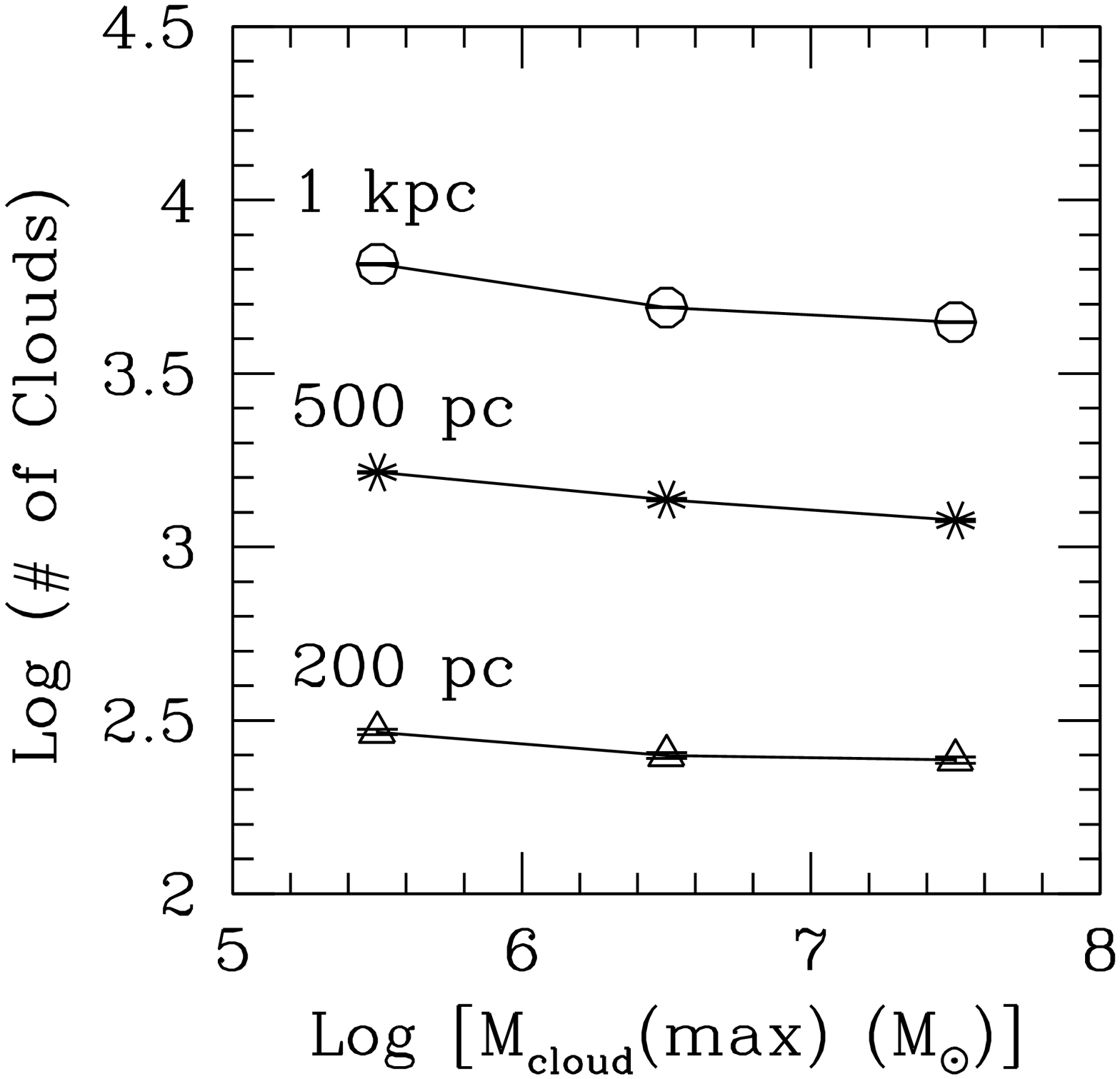}{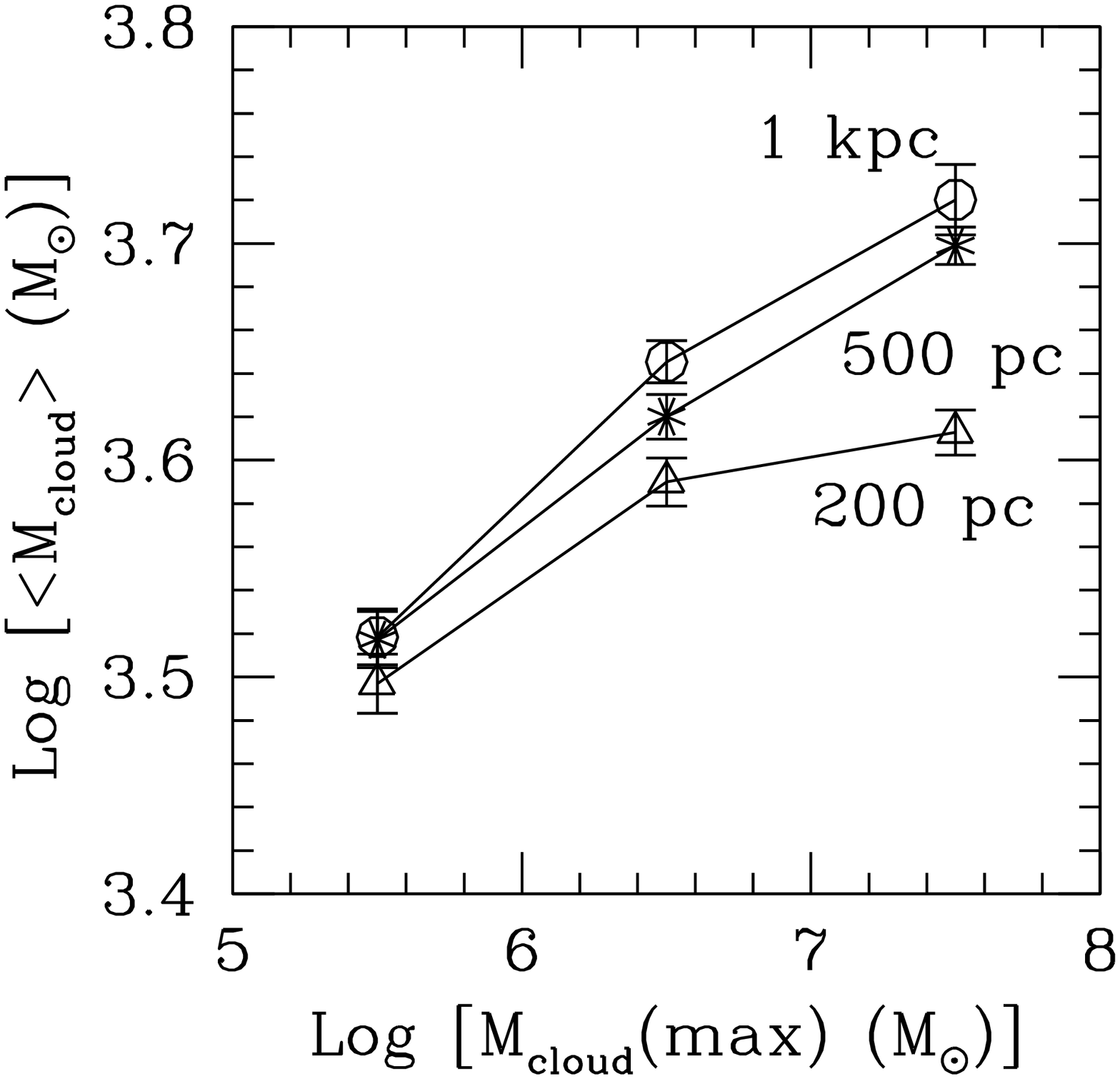}
\caption{The  mean number of clouds  (left) and the mean cloud mass (right) within a region, as a function of the maximum cloud mass M$_{cloud}$(max). 
Three representative region sizes are shown in each panel: 200~pc (triangles), 500~pc (asterisks), and 1~kpc (circles). 
Decreasing the maximum cloud mass decreases the mean cloud mass and 
increases the mean number of clouds in each region. 
\label{fig11}}
\end{figure}

\clearpage 
\begin{figure}
\figurenum{12}
\plottwo{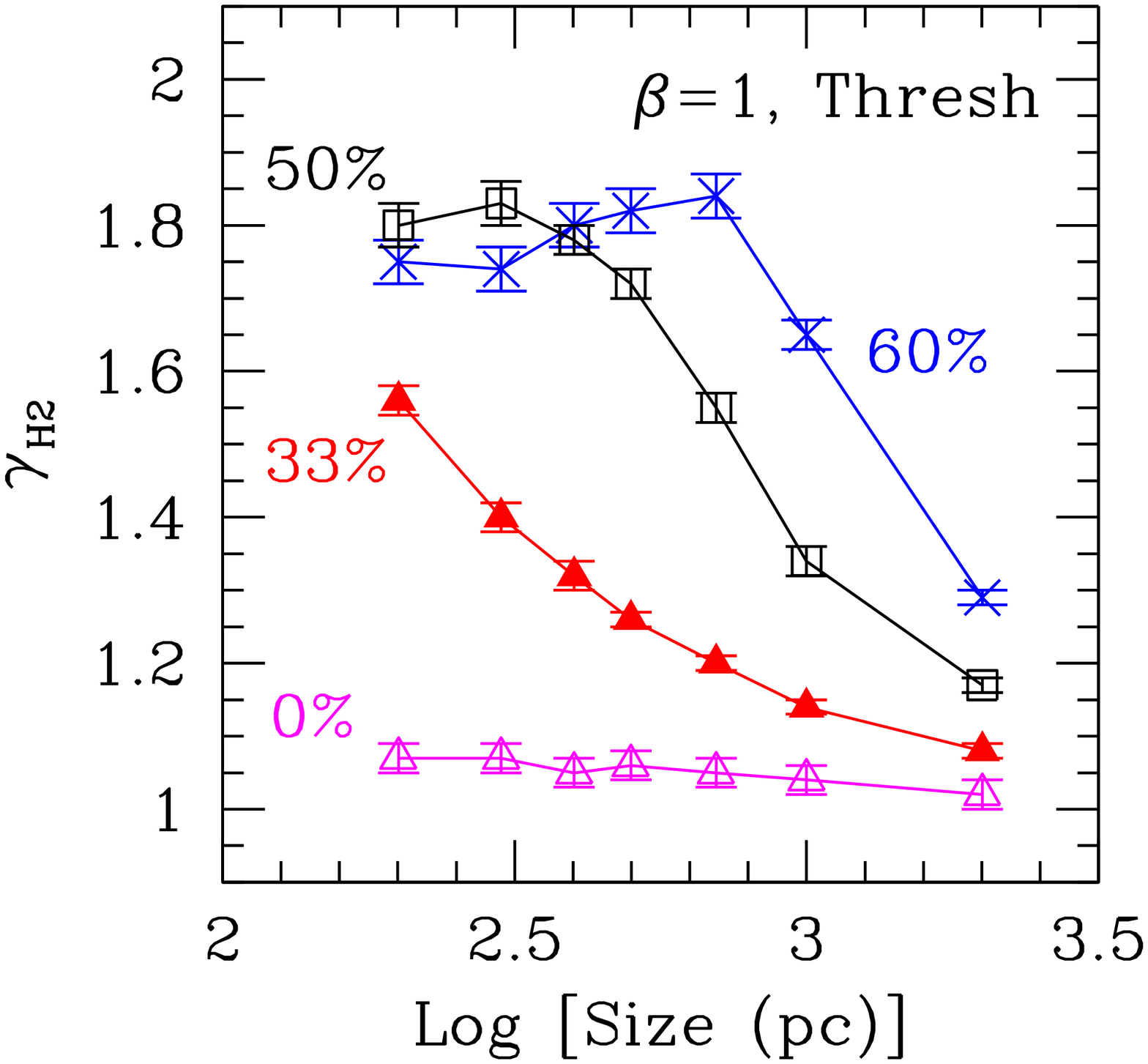}{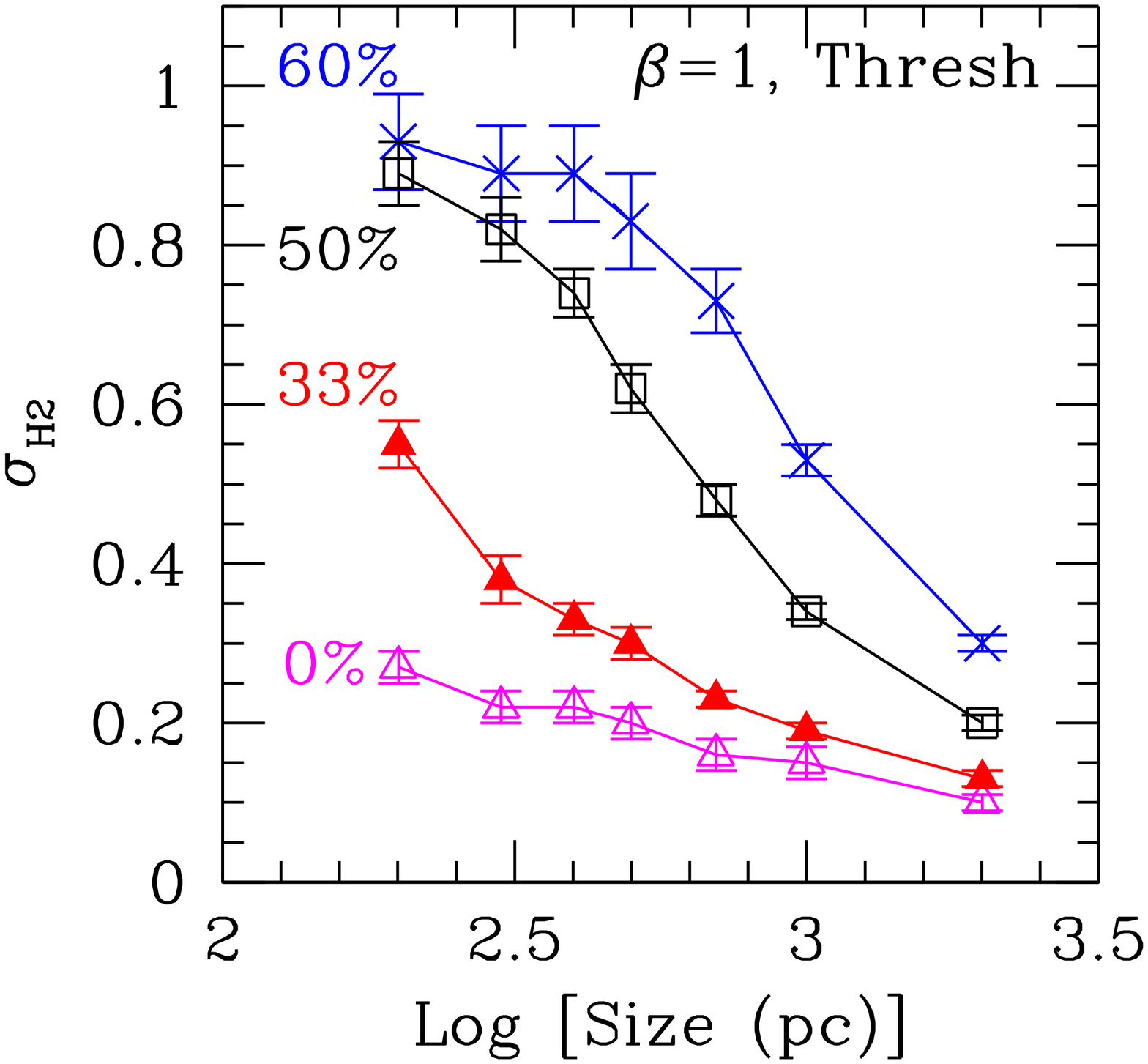}
\caption{
The best fit slope $\gamma_{H2}$ (left panel) and dispersion $\sigma_{H2}$ about the best fitting line (right panel) of the 
Observed SK Law (equation~2) as a function of the sampling region's size, in 
the range 200-2000~pc. The Default Model is used with the addition of a star formation threshold, an exponential distribution of the cloud covering factor 
(equation~6), and $\beta$=1.0. The case with no star formation threshold (from Figure~\ref{fig3}, bottom panels) 
is shown for comparison, with label `0\%' (magenta empty triangles).   
The other percentage numbers in the two panels refer to the fraction of the total SFR selectively removed on average from 
each region, by assigning SFR=0 to clouds with mass below a given M$_{cloud}$(thr). The three cases shown correspond to 1/3 (red filled triangles), 
1/2 (black empty squares), and 60\% (blue crosses) of the total 
star formation removed, and  threshold masses of 10$^{4.3}$~M$_{\odot}$, 10$^{5.1}$~M$_{\odot}$, and 10$^{5.6}$~M$_{\odot}$, respectively. 
\label{fig12}}
\end{figure}

\clearpage 
\begin{figure}
\figurenum{13}
\plottwo{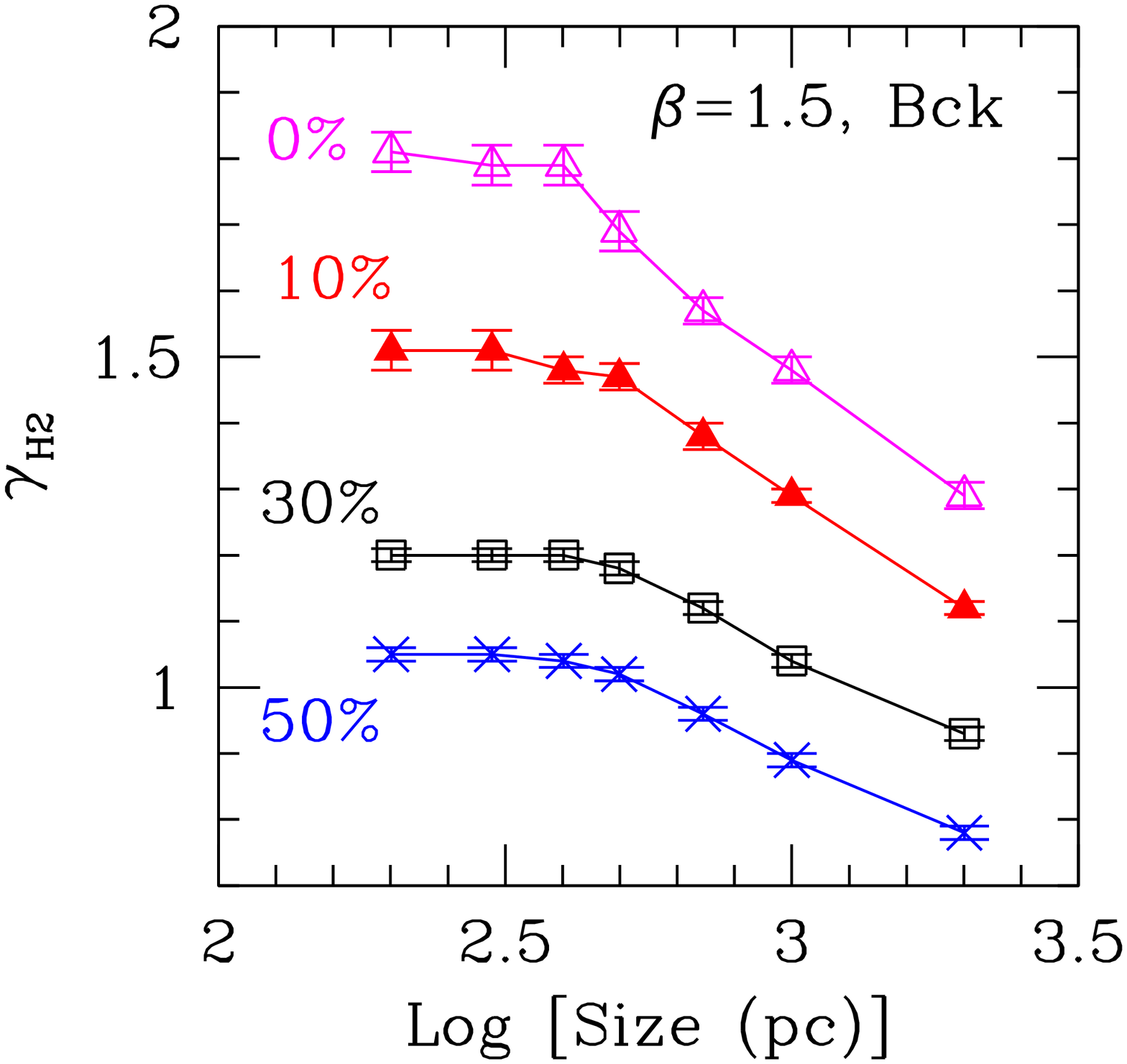}{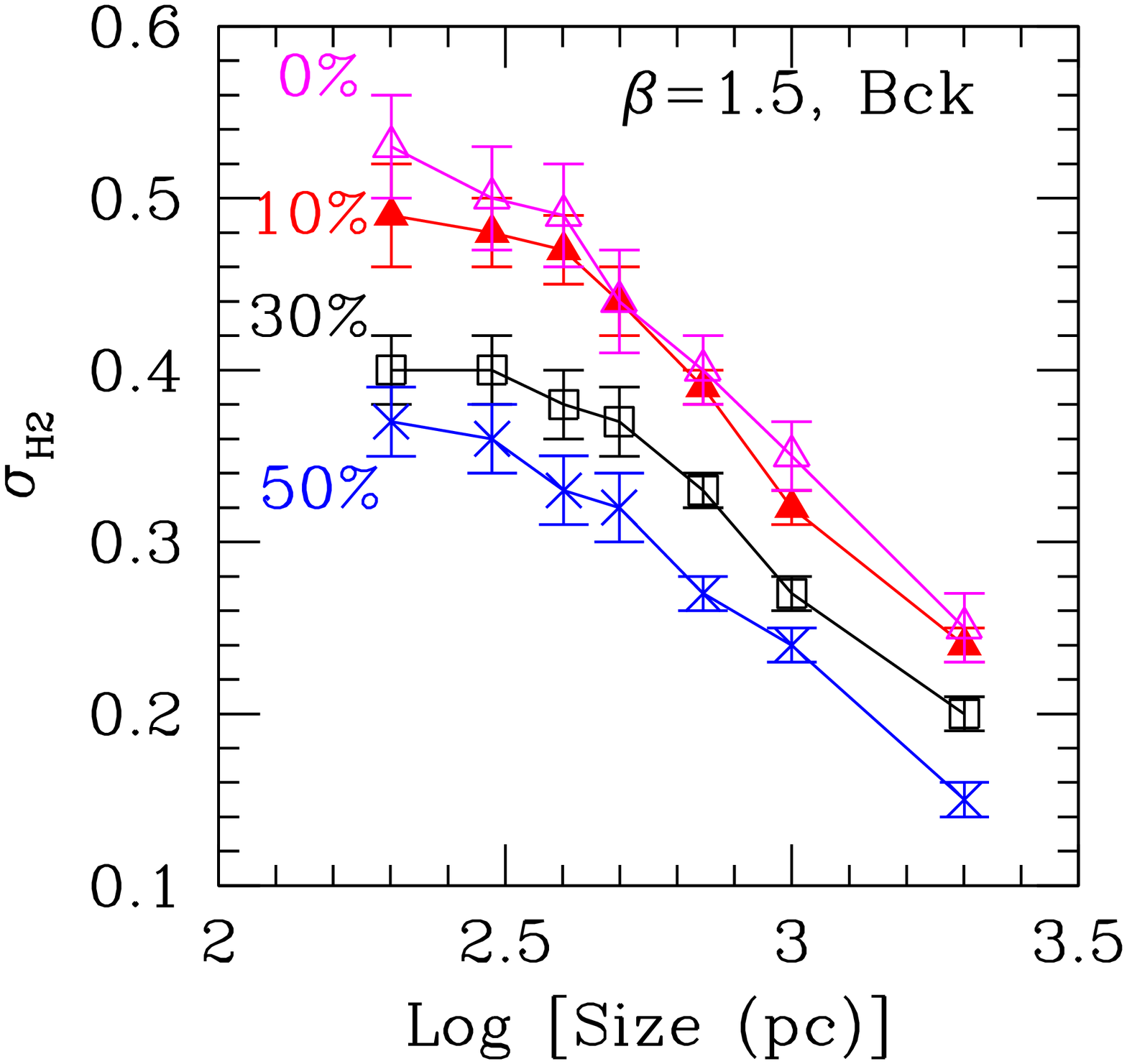}
\caption{ The best fit slope $\gamma_{H2}$ (left panel) and dispersion $\sigma_{H2}$ about the best fitting line (right panel) of the 
Observed SK Law (equation~2) as a function of the sampling region's size, in 
the range 200-2000~pc. The Default Model with the addition of a uniform background to the star formation,  an exponential distribution of the cloud covering factor 
(equation~6), and $\beta$=1.5 is used. The case with no background added to the star formation (from Figure~\ref{fig3}, bottom panels) 
is shown for comparison, with label `0\%' (magenta empty triangles). 
 The percentage numbers in the two panels refer to the fraction of the total SFR that is added into each region as a 
uniform background, which adds to the region's measured SFR surface density without contributing to the gas surface density.  
The three cases in which the added background corresponds to  
10\% (red filled triangles), 30\% (black empty squares), and 50\% (blue crosses) of the total SFR are shown. 
\label{fig13}}
\end{figure}

\clearpage 
\begin{figure}
\figurenum{14}
\plotone{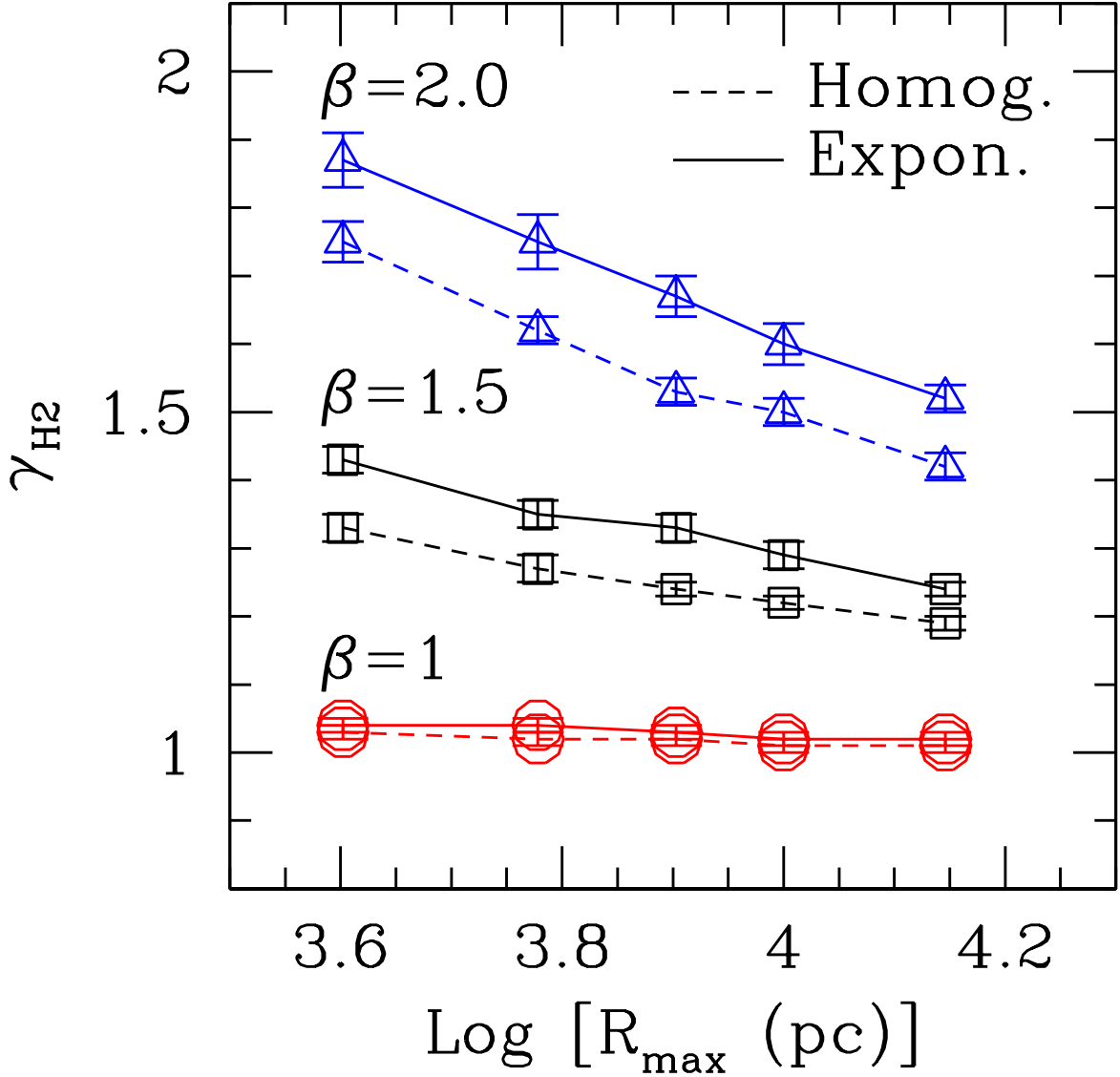}
\caption{The best fit slope $\gamma_{H2}$ of the Observed SK Law for the radial profile simulation. Both cases of a uniform (dash lines) and 
exponential (continuous lines) distribution of cloud covering factors are shown. The Default Model is 
modified to accommodate the increasing area of the annuli in azimuthally--averaged analyses. The measured values of $\gamma_{H2}$ 
are plotted as a function of the maximum radius, R$_{max}$, used. The radial step is, in all cases, 200~pc. 
\label{fig14}}
\end{figure}

\clearpage 
\begin{figure}
\figurenum{15}
\plottwo{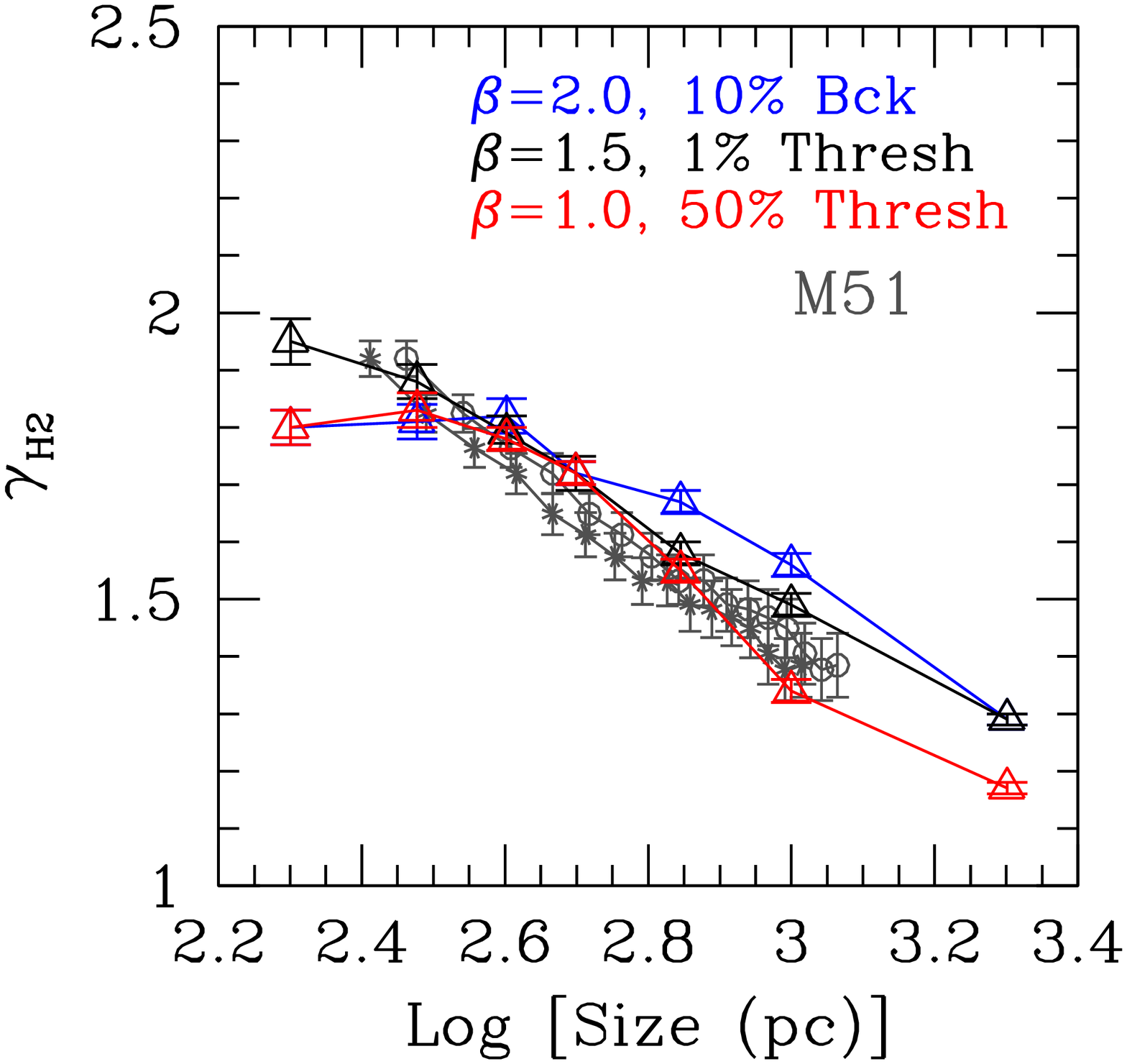}{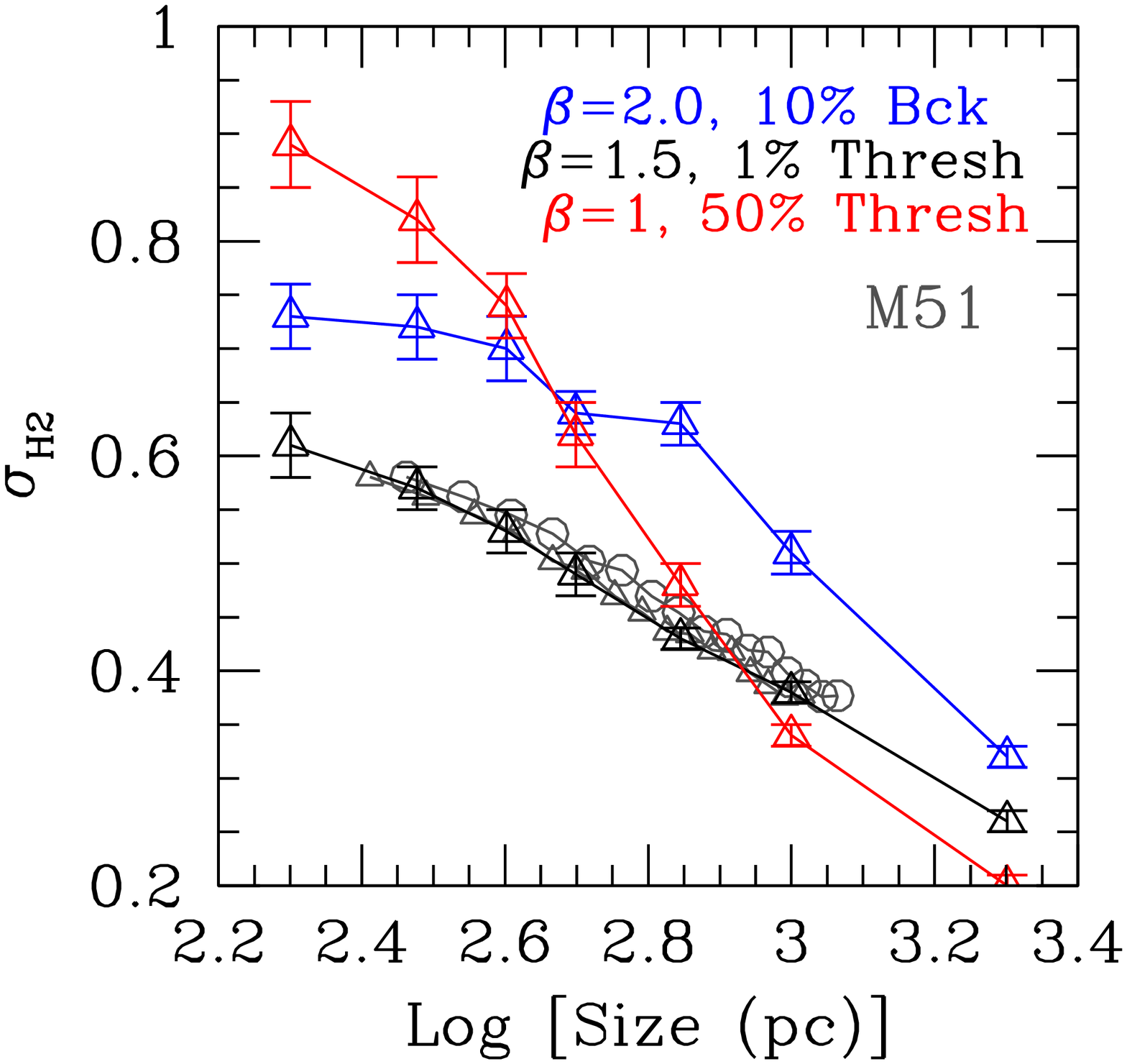}
\plottwo{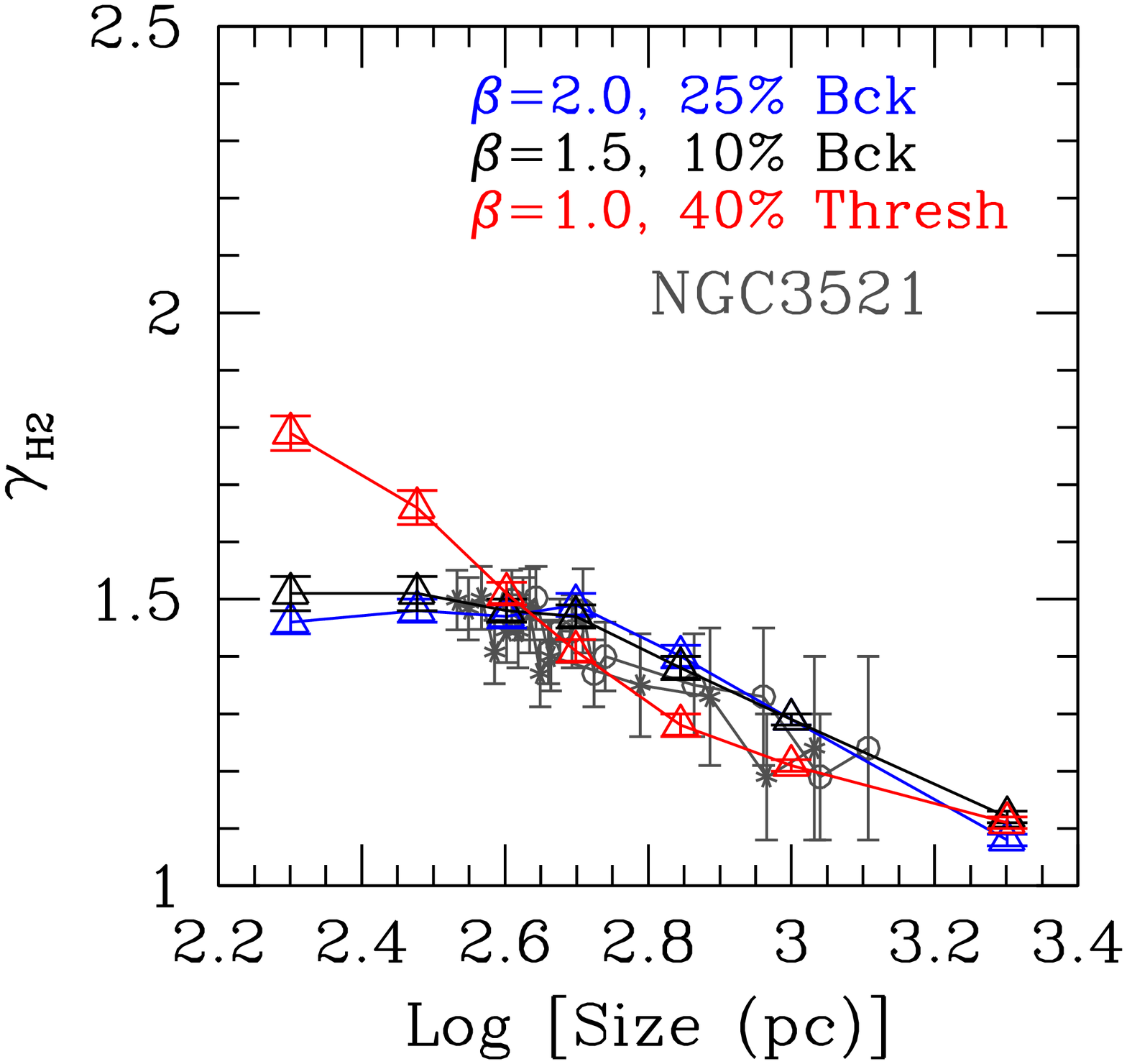}{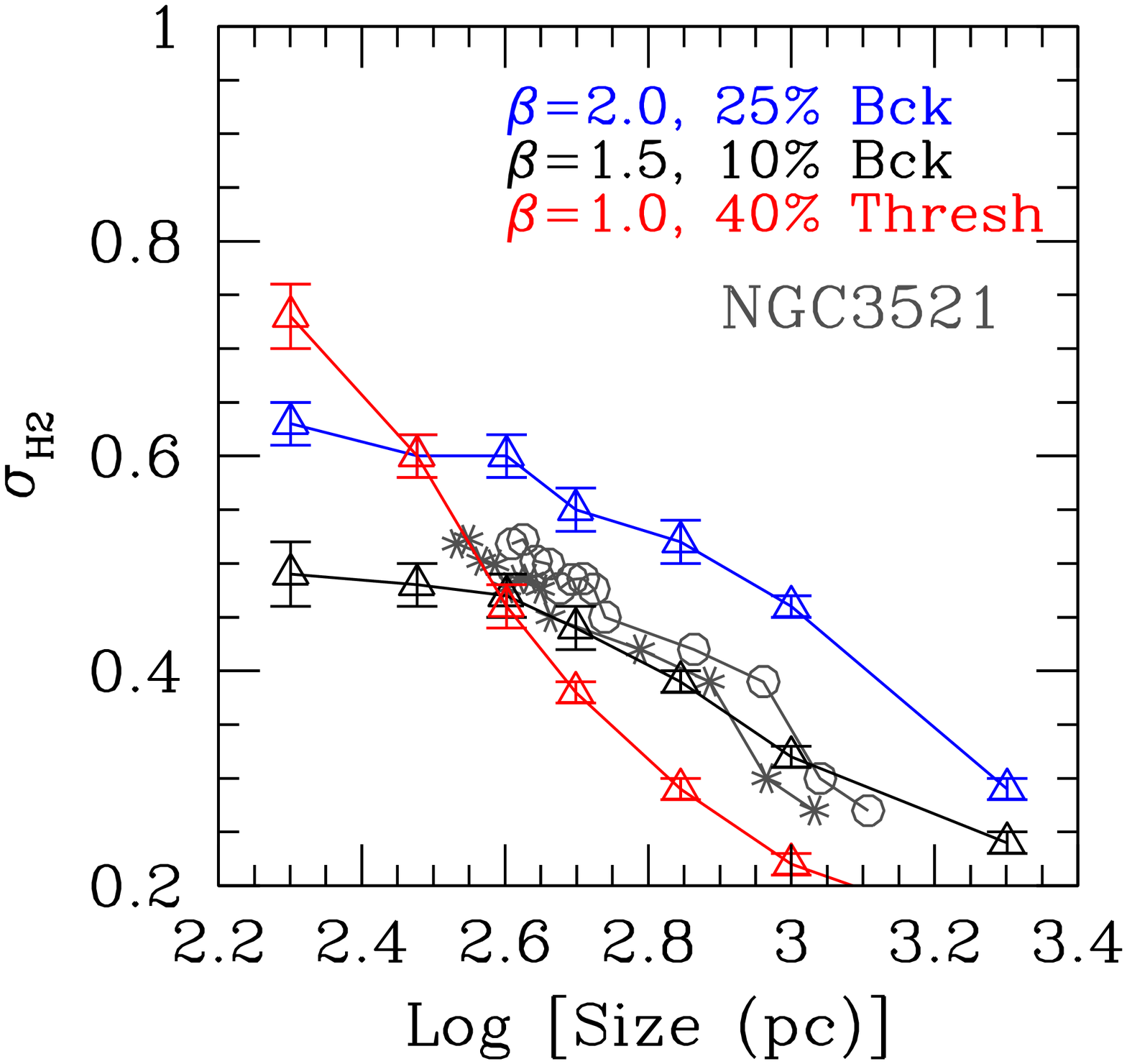}
\caption{The data for $\gamma_{H2}$ (left panels) and $\sigma_{H2}$ (right panels) for the two galaxies M51a (top panels) and NGC3521 (bottom panels) from \citet{Liu2011} are compared with our simulations, using the Default Model with an exponentially decreasing cloud covering factor. 
Two extreme values of the inclination are used: 
20$^{\circ}$--42$^{\circ}$ for M51a and 65$^{\circ}$--73$^{\circ}$ for NGC3521 (grey asterisks and empty circles, respectively). The observed  
linear sizes are the square root of  the de--projected areas in \citet{Liu2011}. For each value of $\beta$, the simulations that most closely approach the 
observed $\gamma_{H2}$ trends are reported. For $\beta$=2.0 (blue), a uniform background is added to the SFR. 
For $\beta$=1.5 (black), a small SFR threshold and a uniform background addition to the SFR are  required for M51a and NGC3521, 
respectively.  For $\beta$=1.0 (red), the largest SFR threshold compatible with the data is shown for both galaxies. 
\label{fig15}}
\end{figure}

\clearpage 
\begin{figure}
\figurenum{16}
\plottwo{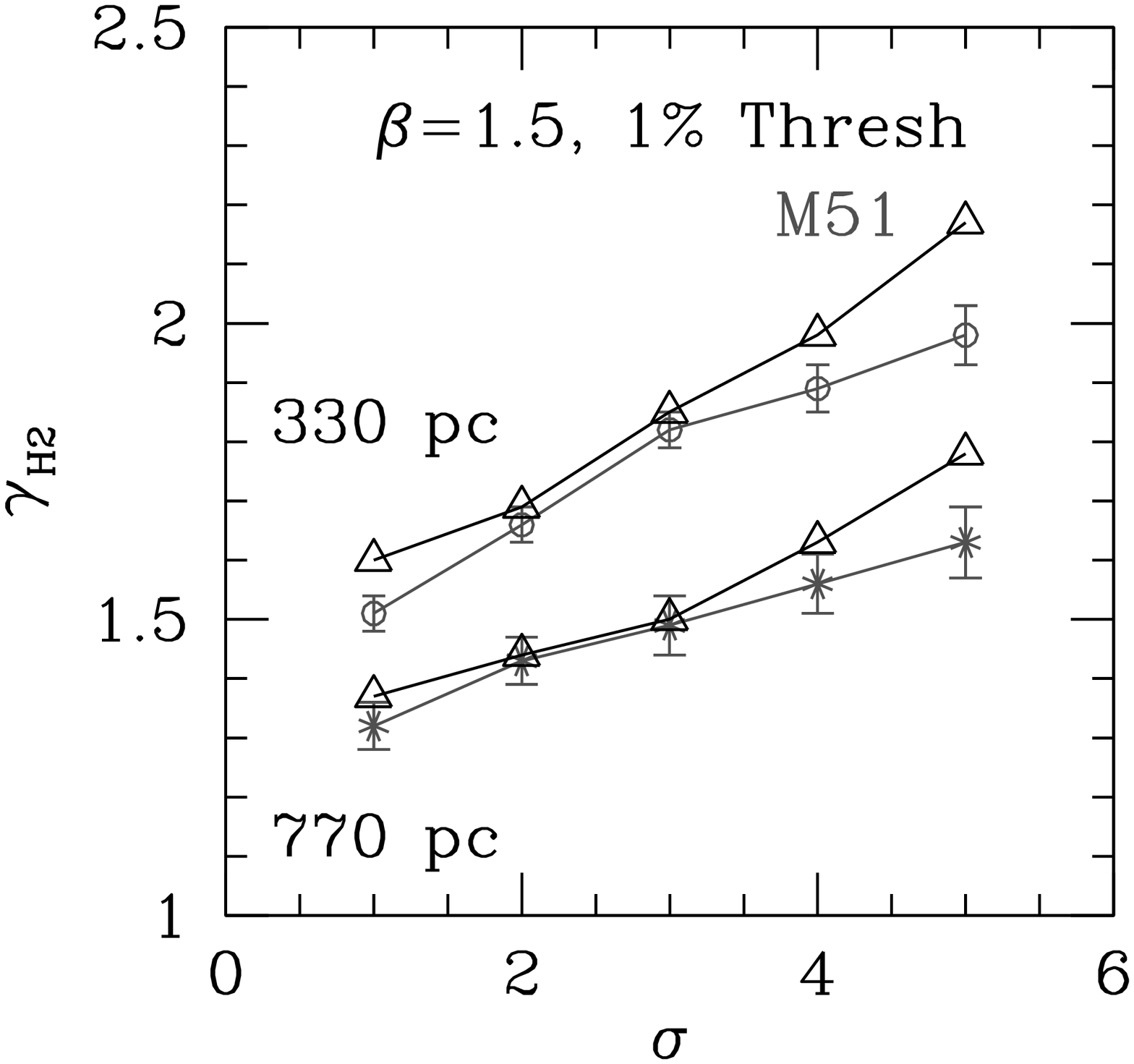}{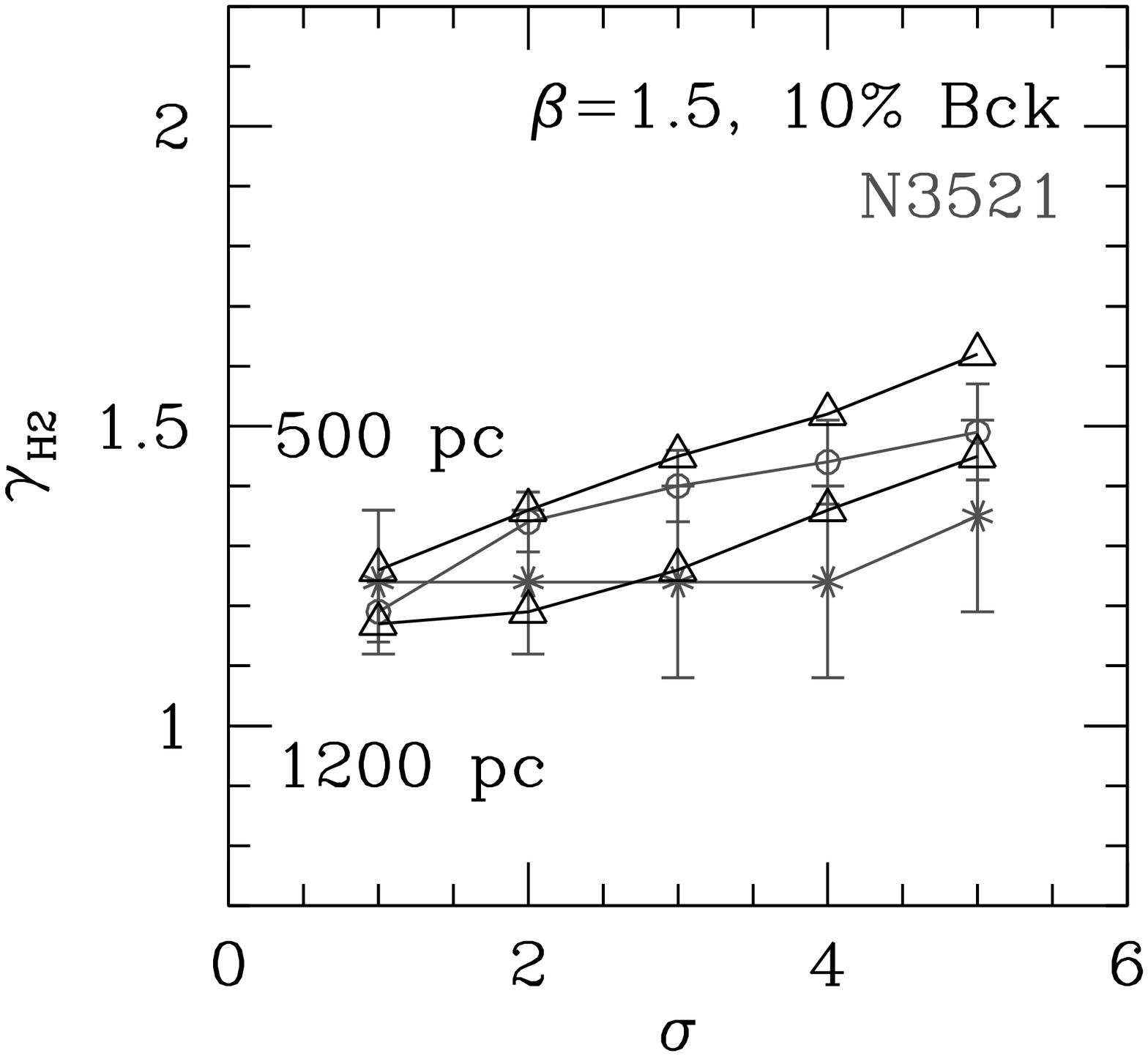}
\caption{Comparison between observed (grey) and model (black) values of $\gamma_{H2}$ for M51a (left panel) and NGC3521 (right 
panel) as a function of the detection limit of the data along $\Sigma_{H2}$, in the range 
1~$\sigma$--5~$\sigma$. The increasing limit corresponds to a decrease in the dynamical range of the data along the molecular gas axis. 
The observational data are from Tables~3 and 4 of \citet{Liu2011}, at the two projected sizes of 300 and 700~pc. These correspond to de--projected 
sizes $\sim$330~pc and $\sim$770~pc for M51a and $\sim$500~pc and $\sim$1,200~pc for NGC3521. The small size is shown with circles and the large size with asterisks. The Default Model is used with an exponentially decreasing cloud covering factor and $\beta$=1.5. 
For M51a, the model's trend is generally consistent  with the observational data, although the values of $\gamma_{H2}$ are over-predicted  at the highest 
detection limits, due to the increased sensitivity of the fits to the decreasing dynamical range. For NGC3521, there is a general agreement between models and observations within the 1.5~$\sigma$ error of the data, but uncertainties 
are large for this galaxy.  A slightly better agreement is attained for both galaxies between observations and models when a uniform distribution of cloud covering 
factors is used in the Default Model.
\label{fig16}}
\end{figure}

\end{document}